%% file: ms.tex
\documentclass[twocolumn]{aastex63} 

\usepackage{graphicx}
\usepackage{xcolor}

\accepted{to AJ on \today}
\newcommand{\nstars}{23 }

\begin{document}

\title{Determining Which Binary Component Hosts the TESS Transiting Planet}

\author[0000-0002-9903-9911]{Kathryn~V.~Lester}
\affil{NASA Ames Research Center, Moffett Field, CA 94035, USA}

\author[0000-0002-2532-2853]{Steve~B.~Howell}
\affil{NASA Ames Research Center, Moffett Field, CA 94035, USA}

\author[0000-0002-5741-3047]{David~R.~Ciardi}
\affiliation{NASA Exoplanet Science Institute, Caltech/IPAC, Pasadena, CA 91125, USA}

\author[0000-0001-7233-7508]{Rachel~A.~Matson}
\affiliation{U.S. Naval Observatory, Washington, D.C. 20392, USA}
	
\correspondingauthor{Kathryn Lester}
\email{kathryn.v.lester@nasa.gov}

\begin{abstract}
 
The NASA TESS mission has discovered many transiting planets orbiting bright nearby stars, and high-resolution imaging studies have revealed that a number of these exoplanet hosts reside in binary or multiple star systems. In such systems, transit observations alone cannot determine which star in the binary system actually hosts the orbiting planet. The knowledge of which star the planet orbits is necessary for determining accurate physical properties for the planet, especially its true radius and mean bulk density. We derived the mean stellar densities for the components of \nstars binary systems using the light curve transit shape and the binary flux ratio from speckle imaging, then tested the consistency with stellar models to determine which component is the host star. We found that 70\% of the TESS transiting planets in our sample orbit the primary star.
\\

\end{abstract}


\section{Introduction} 

The NASA TESS mission has discovered thousands of transiting planets orbiting bright nearby stars \citep{tess, TOIcatalog}, and many more planet candidates will be found as the TESS Extended Mission continues. Follow-up high resolution imaging studies have revealed that a number of these exoplanet host stars reside in binary or multiple star systems \citep[e.g.,][]{furlan17, matson18, howell21, lester21}. In these systems, transit observations alone cannot determine which binary component star actually hosts the orbiting planet. 

Identifying the correct host component is essential for accurate planet properties. Correcting the transit depth for dilution from a companion greatly affects the planet's derived radius and thus the mean density and composition \citep{ciardi15, furlan17b, hirsch17}. Dilution from a bright companion star also affects the host star properties determined spectroscopically \citep{furlan20}, such as the stellar radius used to calculate the planet radii. Lastly, binary components of different effective temperatures have different habitable zones (HZ), so a transiting planet may lie within the HZ if orbiting the primary star but outside the HZ if orbiting the secondary star. Therefore, studies of exoplanet demographics and Earth-like, habitable planets must take into account whether each planet hosts resides in a multistar system \citep{saval20}, and if so, determine the binary properties and which component actually hosts the planet.

\citet{seager03} derived a method to determine the stellar density of a planet host star using the transit geometry, which can be extended to planet hosts in multistar systems if the transit depth is corrected for the flux contribution of each component. \citet{payne18} used this method to identify the planet host star in 8 K2 binary systems and found that the brighter primary star was more likely to host the planet in 5 out of 8 systems. They also found that systems with fainter companions were more likely to have primary host stars. 

In this work, we extend their study to \nstars binary systems found to host transiting planets by TESS. We introduce our sample in Section~\ref{sample}, then describe the transit modeling and stellar density calculations in Section~\ref{sect3}. We identify the planet host components and present our results in Section~\ref{results}, and discuss the implications of our study in Section~\ref{disc}.

\break

\section{Sample \label{sample}} 

\subsection{Binary TOI's} 
We compiled a list of over 300 TOI exoplanet host candidates found to be binary systems through speckle interferometry at Gemini North \& South \citep{lester21}, SOAR \citep{ziegler20, ziegler21}, and WIYN \citep{howell21}. Systems suitable for this stellar density analysis must pass several requirements by having:
\begin{itemize}
\setlength\itemsep{0em}
\item binary separation $r < 1.2$", so the companions are well characterized by speckle imaging
\item at least two consecutive transits, so the orbital period can be constrained \citep{seager03}
\item flat bottom transits, so the contact times can be measured \citep{seager03}
\item transits with sufficient signal-to-noise ratio (SNR $> 3$) to model
\item planet candidate or confirmed planet status\footnote{as of December 2021 on https://exofop.ipac.caltech.edu/tess/} on the Exoplanet Follow-up Observing Program (ExoFOP) website to avoid background eclipsing binaries. 
\end{itemize}
Of the 300+ TOI binaries evaluated, 106 systems had angular separations less than 1.2", and \nstars systems passed all of the requirements for this analysis. Our final sample is listed in Table~\ref{atmospar} and includes 20 binary systems and 3 triple systems. Most of the rejected systems had V-shaped (grazing) transits or were found to be false positives through follow-up photometry by the community. Planets around M-type hosts stars may also create V-shaped transits due to different limb darkening values than Solar type stars, but we had very few M-type binaries in our sample to begin with.

\subsection{Stellar Parameters\label{params}} 
We adopted the effective temperatures ($T_{\rm eff}$) and uncertainties of the primary components listed on ExoFOP from either the TESS Input Catalog \citep[TIC,][]{tic81} or a spectroscopic analysis (when available). We then used magnitude difference ($\Delta m$) between the binary components from speckle imaging to calculate the effective temperatures of the secondary and tertiary components using a polynomial fit to the Modern Mean Dwarf Stellar Color and Effective Temperature Sequence \citep{pecaut13} as described in \citet{matson19}.
For each primary star, we calculated the $I$-band magnitude differences of all stars with cooler effective temperatures, fit a seventh order polynomial to the these values as a function of $T_{\rm eff}$, and interpolated the $T_{\rm eff}$ of the companions at the observed $\Delta m$ values. We estimated the uncertainties in $T_{\rm eff}$ of the secondary or tertiary stars based on the polynomial fit and the uncertainty in $\Delta m$, then combined this value in quadrature with the uncertainties in the primary star's $T_{\rm eff}$.
We adopted $\Delta m$ uncertainties of 0.2~mag \citep{howell21} for all systems, unless a specific value was listed on ExoFOP. Metallicity measurements were not available for most systems, so we adopted solar values where needed. The stellar properties of our binary TOI's are listed in Table~\ref{atmospar}, with the TIC number, the effective temperature of each component, the metallicity, the binary angular separation(s), the magnitude difference(s) from speckle imaging, and the corresponding references. 

\input{table_atmospar.txt}

\section{Stellar Densities \label{sect3}} 
\subsection{TESS Photometry} 
\begin{figure*}
\centering
\includegraphics[width=\textwidth]{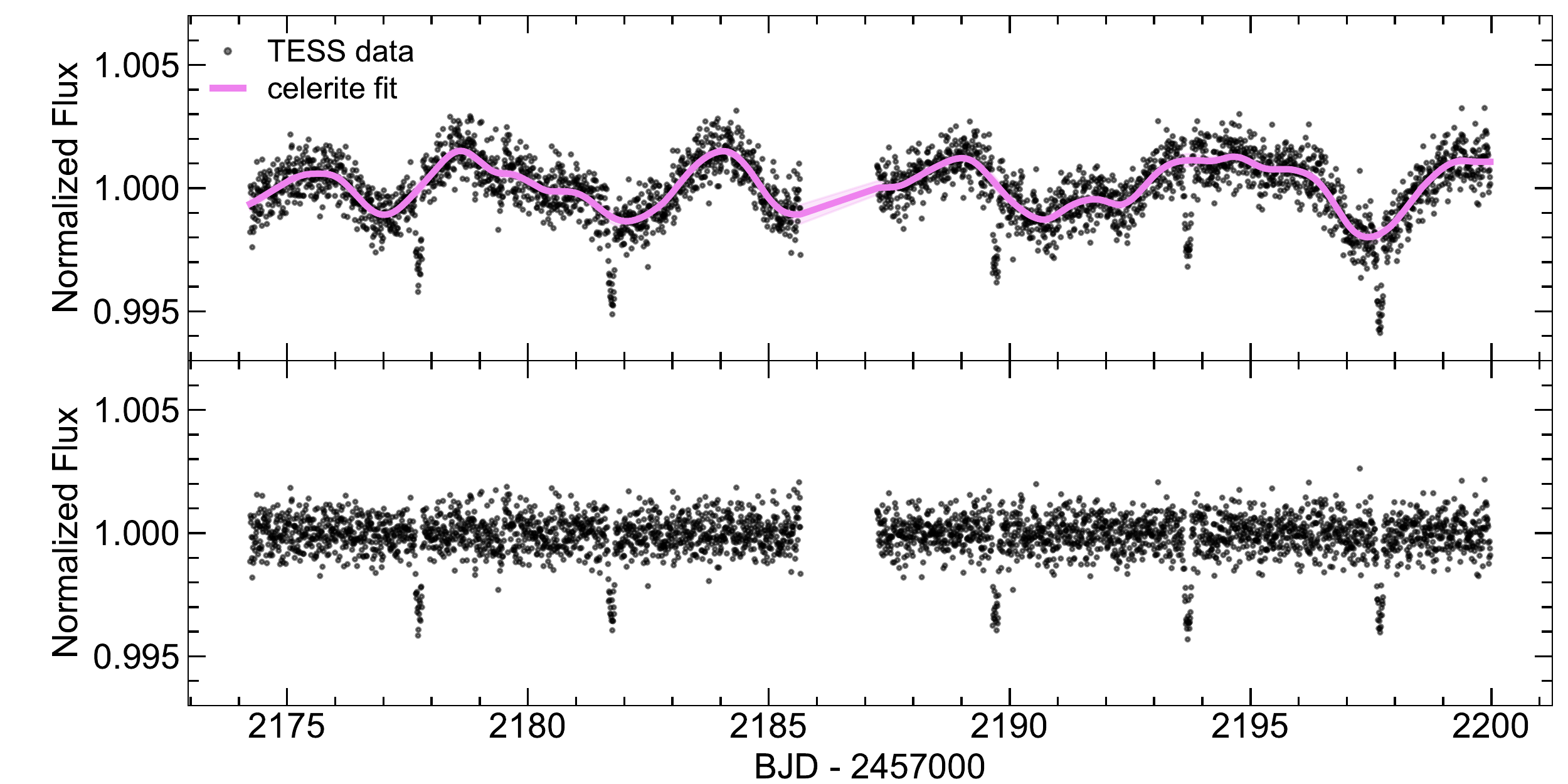}
\caption{Removal of stellar variability signal for TOI 322. The top panel shows the normalized TESS photometry from sector 32 (black points) and the \texttt{celerite} fit to the out-of-transit variability (pink line). The bottom panel shows the flattened, transit light curve on the same scale.
\label{celerite}}
\end{figure*}

We downloaded the 2~minute Pre-Search Data Conditioning Simple Aperture Photometry \citep[PDCSAP][]{jenkins16} photometry for each TOI from the Mikulski Archive for Space Telescopes (MAST), removed the data with poor quality flags, then normalized each light curve by dividing by the mean count value. Several systems (TOI 159, 309, 322, 487, 697, 1307, 1740) showed large scale stellar variability that would hinder transit fitting. We adopted the preliminary ephemerides from ExoFOP to mask out the planet transits for each system, then used the \texttt{celerite}\footnote{\href{https://github.com/exoplanet-dev/celerite2}{https://github.com/exoplanet-dev/celerite2}} Gaussian processes code \citep{celerite1, celerite2} to fit and remove the stellar variability. Figure~\ref{celerite} shows an example of the \texttt{celerite} fit and flattened light curve for TOI 322. Finally, we binned each light curve into 10 minute intervals and kept only the data points within $10-12$ hours of mid-transit in order to save on computation time.

\subsection{Transit Modeling} 
We fit the transit geometry using the \texttt{EXOFAST} \citep{exofast} online tool\footnote{\href{https://exoplanetarchive.ipac.caltech.edu/cgi-bin/ExoFAST/nph-exofast}{https://exoplanetarchive.ipac.caltech.edu/cgi-bin/ExoFAST/nph-exofast}} from the NASA Exoplanet Science Institute (NExSci). We adopted the transit solutions listed on ExoFOP as initial parameters and ran the MCMC modeling mode to produce robust best-fit transit parameters and uncertainties. We held the orbital eccentricity fixed to zero for all systems. The online tool does not have a TESS bandpass, so we used the $I$-band option as it has the most similar central wavelength. \texttt{EXOFAST} also incorporates limb darkening based on the stellar effective temperatures from Section~\ref{params}. 

\begin{figure*}
\includegraphics[width=0.329\textwidth]{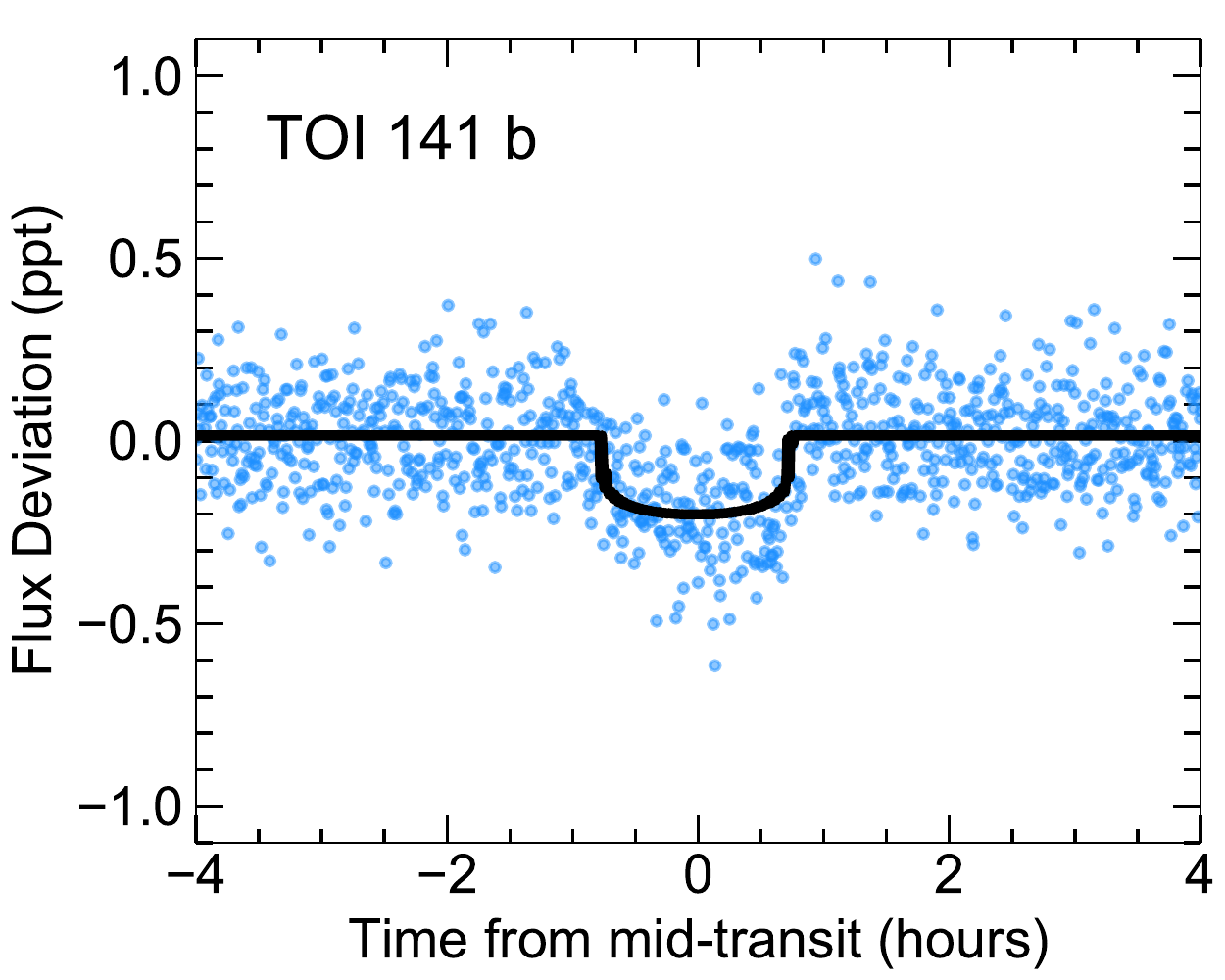}
\includegraphics[width=0.329\textwidth]{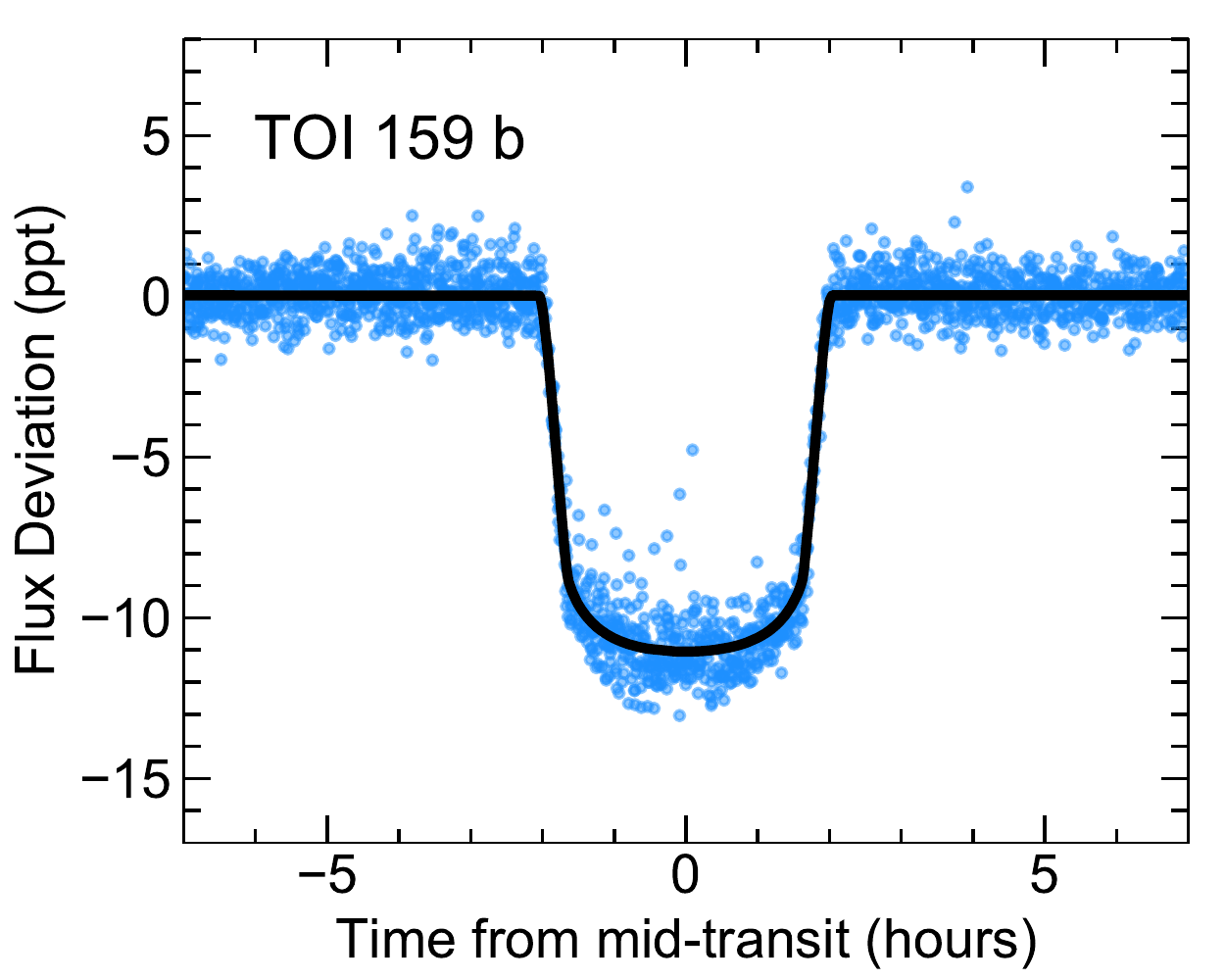}
\includegraphics[width=0.329\textwidth]{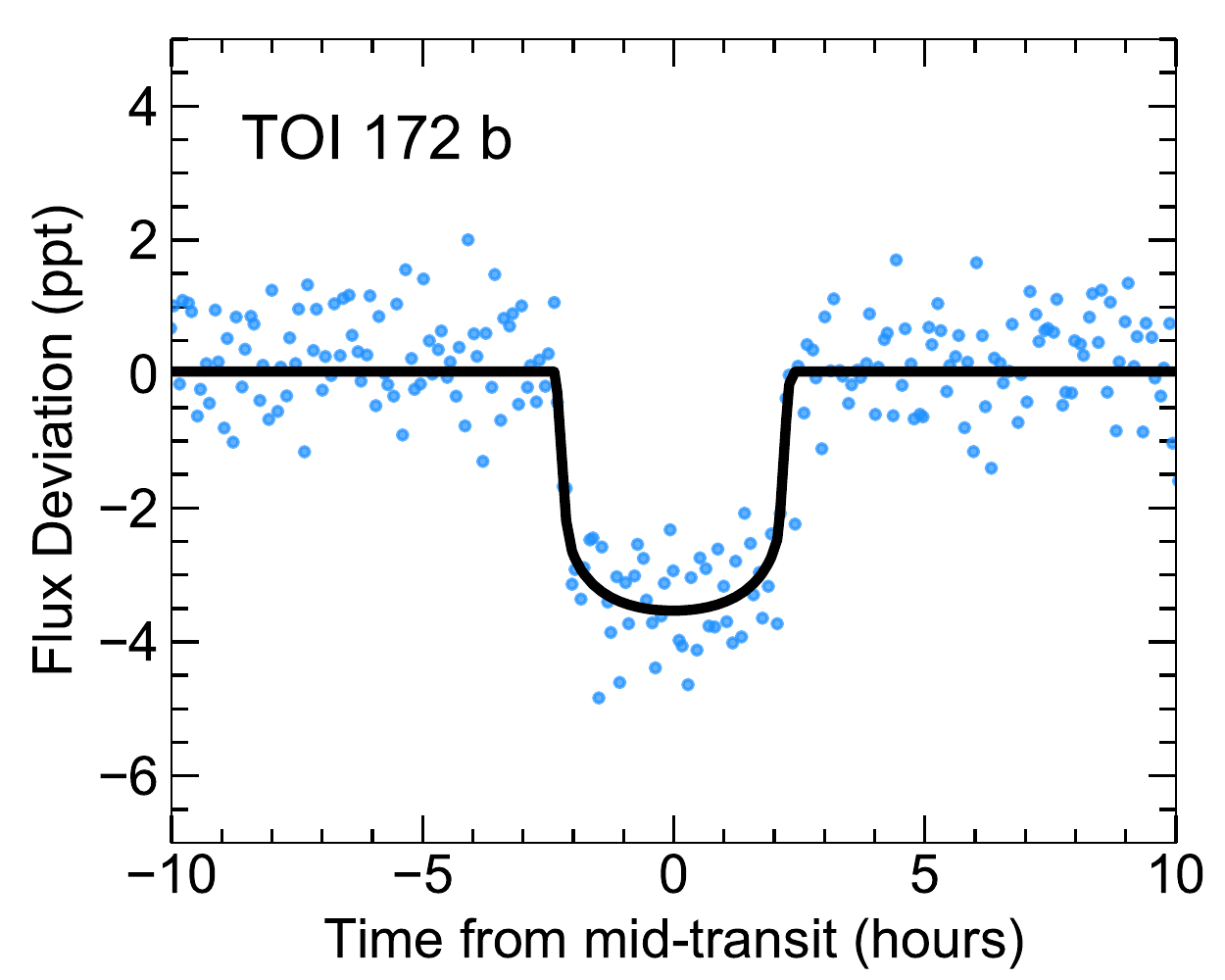}
\includegraphics[width=0.329\textwidth]{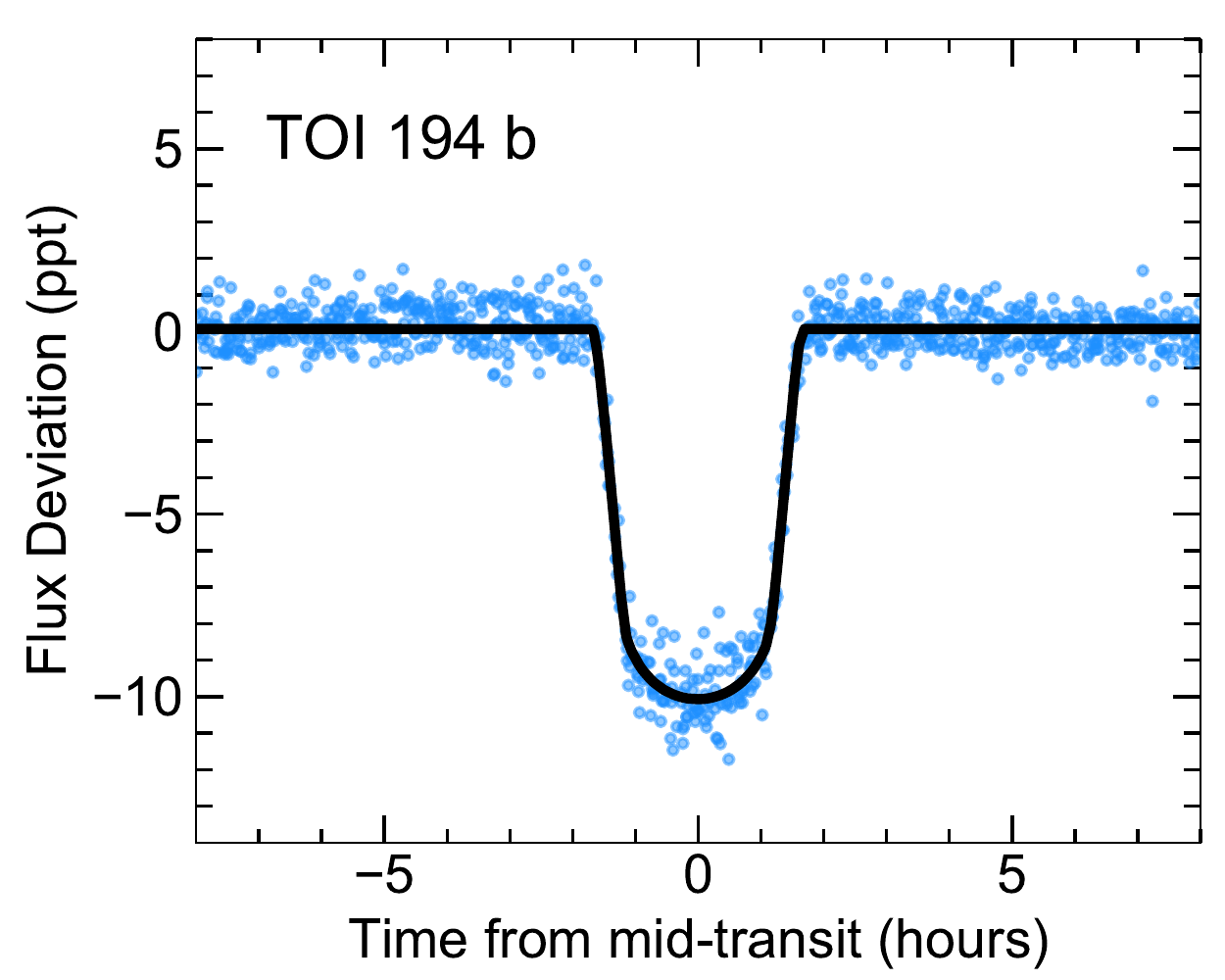}
\includegraphics[width=0.329\textwidth]{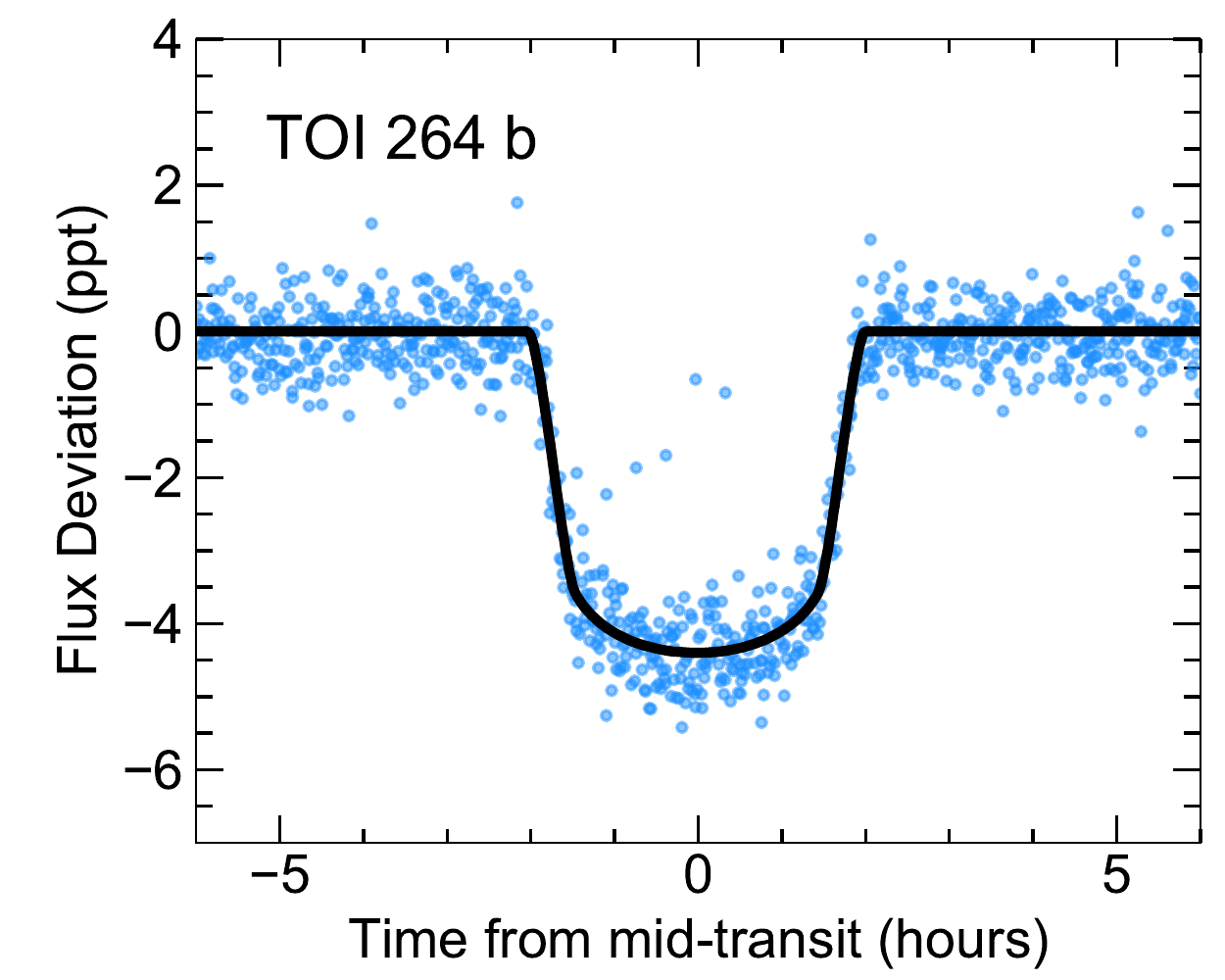}
\includegraphics[width=0.329\textwidth]{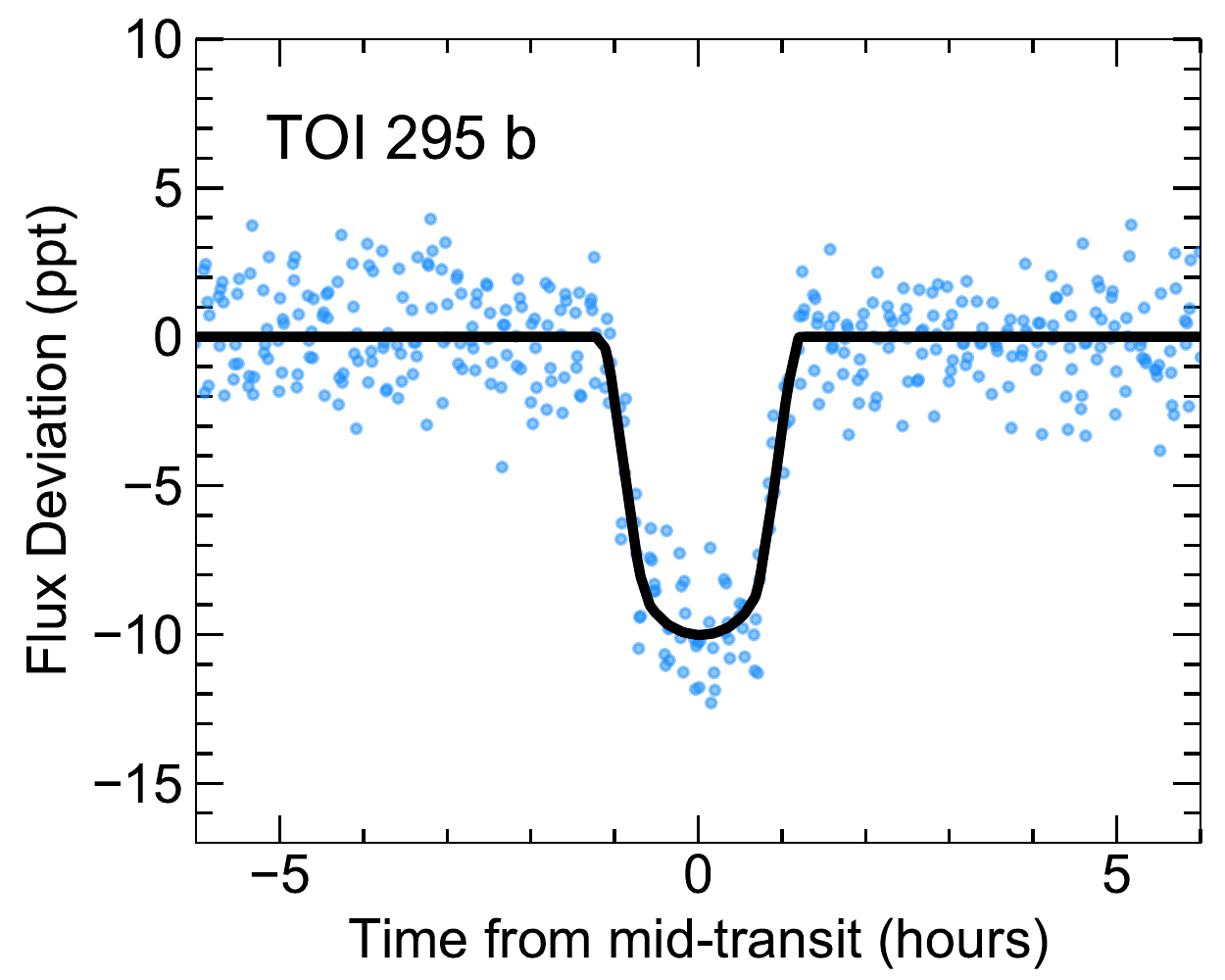}
\includegraphics[width=0.329\textwidth]{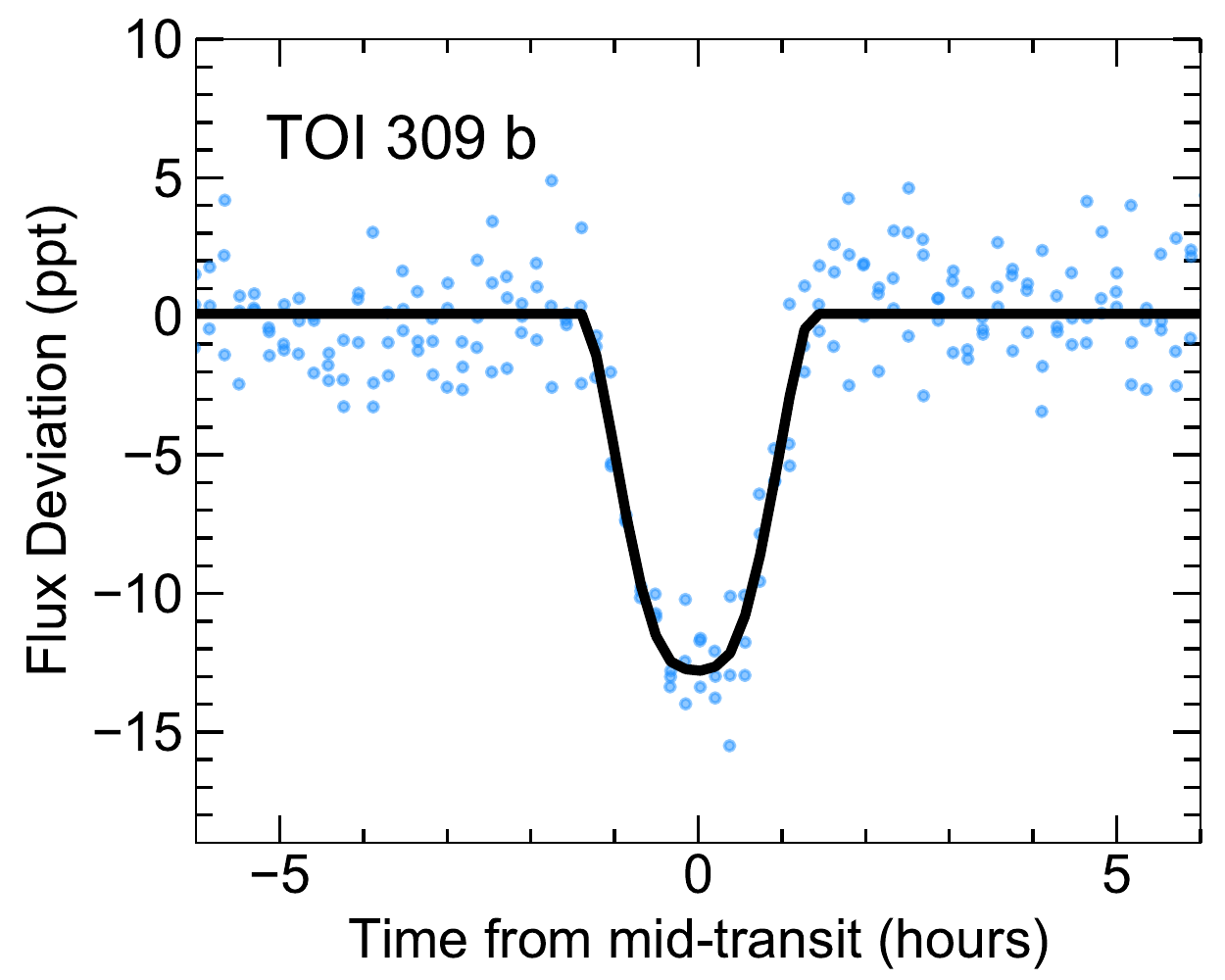}
\includegraphics[width=0.329\textwidth]{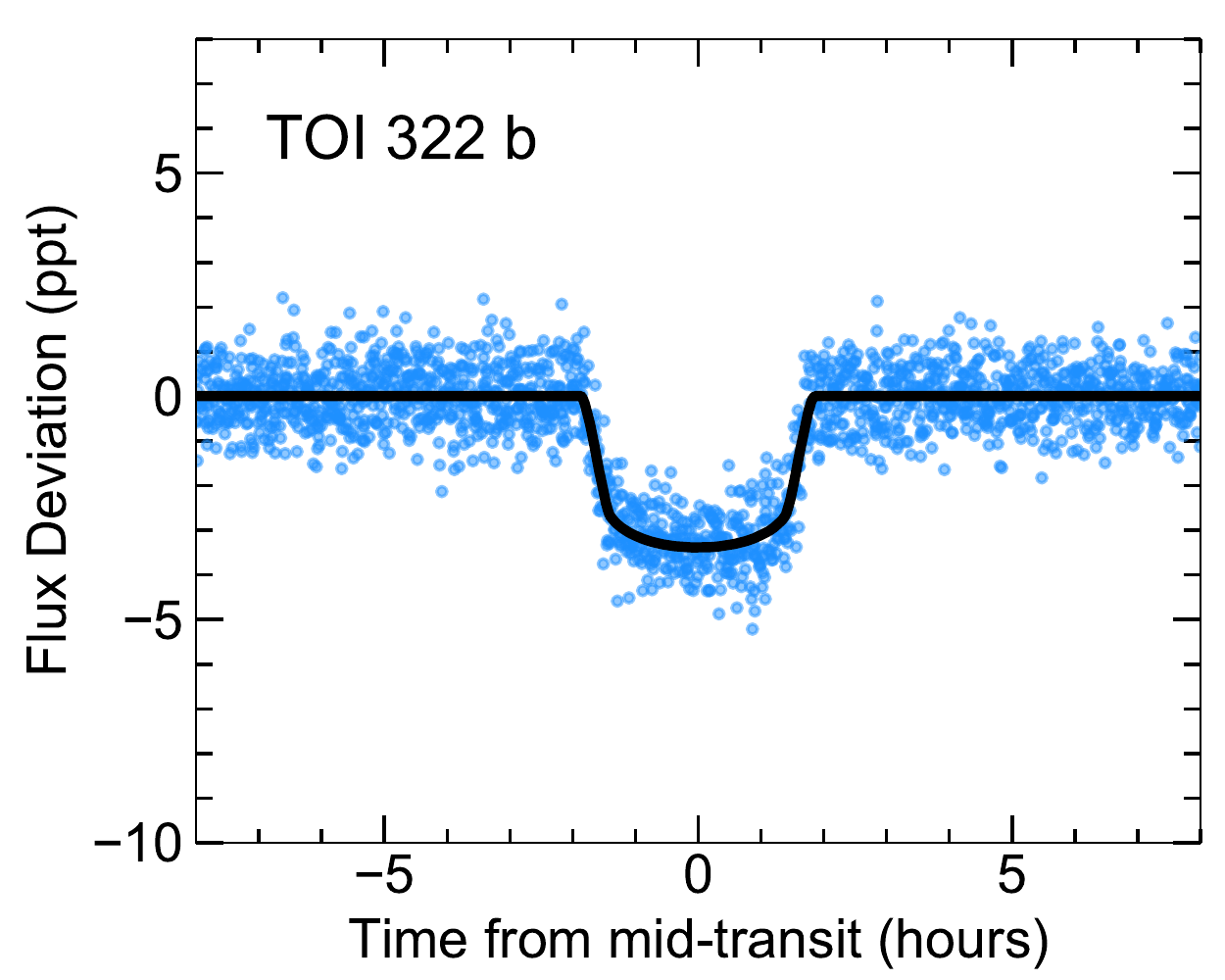}
\includegraphics[width=0.329\textwidth]{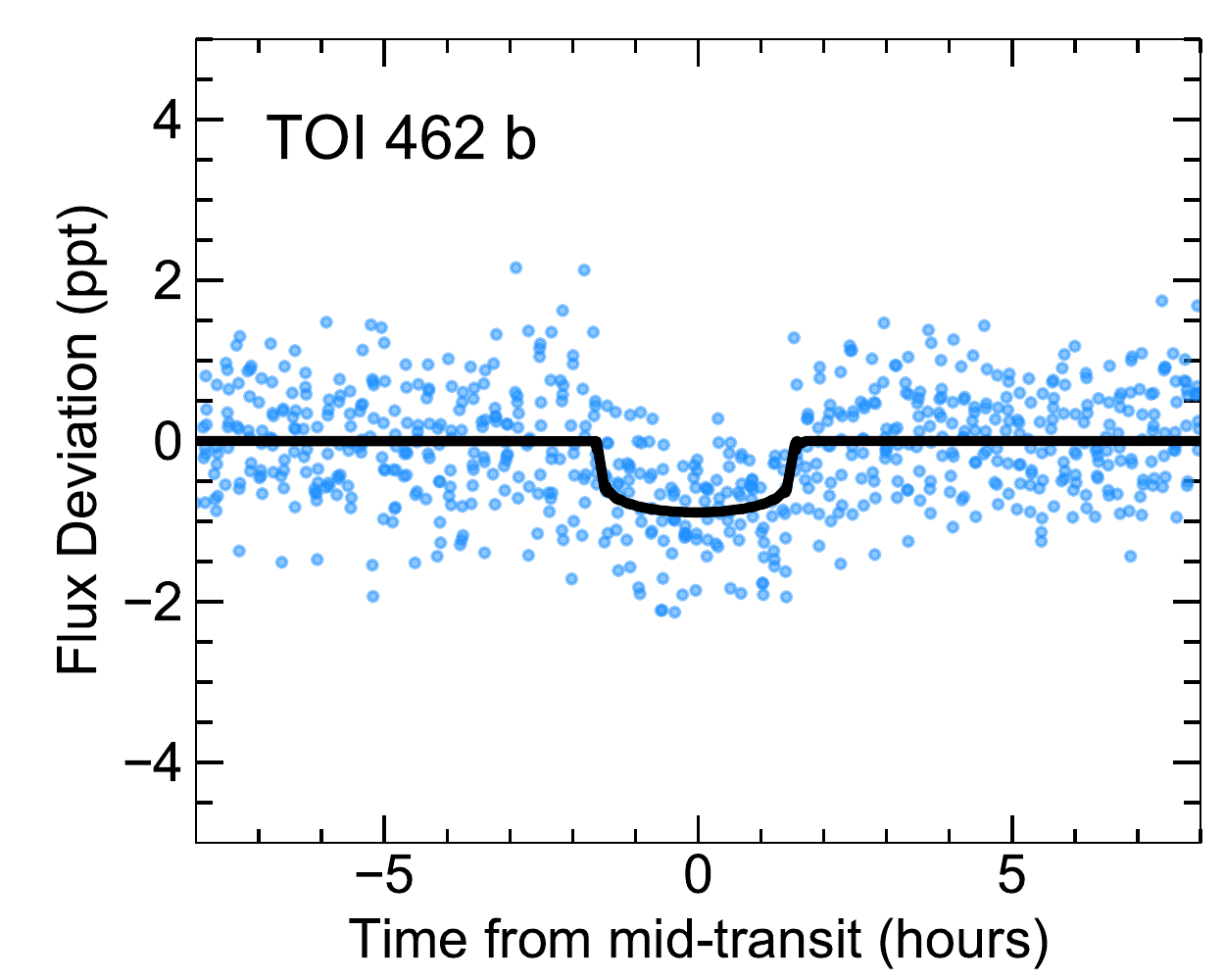}
\includegraphics[width=0.329\textwidth]{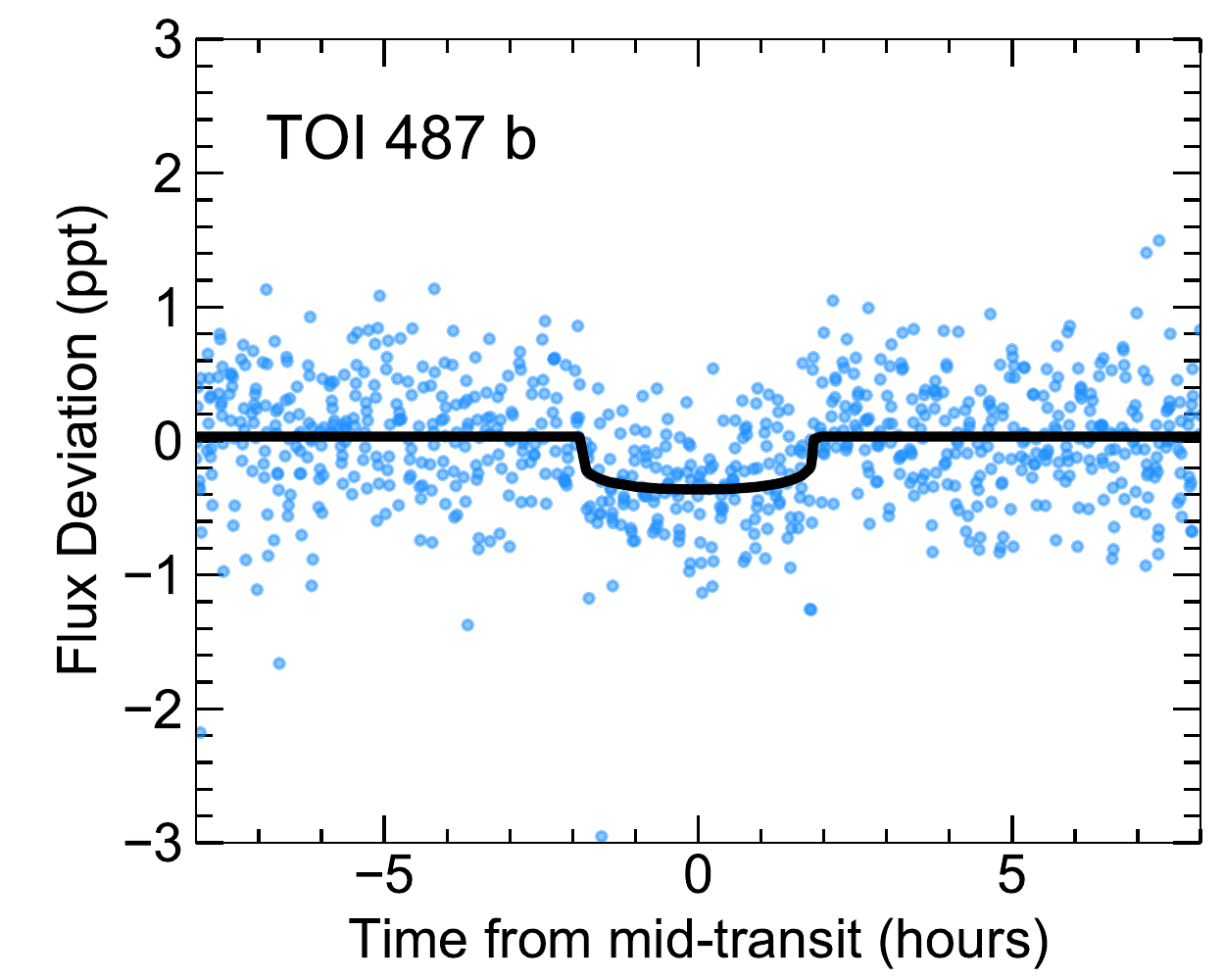}
\includegraphics[width=0.329\textwidth]{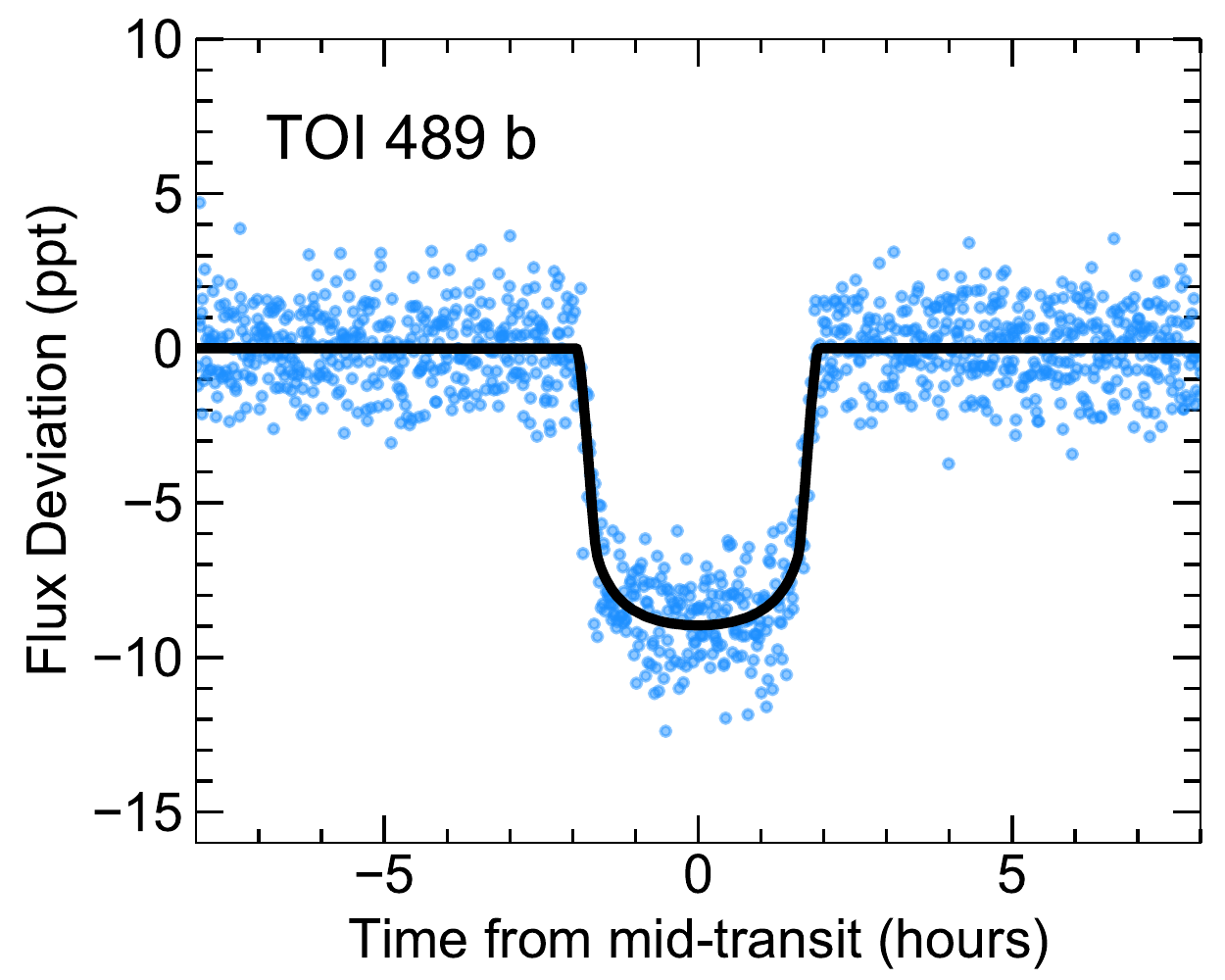}
\includegraphics[width=0.329\textwidth]{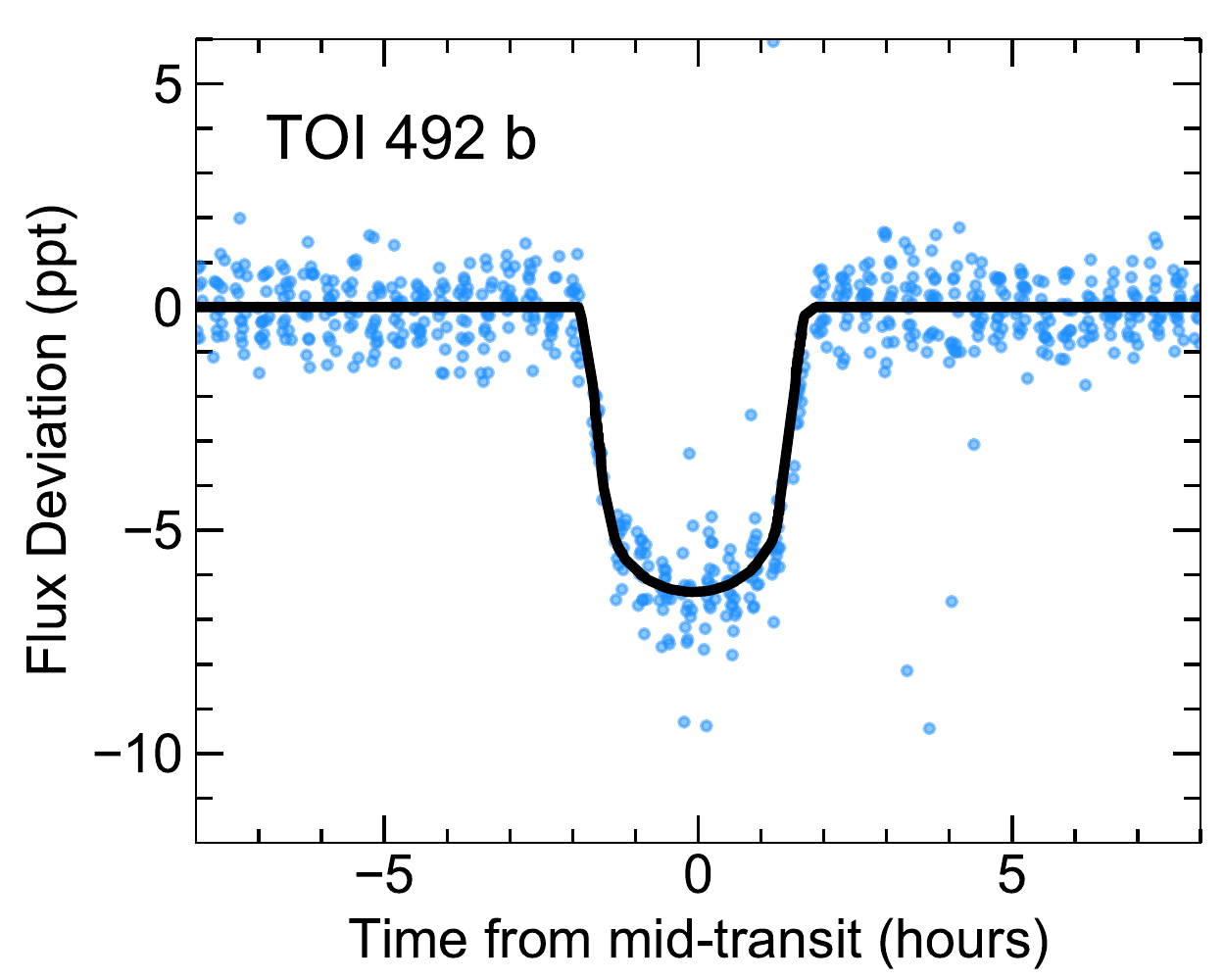}
\caption{Transit light curves for TOI 141 -- 492. The phase folded TESS photometry is shown as the blue circles and the best fit \texttt{EXOFAST} transit model is shown as the solid black line. 
\label{group1}}
\end{figure*}

\begin{figure*}
\includegraphics[width=0.329\textwidth]{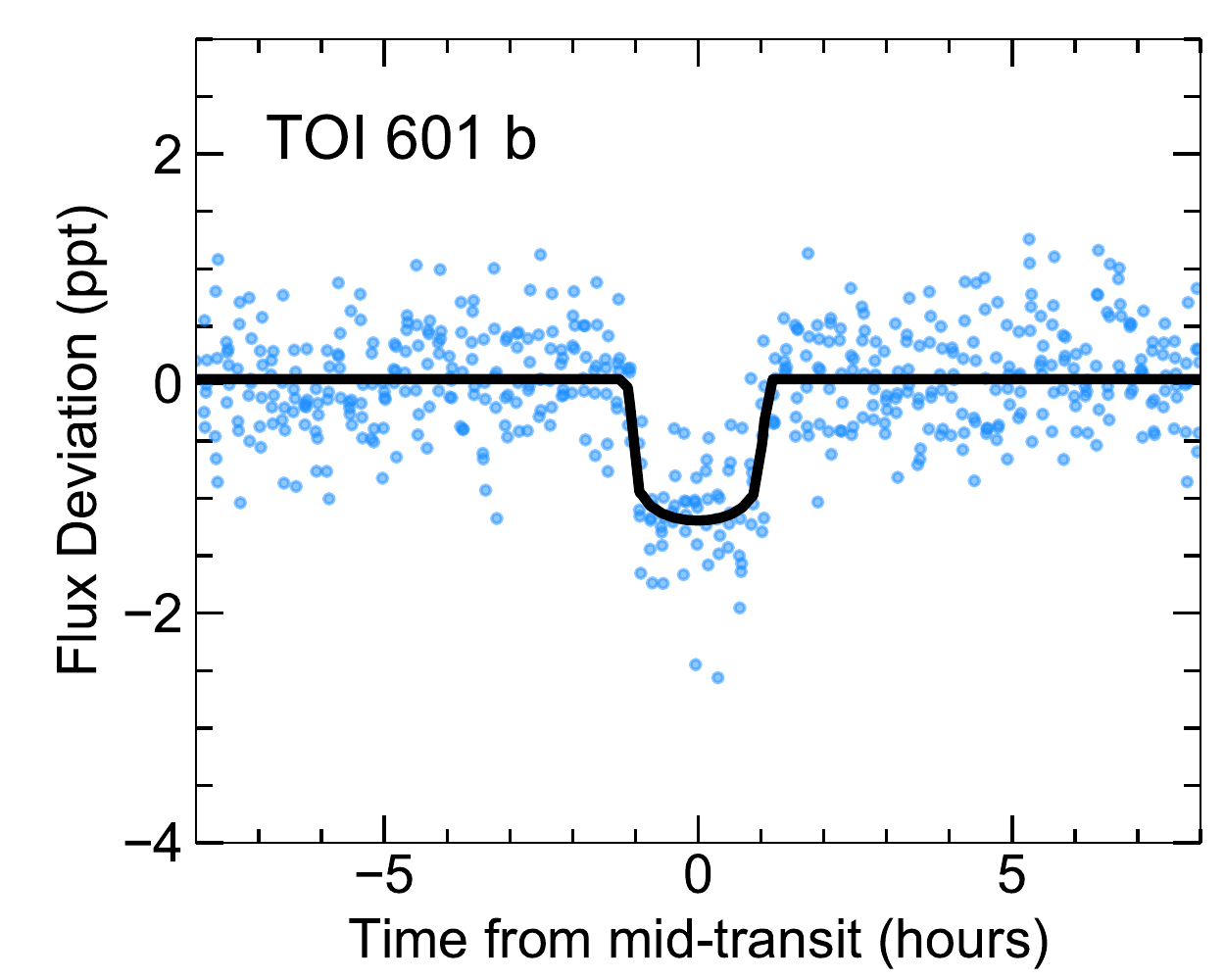}
\includegraphics[width=0.329\textwidth]{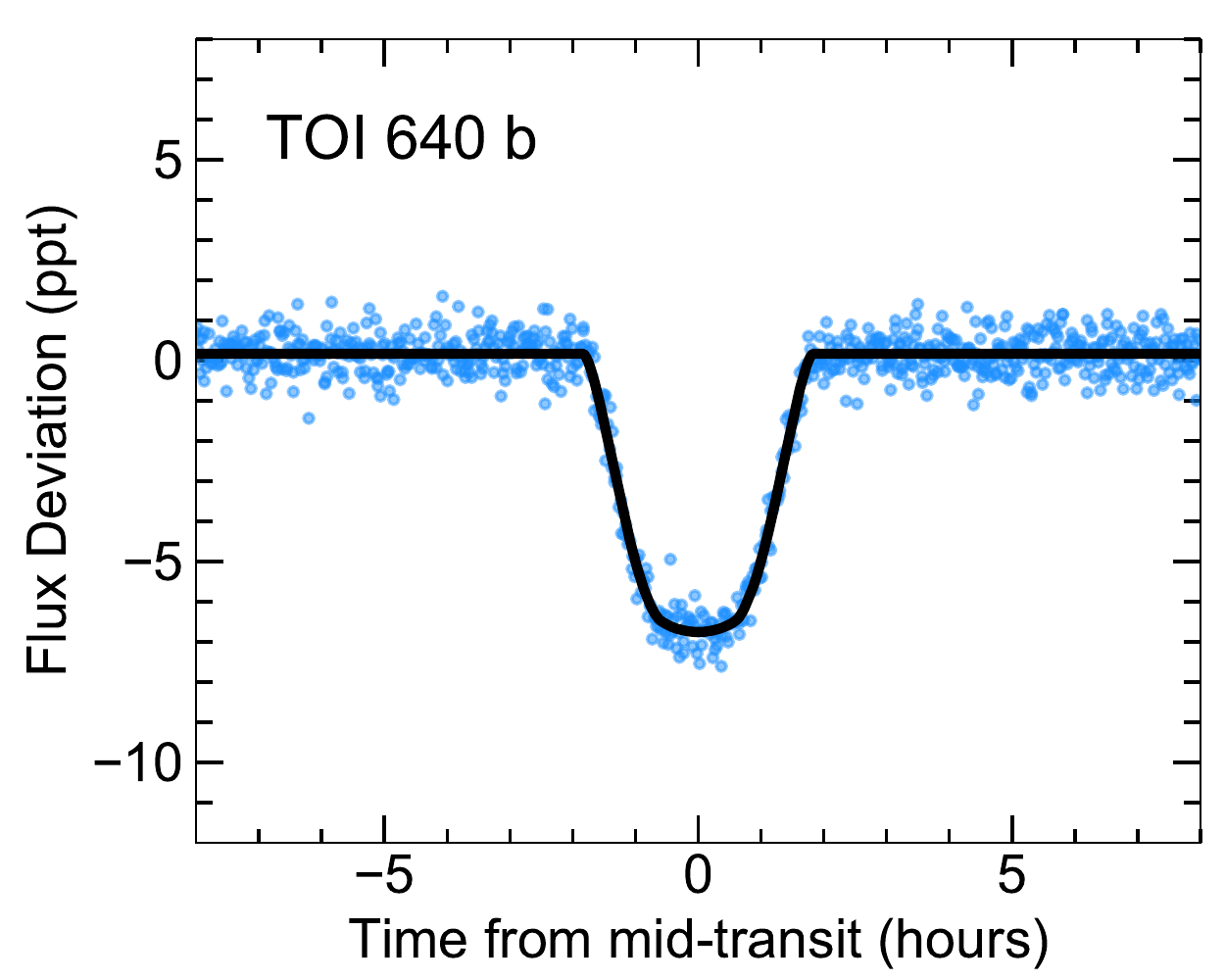}
\includegraphics[width=0.329\textwidth]{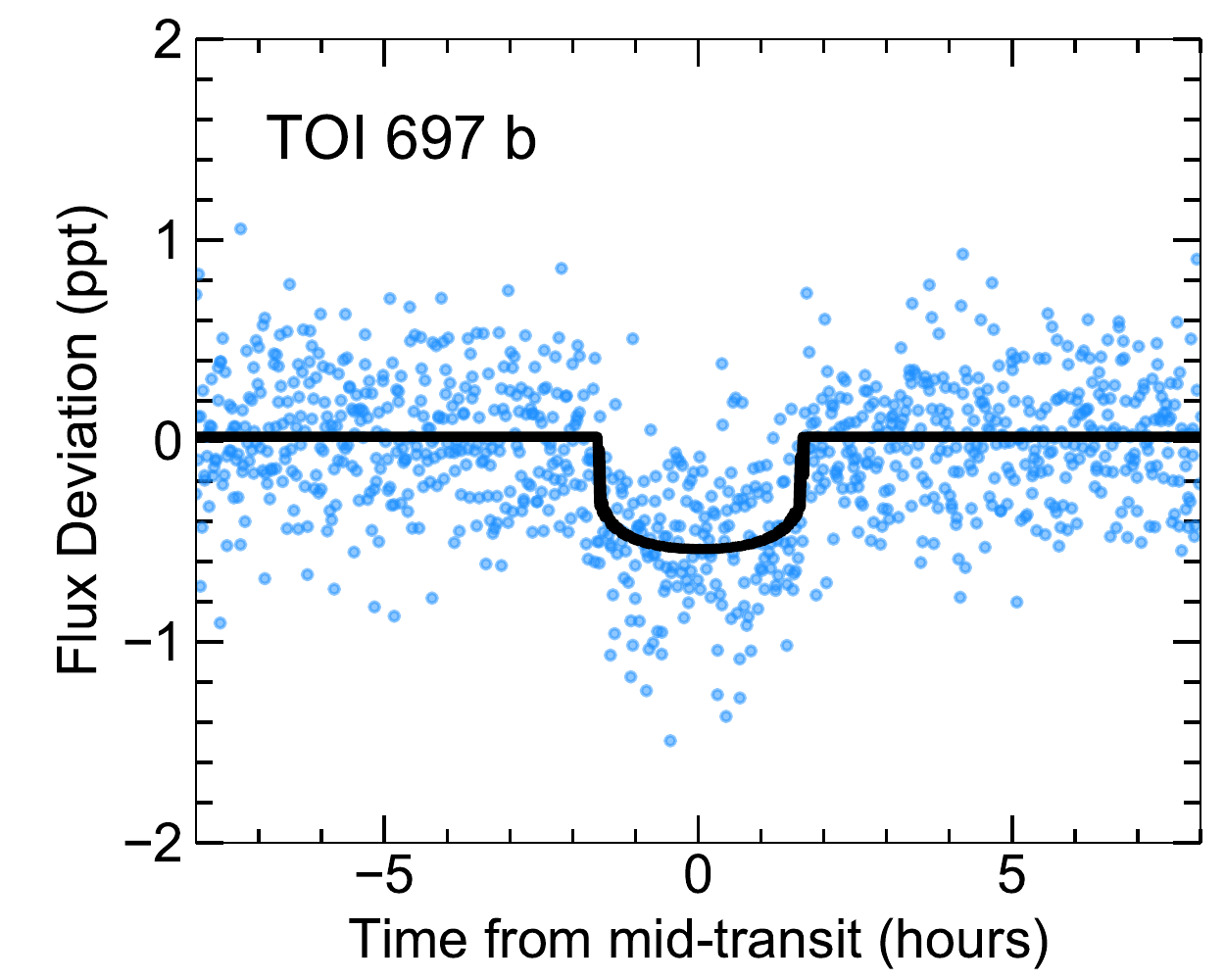}
\includegraphics[width=0.329\textwidth]{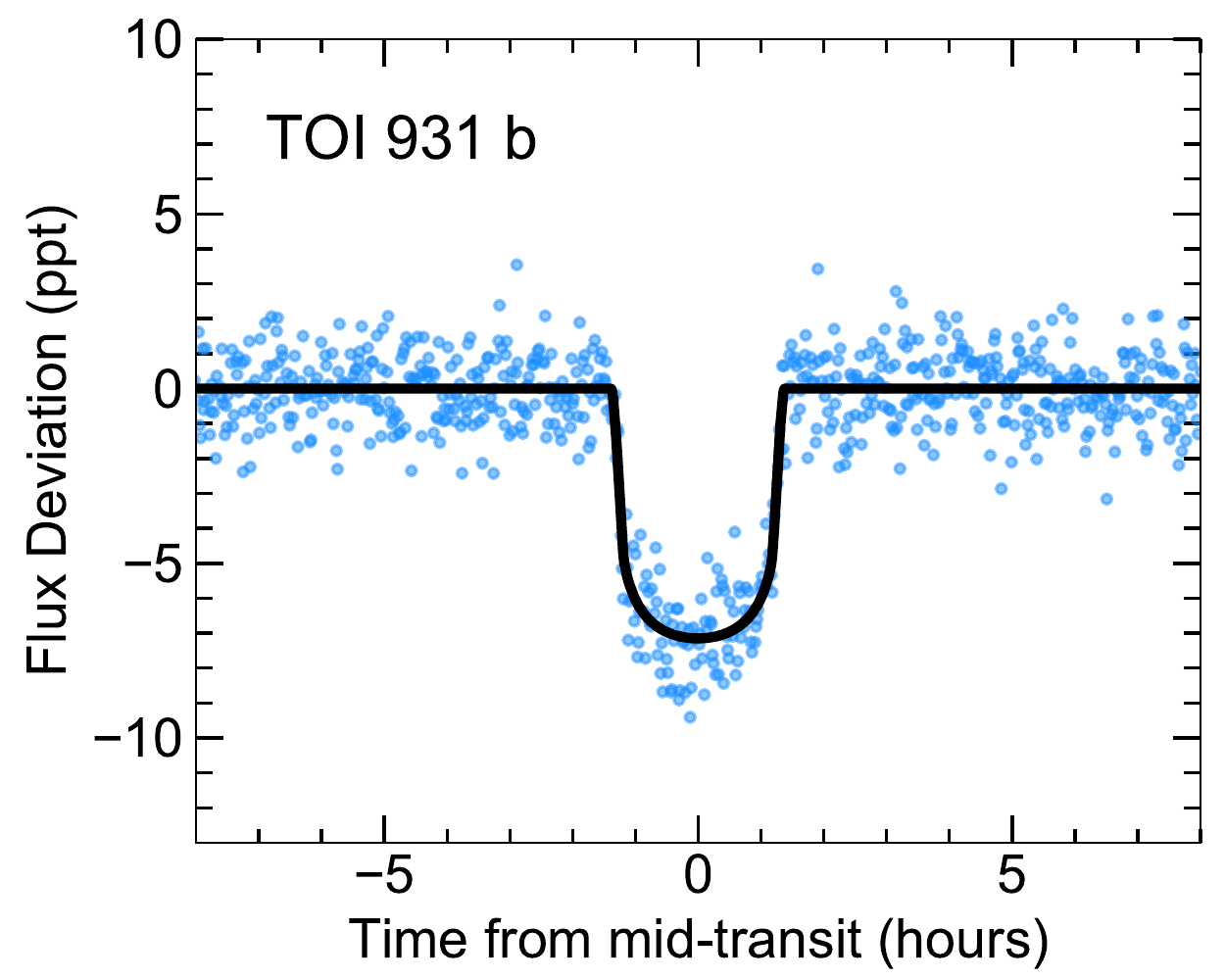}
\includegraphics[width=0.329\textwidth]{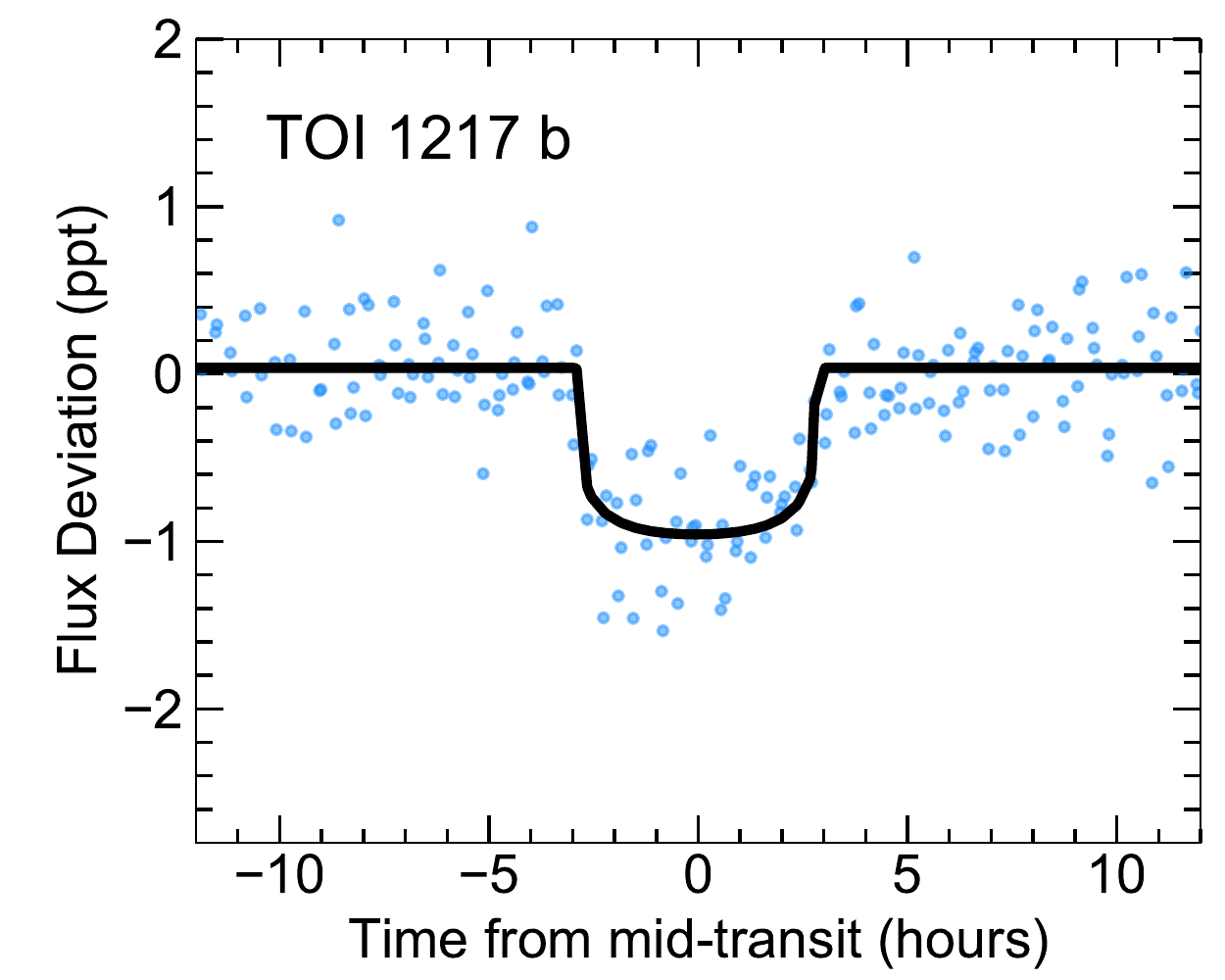}
\includegraphics[width=0.329\textwidth]{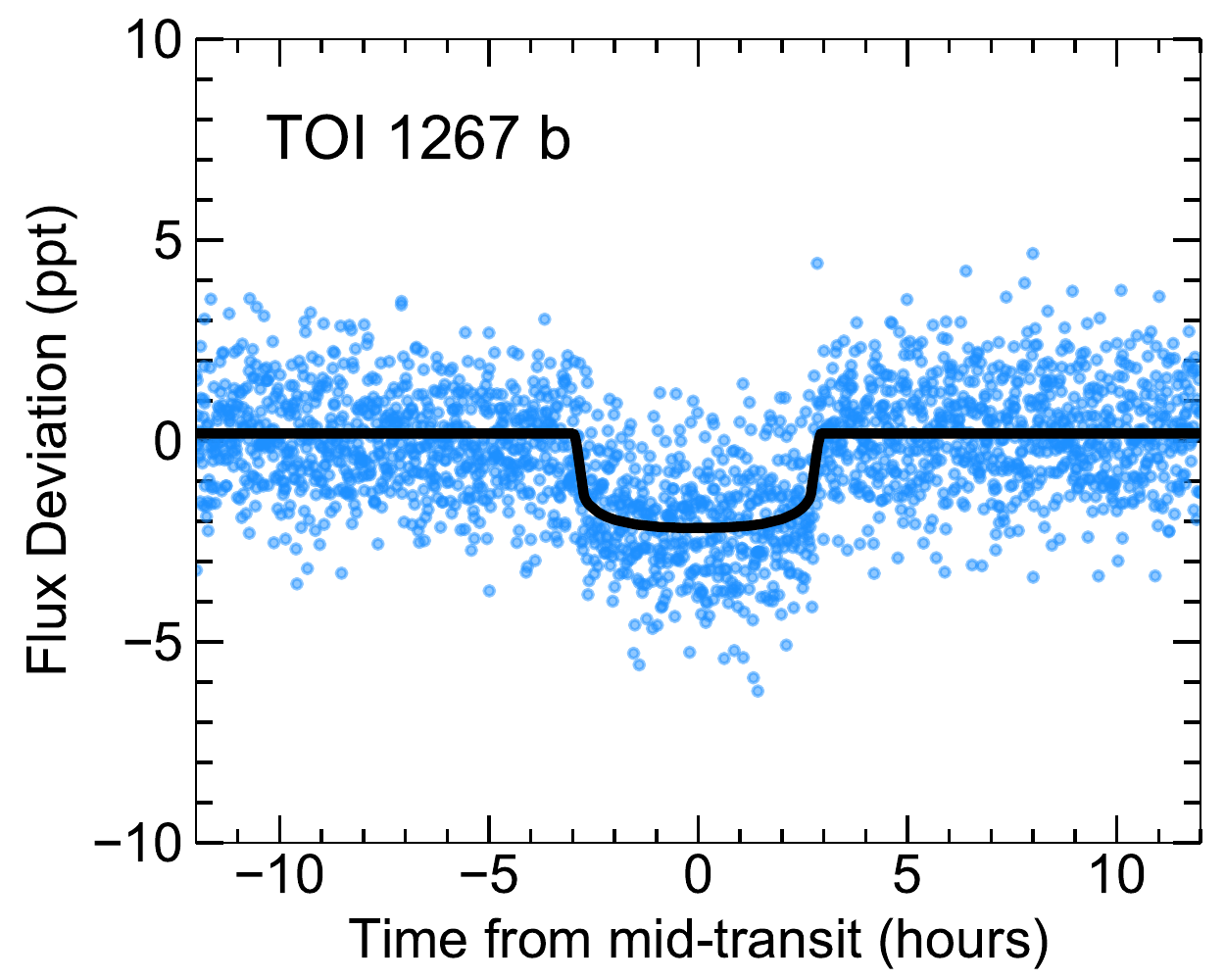}
\includegraphics[width=0.329\textwidth]{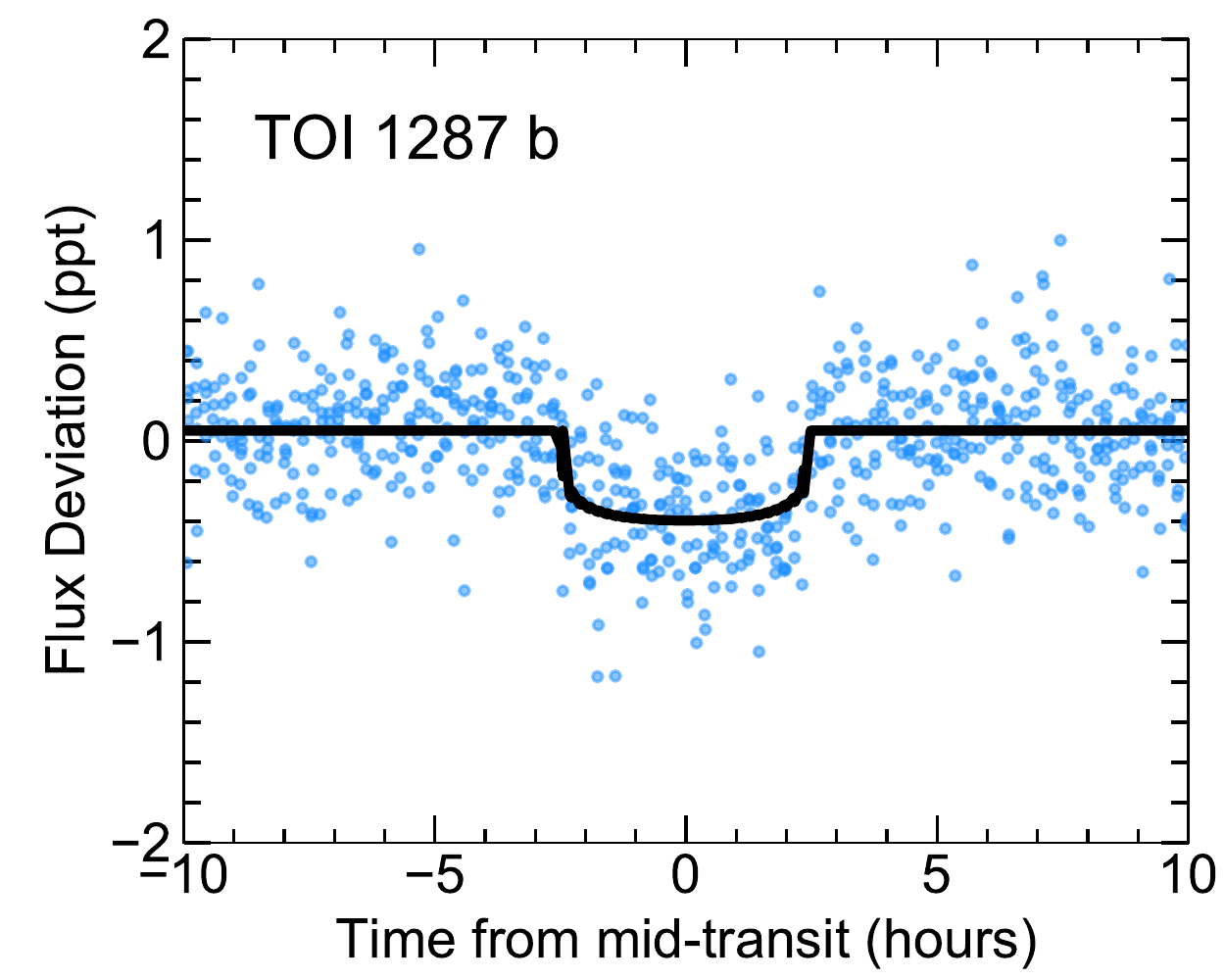}
\includegraphics[width=0.329\textwidth]{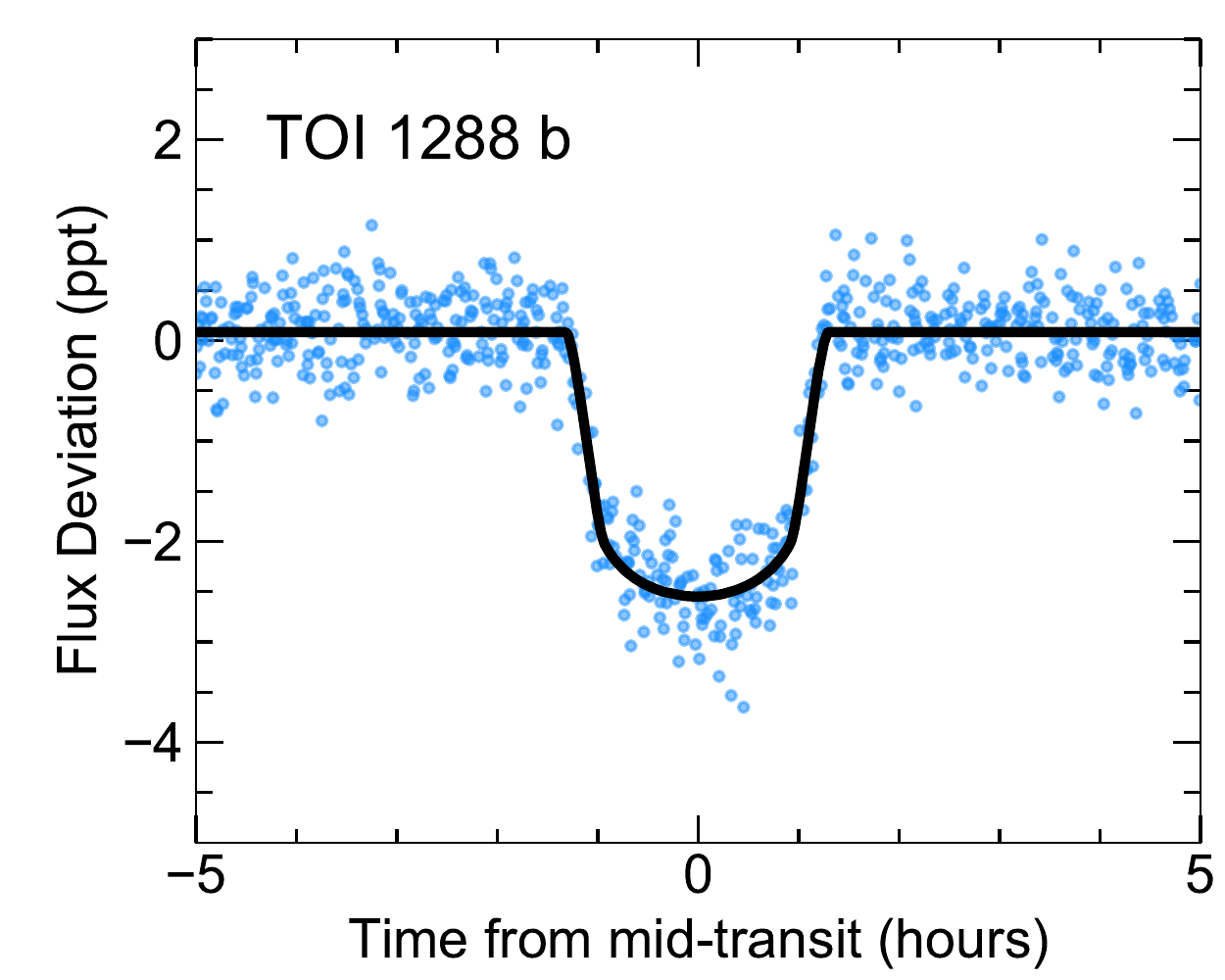}
\includegraphics[width=0.329\textwidth]{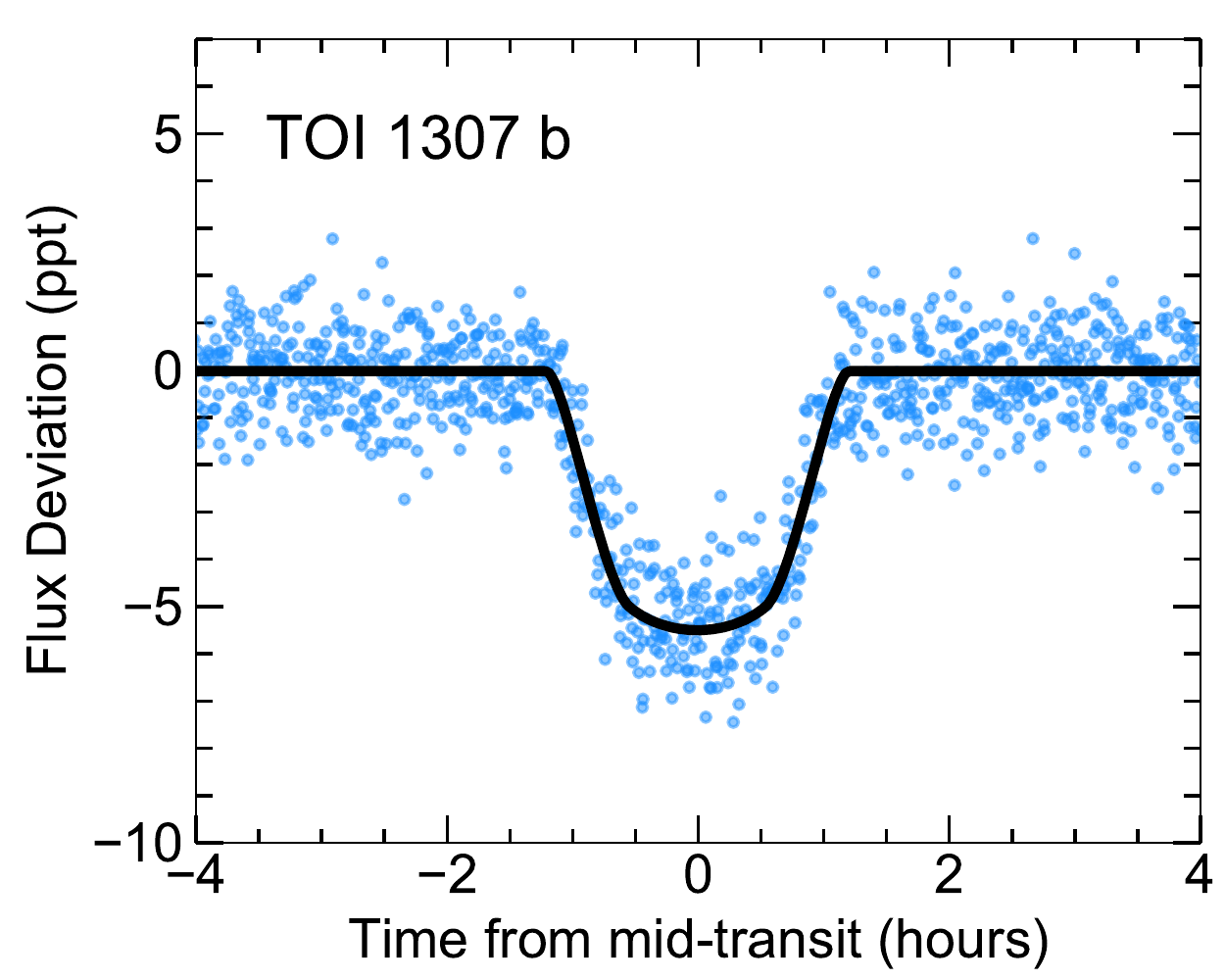}
\includegraphics[width=0.329\textwidth]{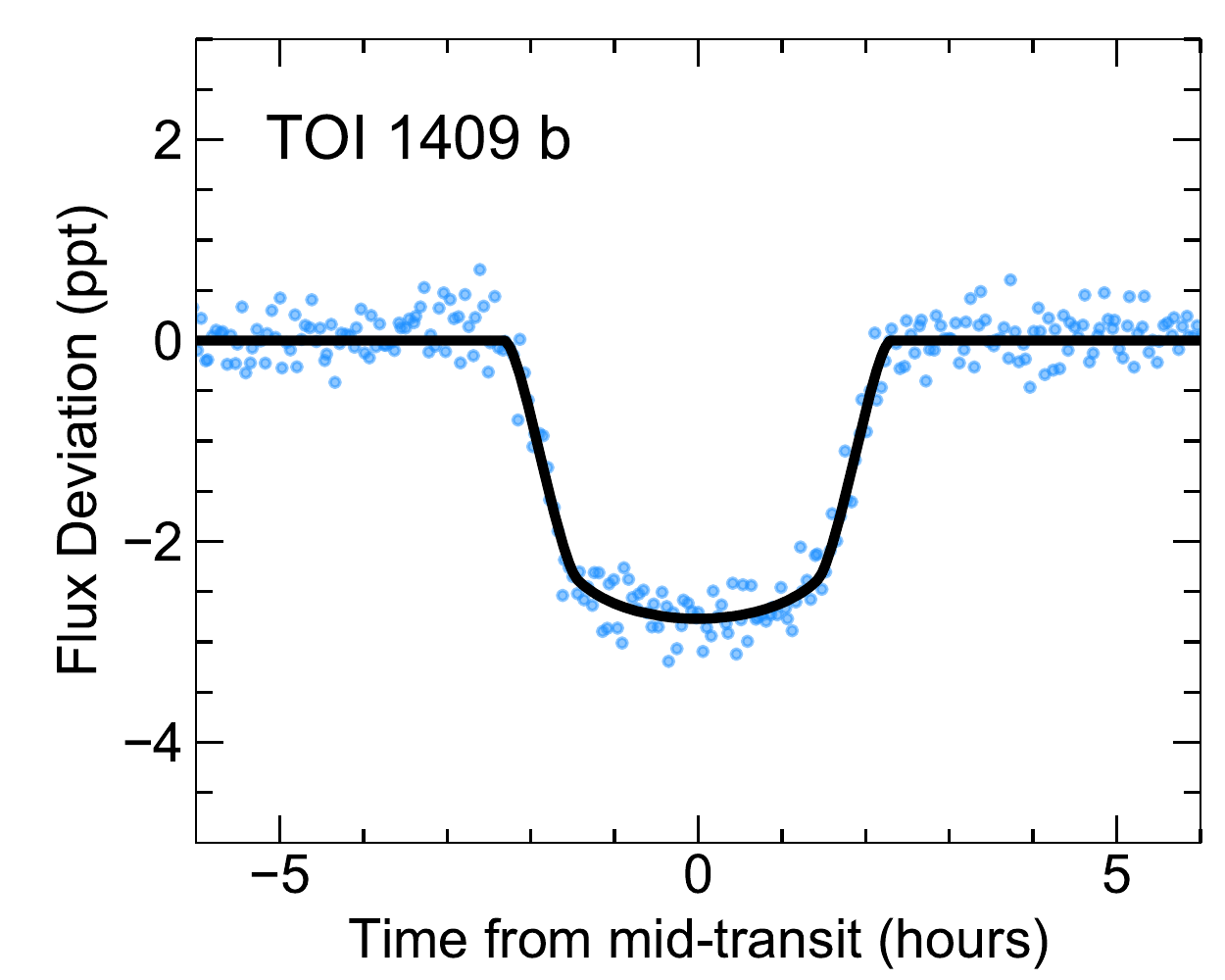}  
\includegraphics[width=0.329\textwidth]{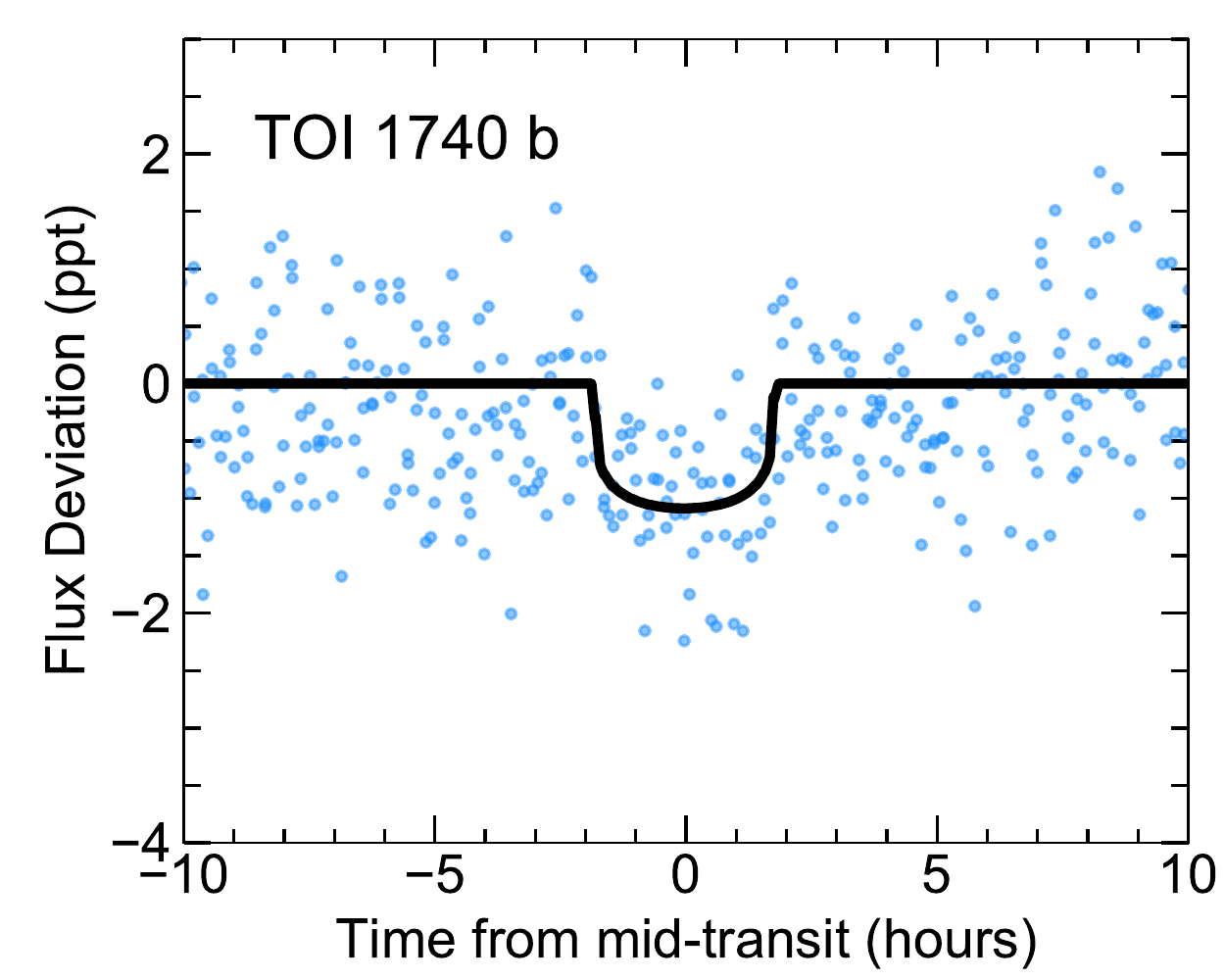}
\caption{Transit light curves for TOI 601 -- 1740. The phase folded TESS photometry is shown as the blue circles and the best fit \texttt{EXOFAST} transit model is shown as the solid black line.
\label{group2}}
\end{figure*}

We did not include any radial velocities in the transit fitting procedure, because our goal was to apply this method to as many TOI systems as possible rather than completing an in-depth analysis on a smaller number of systems. Therefore, our fits do not have the same precision as the combined photometry/radial velocity fits from the literature, but are sufficient for our stellar density calculations. Our best-fit transit geometry parameters are listed in Table~\ref{transitpar}, including the planet period ($P$), transit depth ($\delta$), total transit duration ($t_T$), impact paramter ($b$), planet radius ($R_{pl}$), and transit SNR. Because the host star identity is not yet known, the transit depth and planet radius listed here have not been corrected for the flux for the companion(s); we discuss the planet radius corrections in Section~\ref{results}. Figures~\ref{group1} and \ref{group2} show the phase folded light curve and the best-fit transit model for all of our systems. 

\input{table_transitpar.txt}

\break

\subsection{Observed Stellar Densities} 
We calculated the mean stellar density for each component ($\rho_1, \rho_2, \rho_3$) following Equations 8-11 in \citet{payne18}, which correct the observed transit depth for the flux contamination from the other components. For example, if the planet orbits the primary star in a binary system, the primary's stellar density is
\begin{equation}
\rho_1 = \frac{3\pi}{P^{2}G} \Bigg ( 
\frac{\bigg(1 + \sqrt{(1+ \frac{f_2}{f_1})\delta}~ \bigg)^{2} - b^2\big[1 - \sin ^{2}(t_{T}\pi/P)\big]} {\sin ^{2}(t_{T}\pi/P)}
\Bigg )^{3/2}
\end{equation}
where $P$, $\delta$, $t_T$, and $b$ are the transit parameters listed in the previous section, $f_2/f_1$ is the binary flux ratio, and $G$ is the gravitational constant. If the planet orbits the secondary star, the flux ratio term is inverted and the secondary's stellar density is
\begin{equation}
\rho_2 = \frac{3\pi}{P^{2}G} \Bigg ( 
\frac{\bigg(1 + \sqrt{(1+ \frac{f_1}{f_2})\delta}~ \bigg)^{2} - b^2\big[1 - \sin ^{2}(t_{T}\pi/P)\big]} {\sin ^{2}(t_{T}\pi/P)}
\Bigg )^{3/2}.
\end{equation}
The Gemini and WIYN speckle data were taken in the 832~nm filter and the SOAR data were taken in the $I$-band filter. Both filters are similar to the TESS bandpass so we calculated the flux ratio in the TESS bandpass directly from the speckle magnitude difference. 

\begin{figure}
\centering
\includegraphics[width=0.5\textwidth]{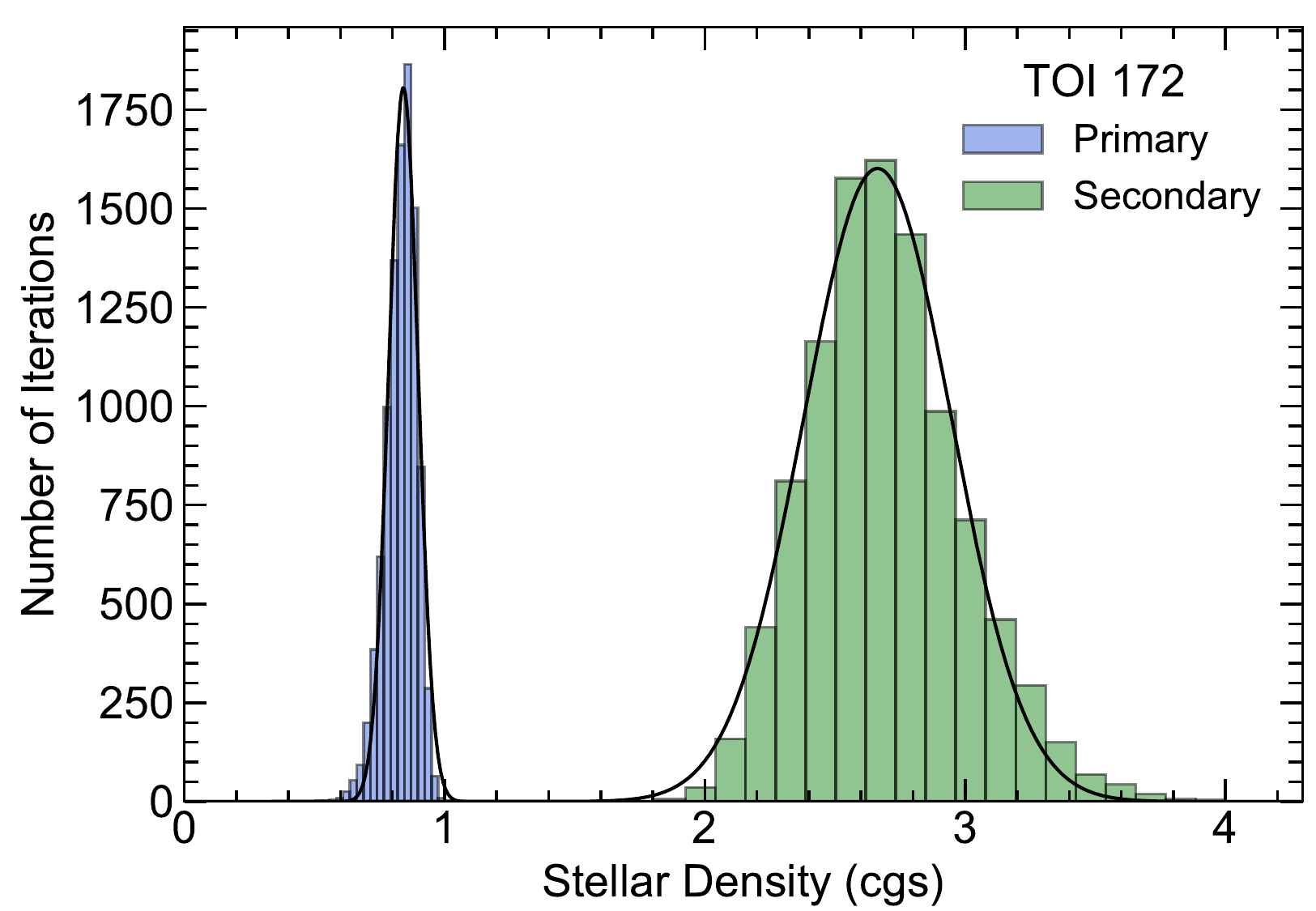}
\caption{Example density histograms for TOI~172 from 10,000 iterations. The histograms for the primary and secondary component are shown in blue and green, respectively. The Gaussian fits, used to determine the $1\sigma$ uncertainties, are shown in solid black.
\label{rhohist}}
\end{figure}

To measure the stellar density uncertainties, we varied all input parameters within their uncertainties over 10,000 iterations, each time calculating the stellar density if the planet orbits the primary, secondary, and tertiary (if applicable). We then created a histogram of the resulting densities for each component as shown in Figure~\ref{rhohist}, fit a Gaussian to each histogram, and adopted the width as the $1\sigma$ uncertainties for each component. Our results are listed in Table~\ref{densities} with the stellar density and uncertainty of each binary component. We found the biggest source of uncertainty in the stellar densities to be the transit impact parameter, especially if the transit is near grazing and the impact parameter's uncertainties extend beyond $b=1$.

\section{Results \label{results}} 

\subsection{Determining Host Component} 
\begin{figure*}
\includegraphics[width=0.329\textwidth]{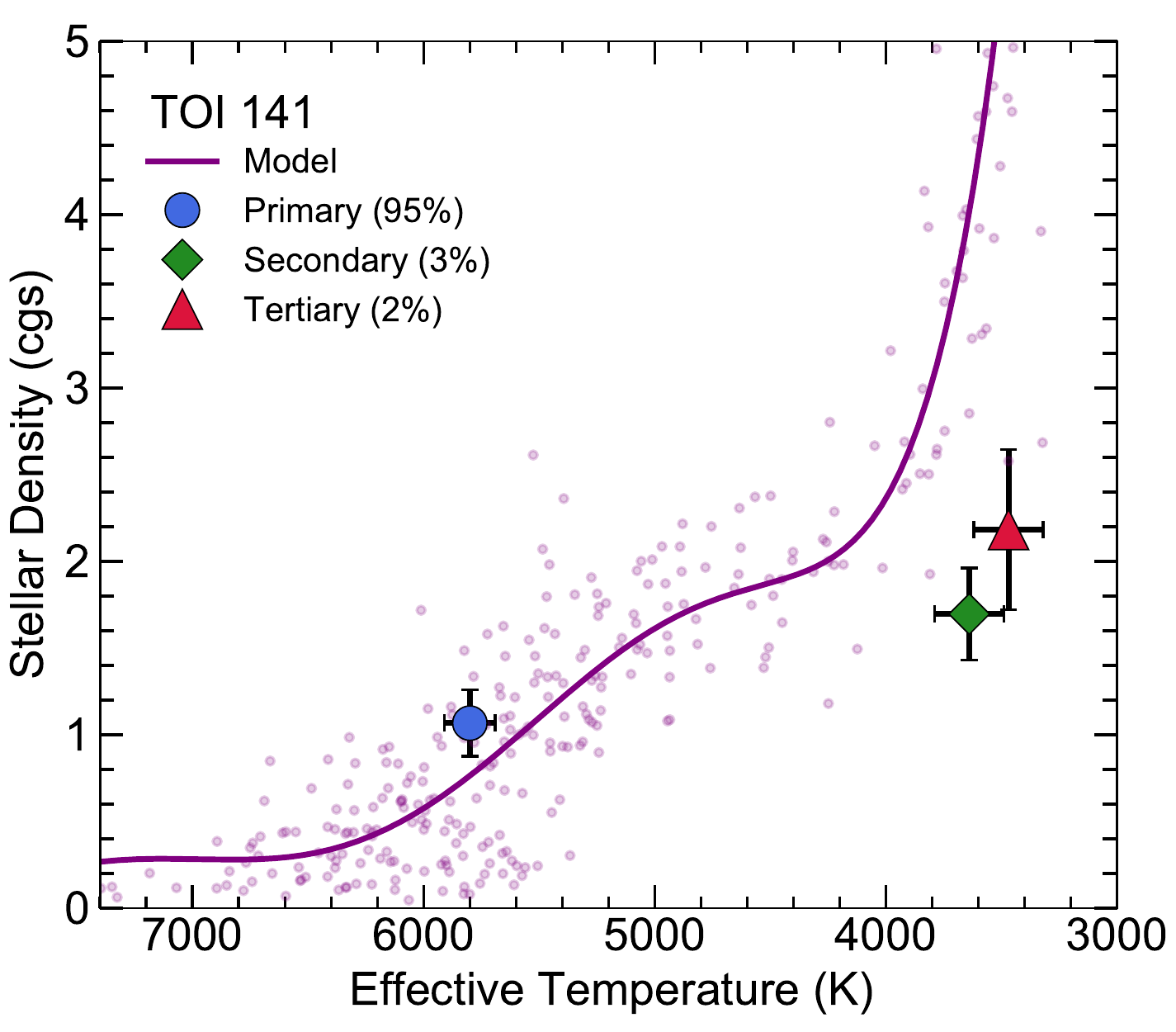}
\includegraphics[width=0.329\textwidth]{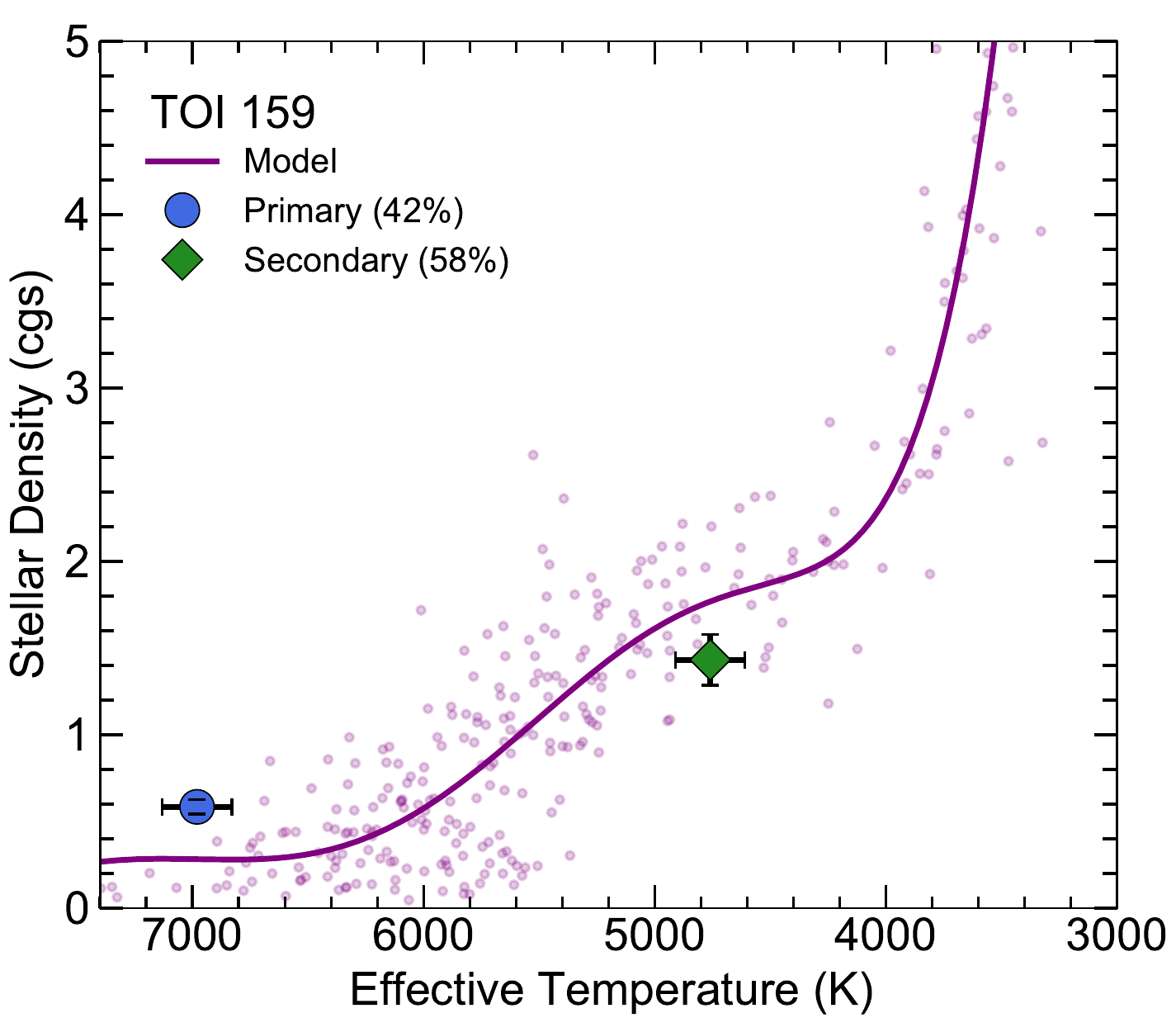}
\includegraphics[width=0.329\textwidth]{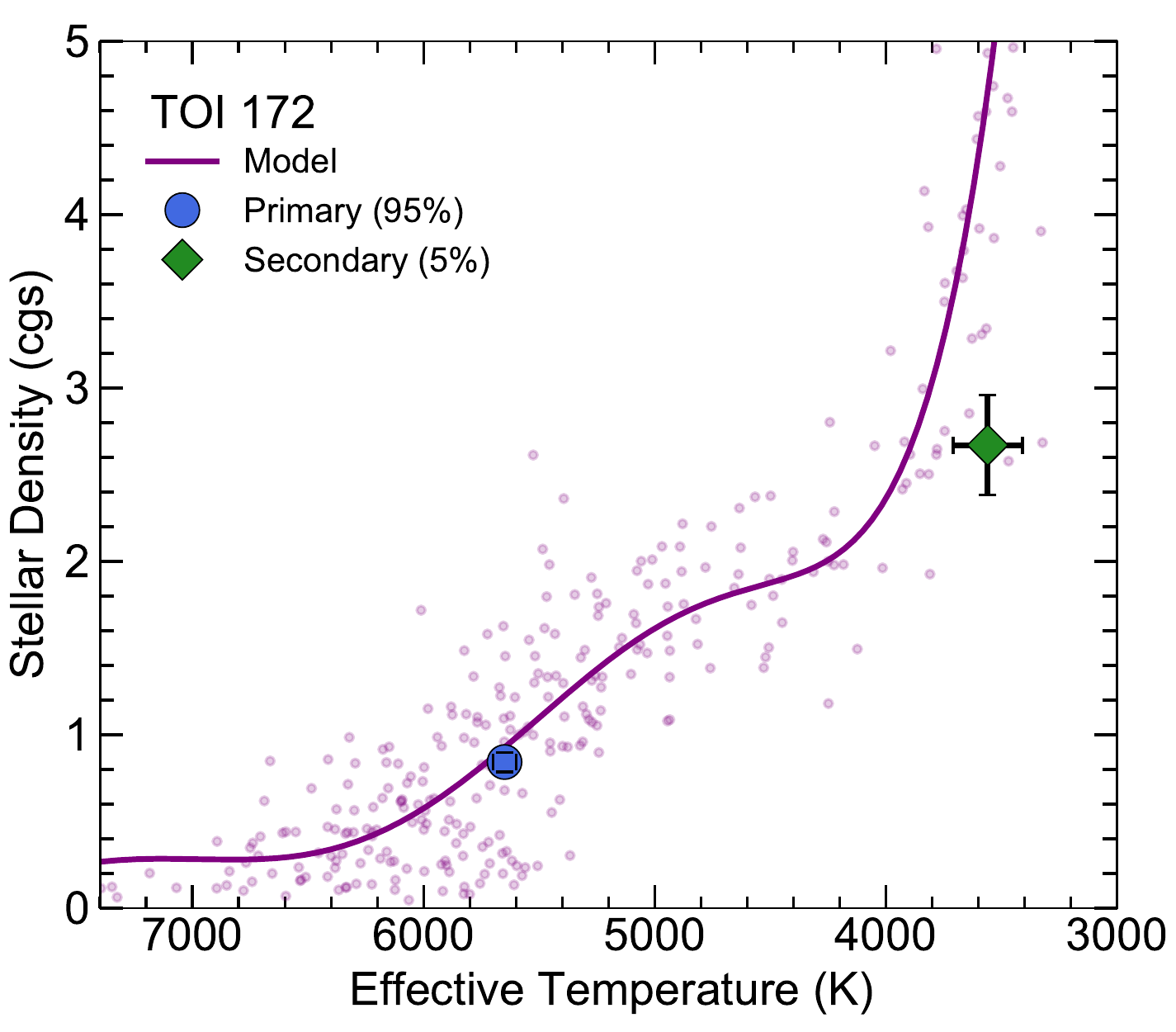}
\includegraphics[width=0.329\textwidth]{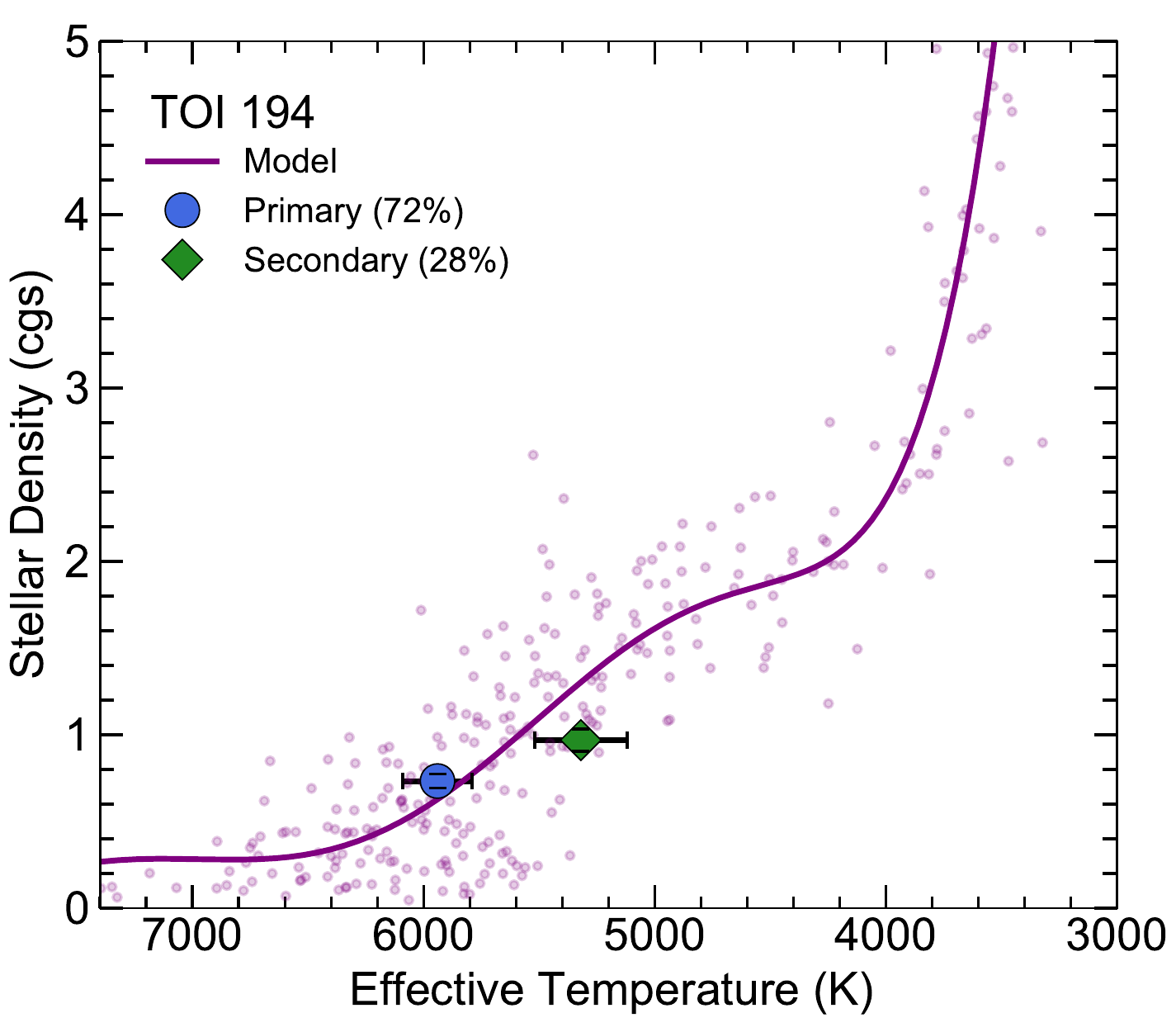}
\includegraphics[width=0.329\textwidth]{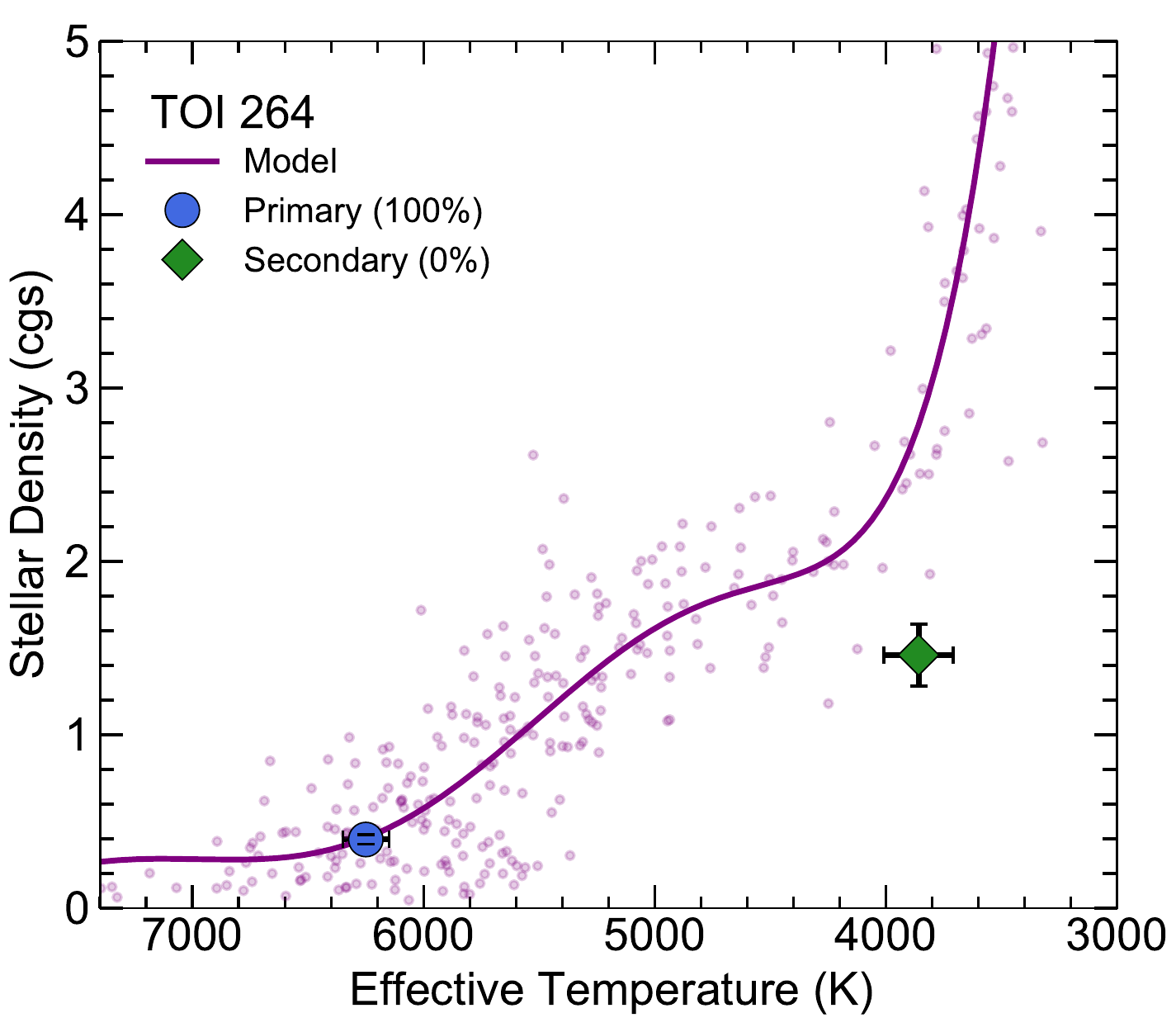}
\includegraphics[width=0.329\textwidth]{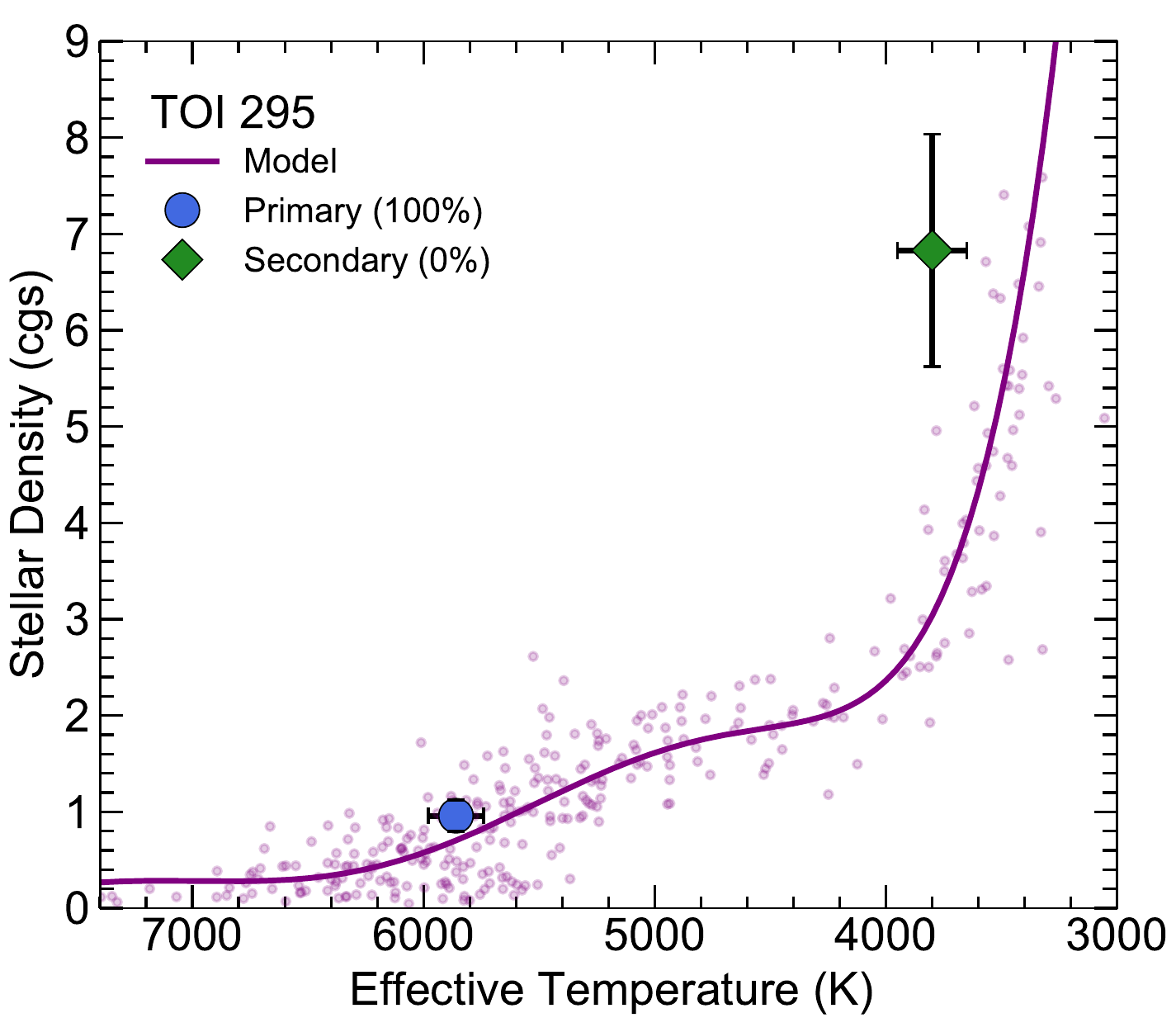}
\includegraphics[width=0.329\textwidth]{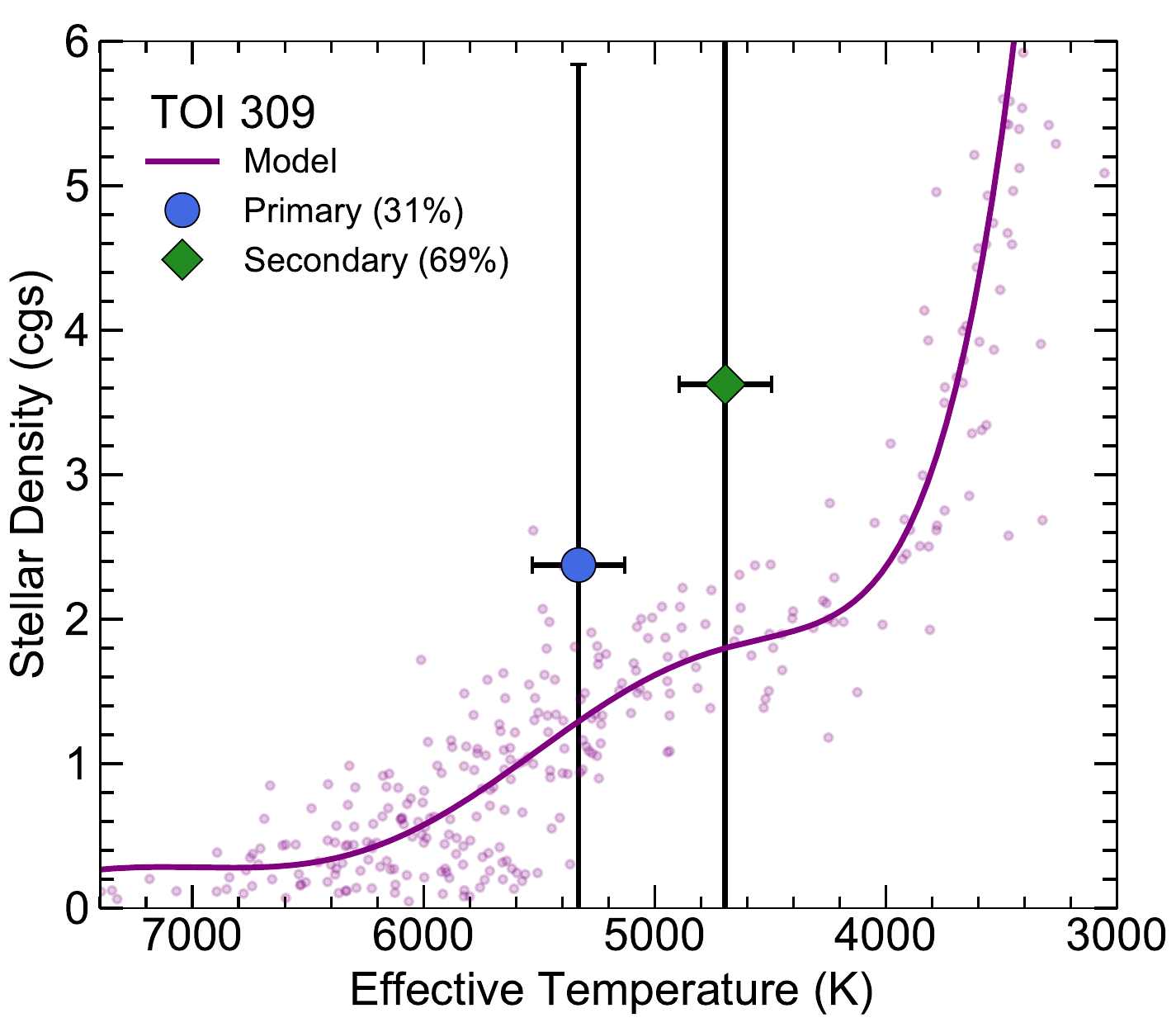}
\includegraphics[width=0.329\textwidth]{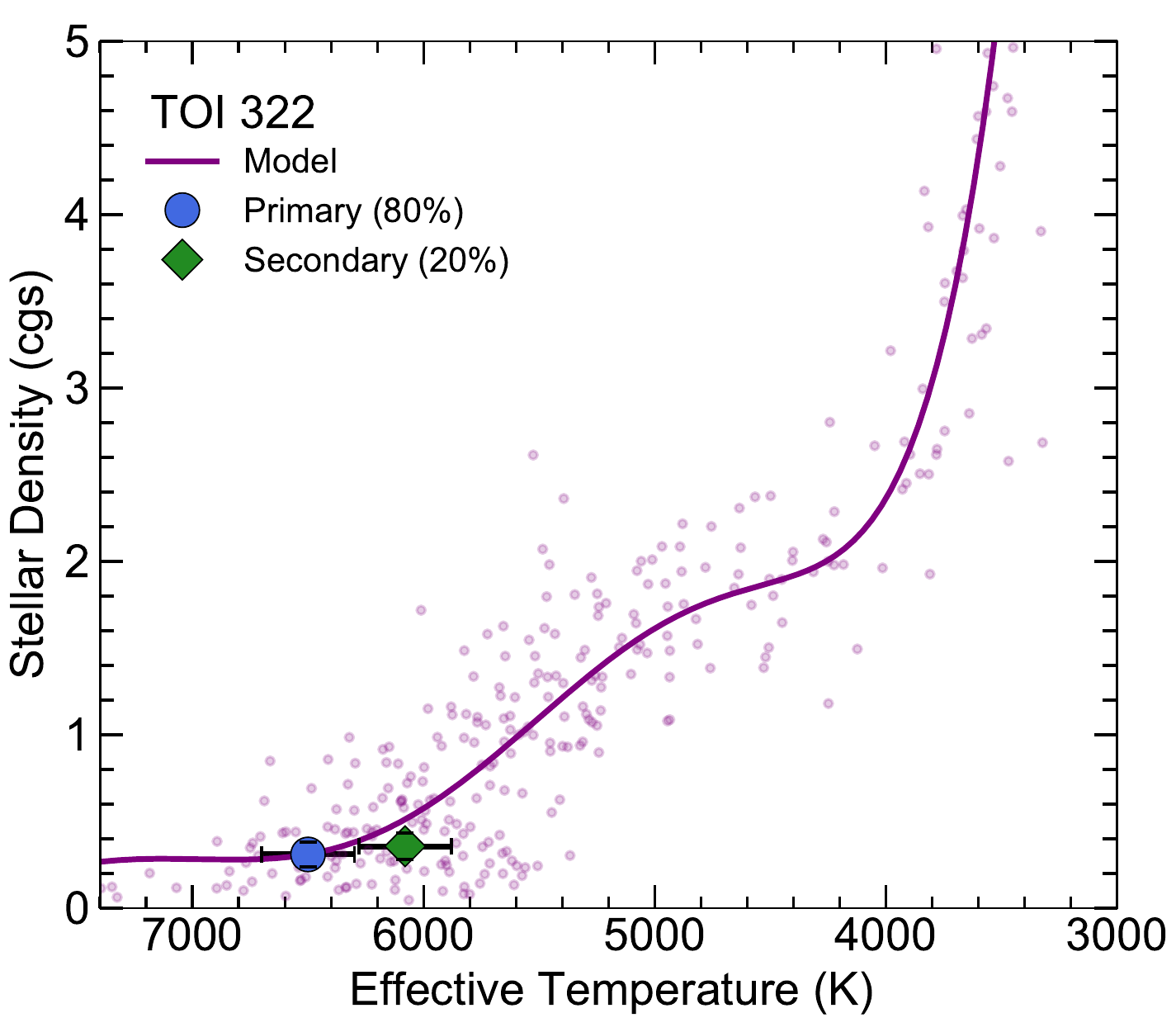}
\includegraphics[width=0.329\textwidth]{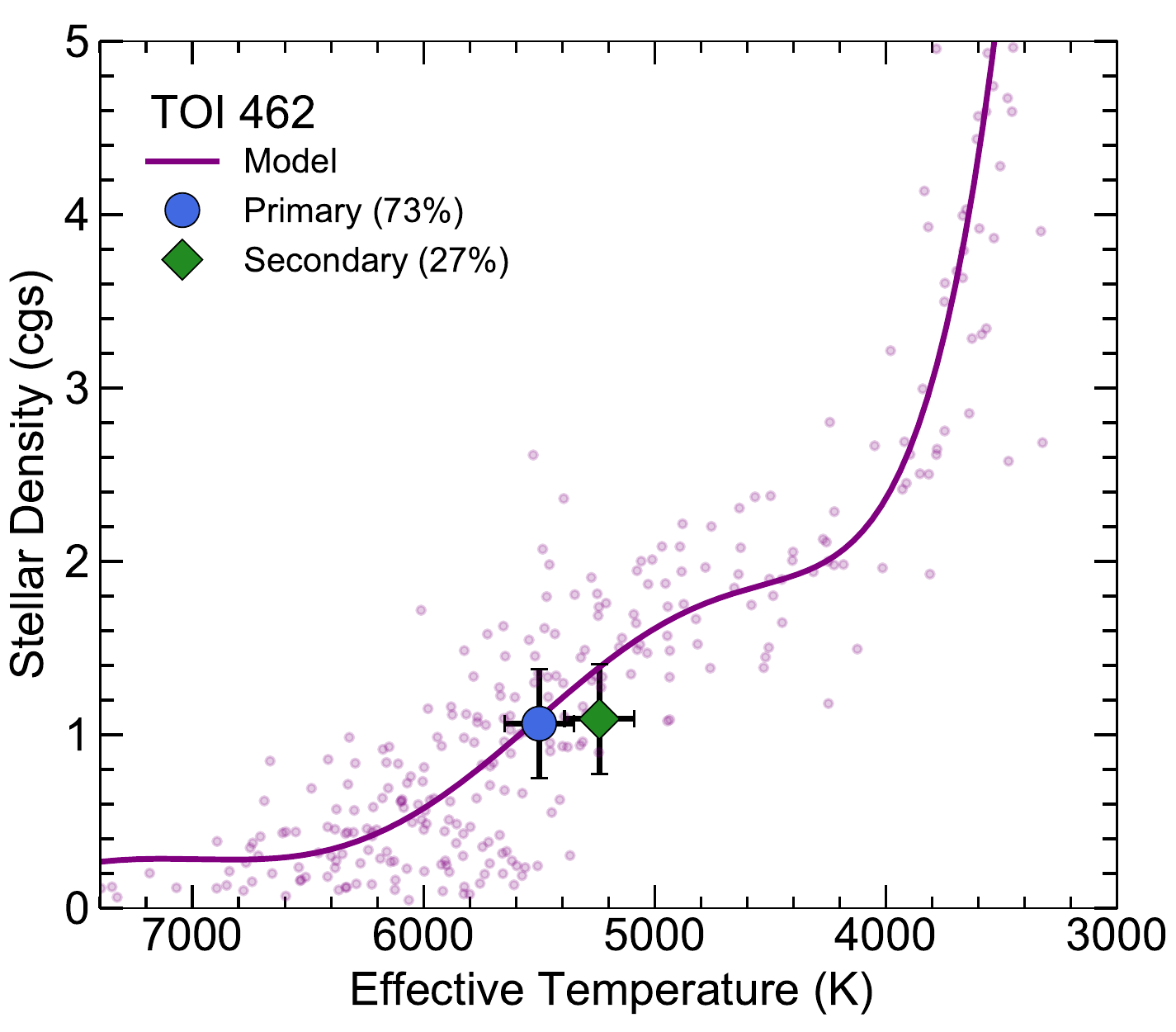}
\includegraphics[width=0.329\textwidth]{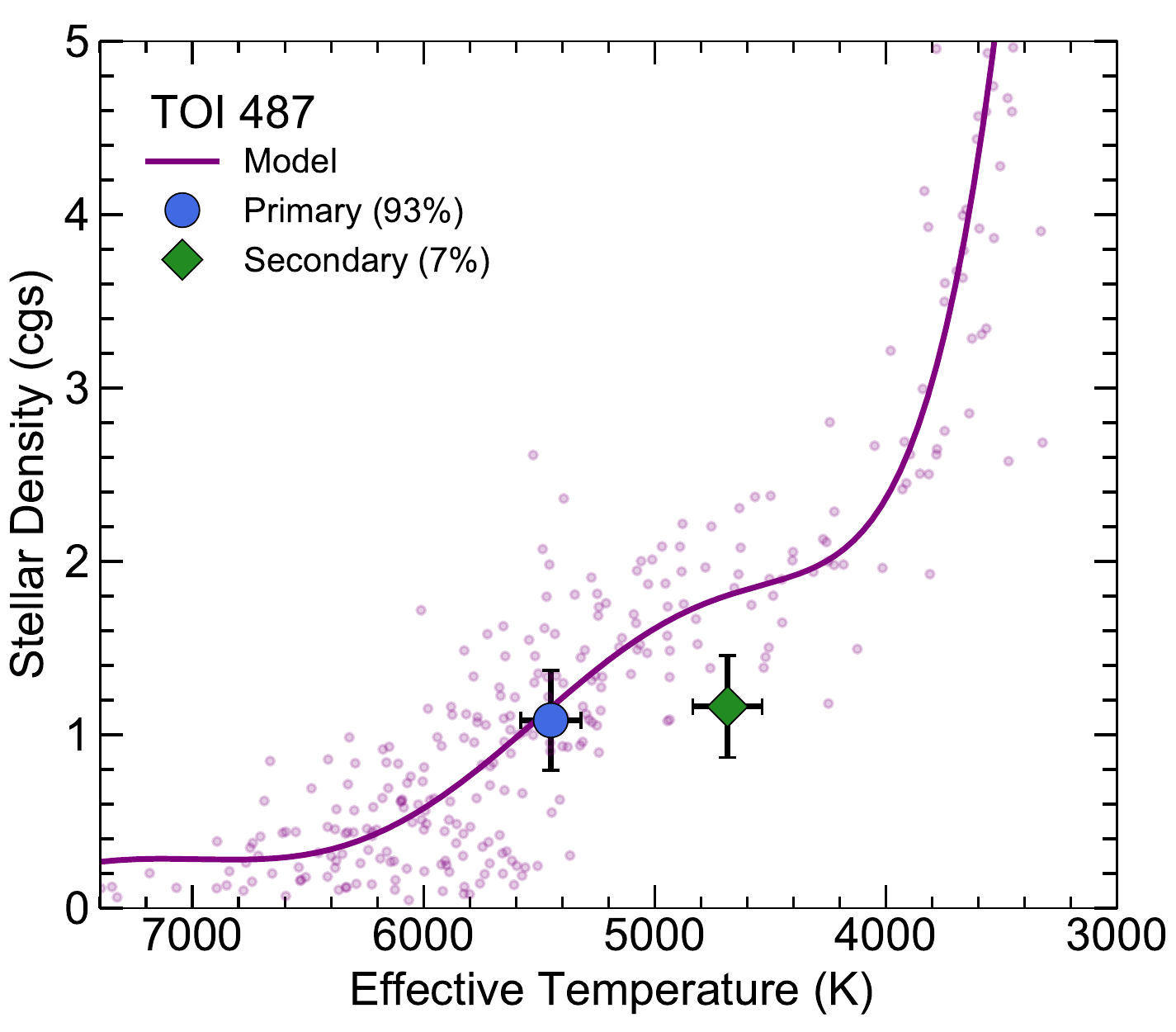}
\includegraphics[width=0.329\textwidth]{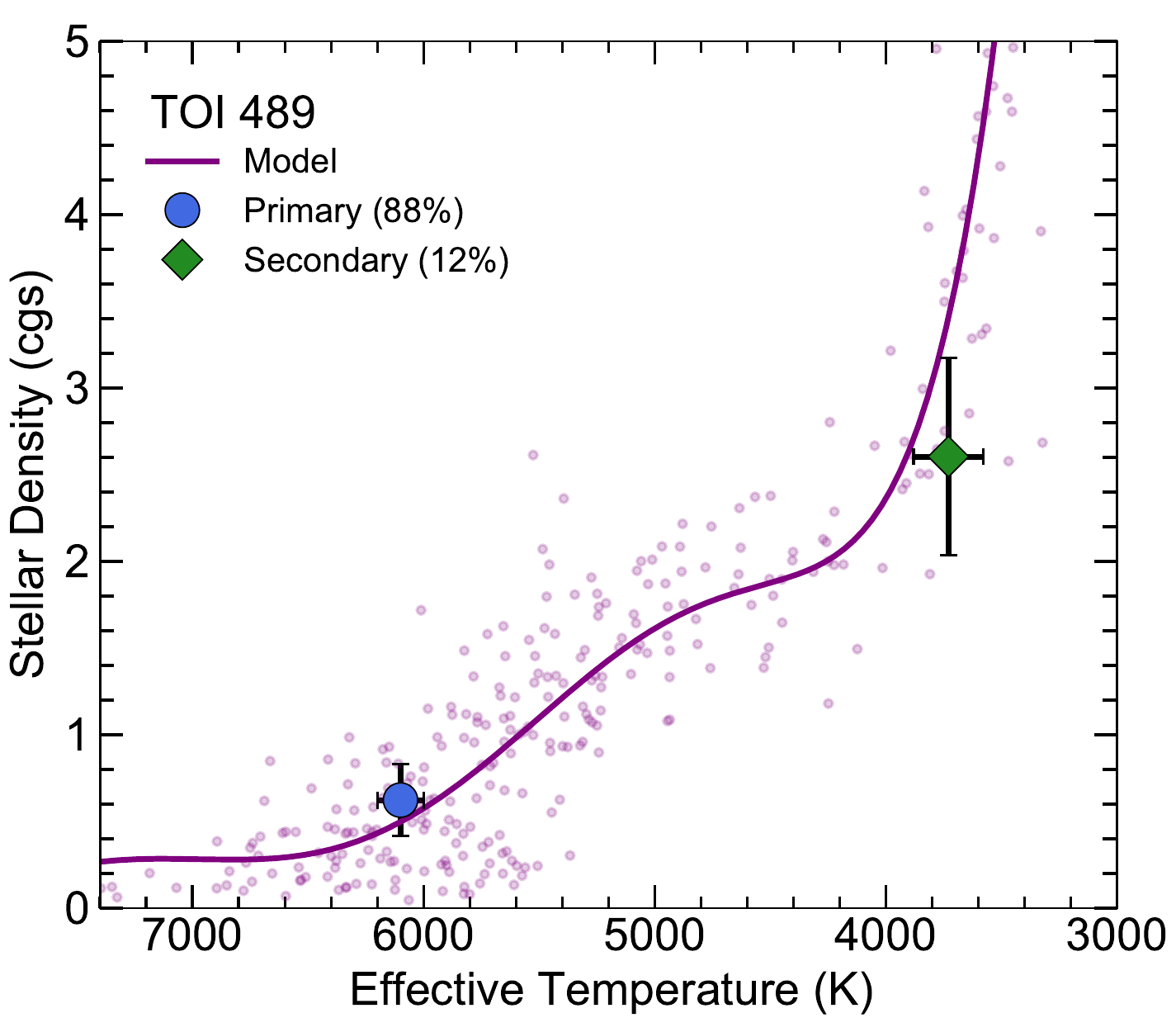}
\includegraphics[width=0.329\textwidth]{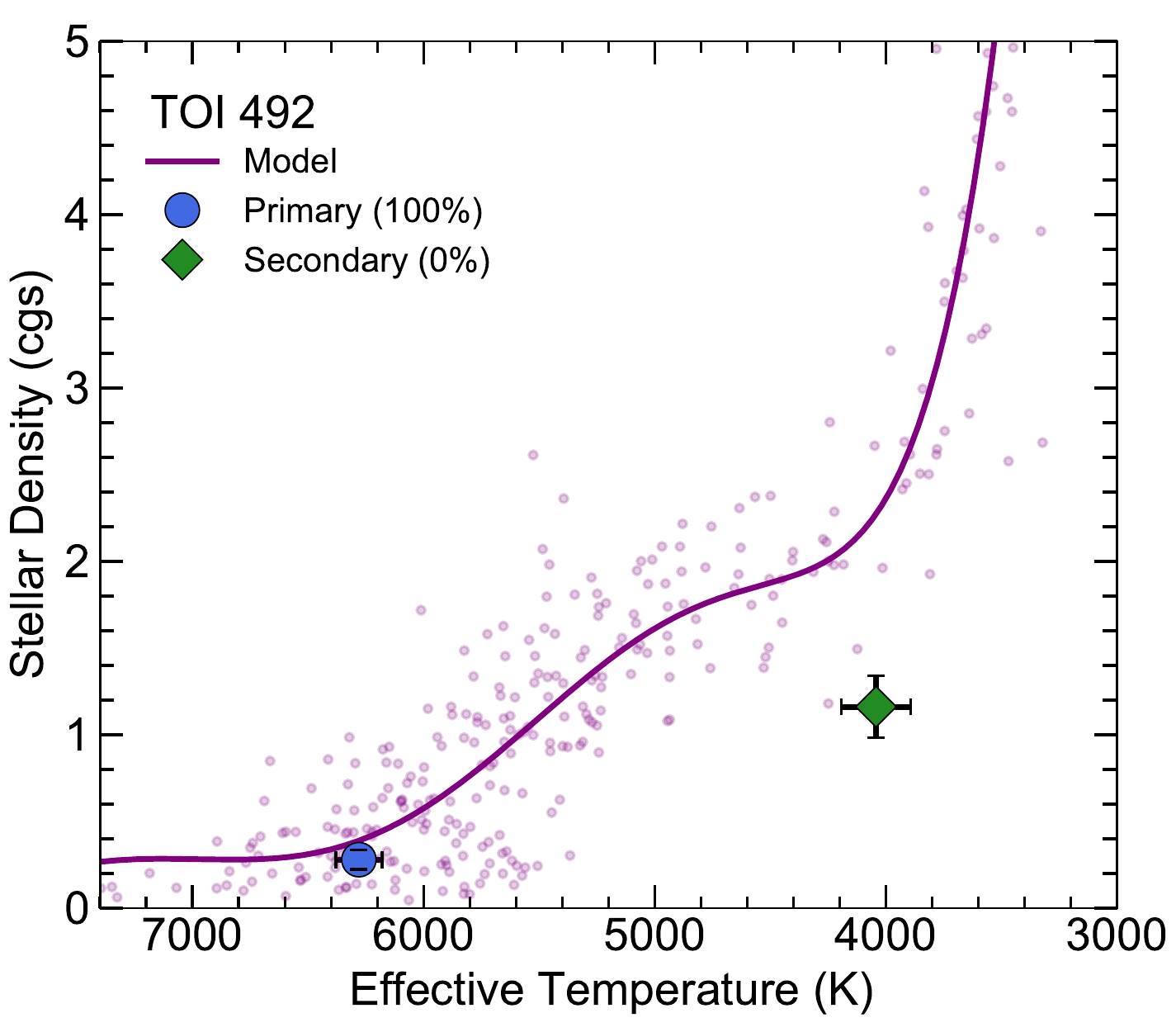}
\caption{Stellar density vs. effective temperature plots for TOI 141 -- 492. The dervied value for the primary component is shown as the blue circle, the secondary component is shown as the green diamond, and any tertiary components are shown as red triangles. The solid purple line shows the  model density-$T_{\rm eff}$ relation built from the single TOI's (purple points).  The host probabilities of each component are also listed.
\label{group3}}
\end{figure*}

\begin{figure*}
\includegraphics[width=0.329\textwidth]{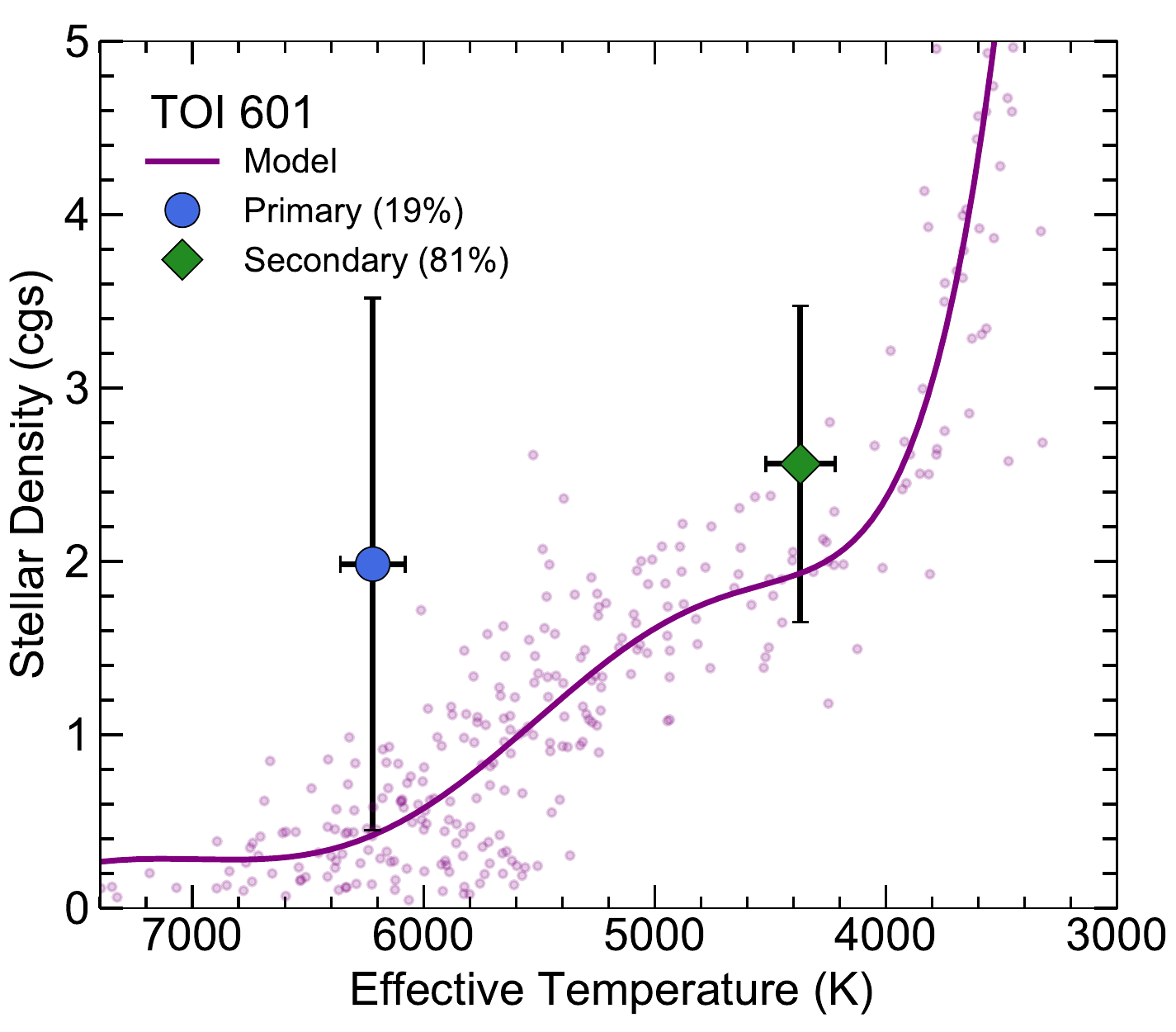}
\includegraphics[width=0.329\textwidth]{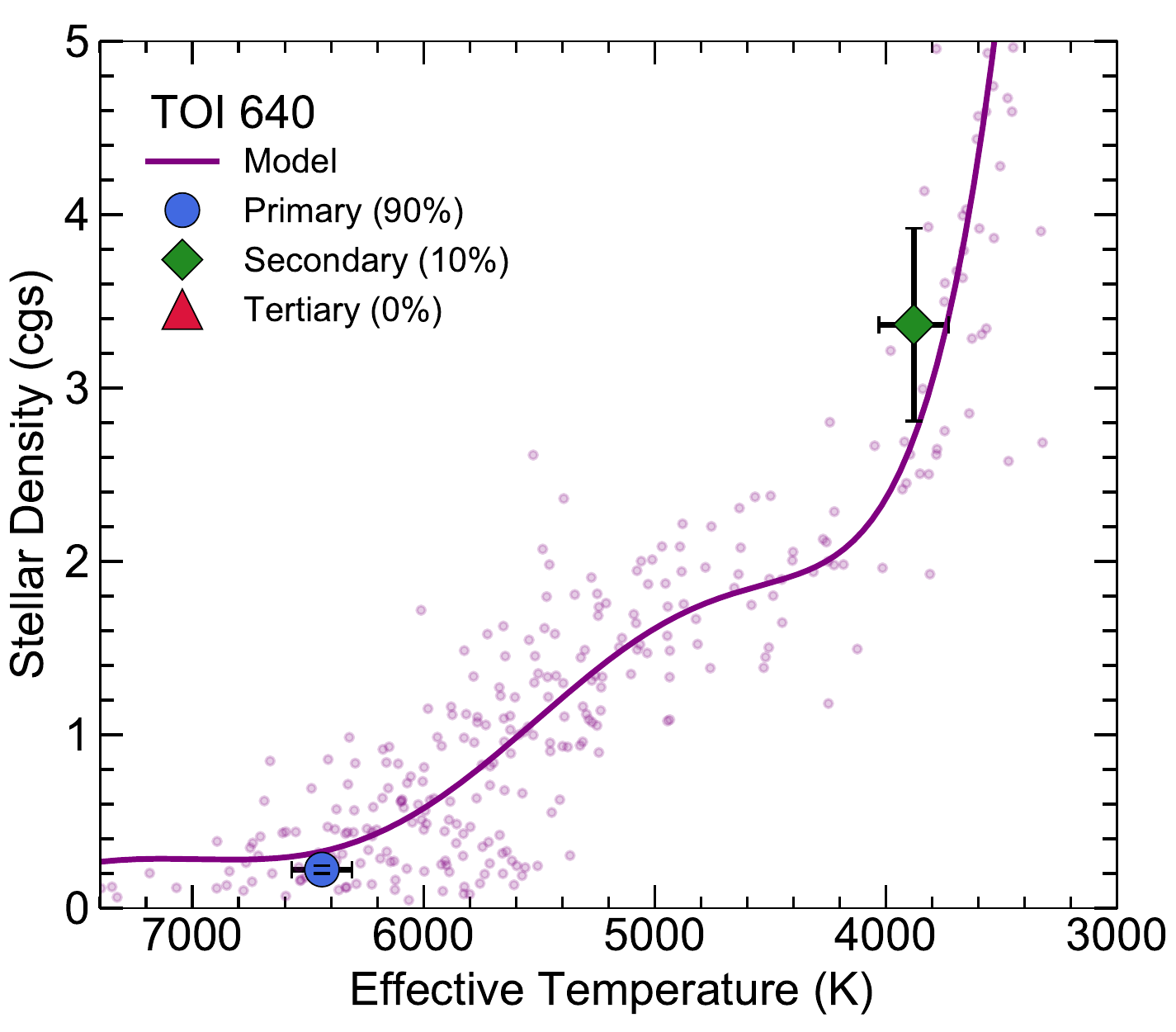}
\includegraphics[width=0.329\textwidth]{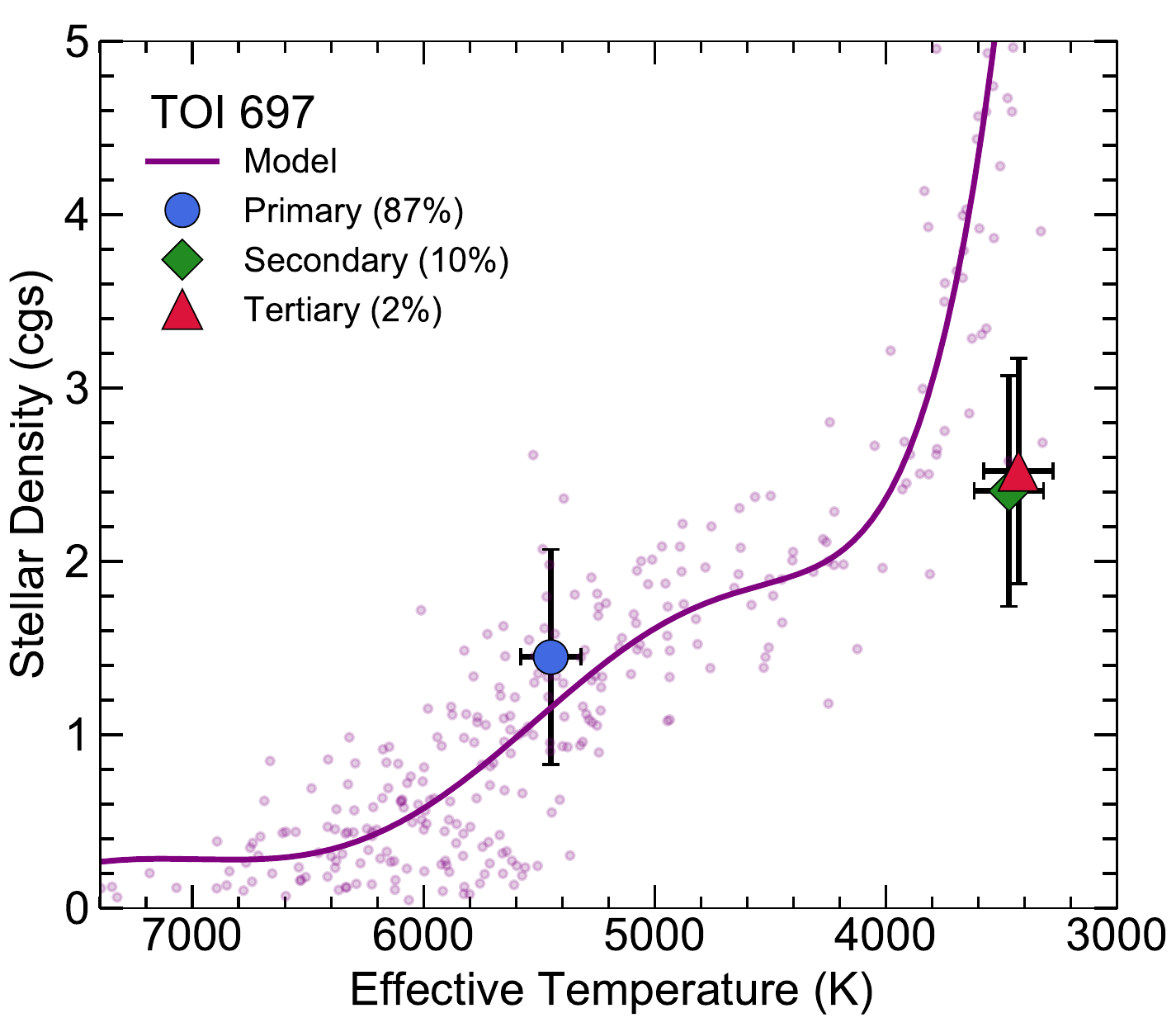}
\includegraphics[width=0.329\textwidth]{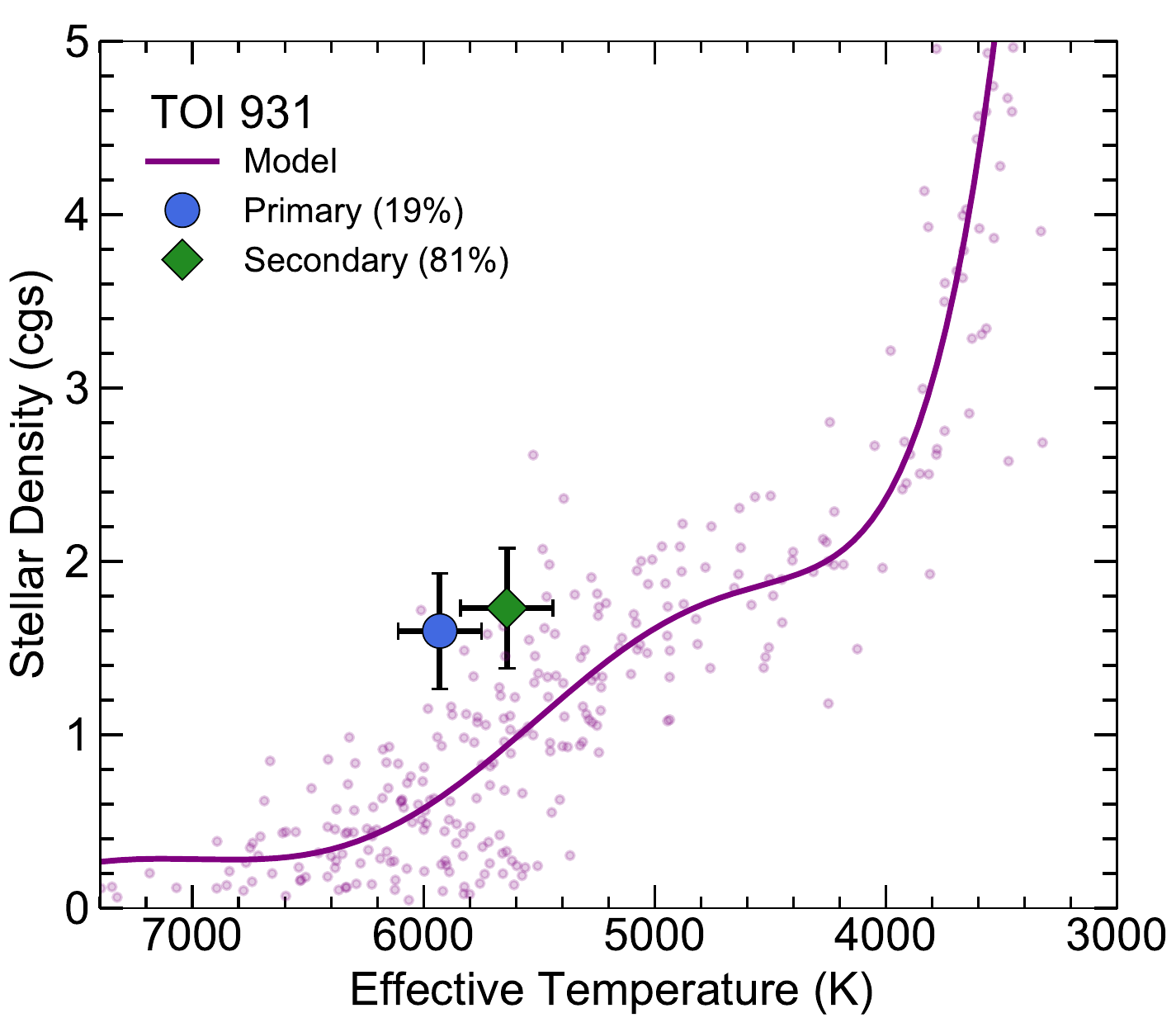}
\includegraphics[width=0.329\textwidth]{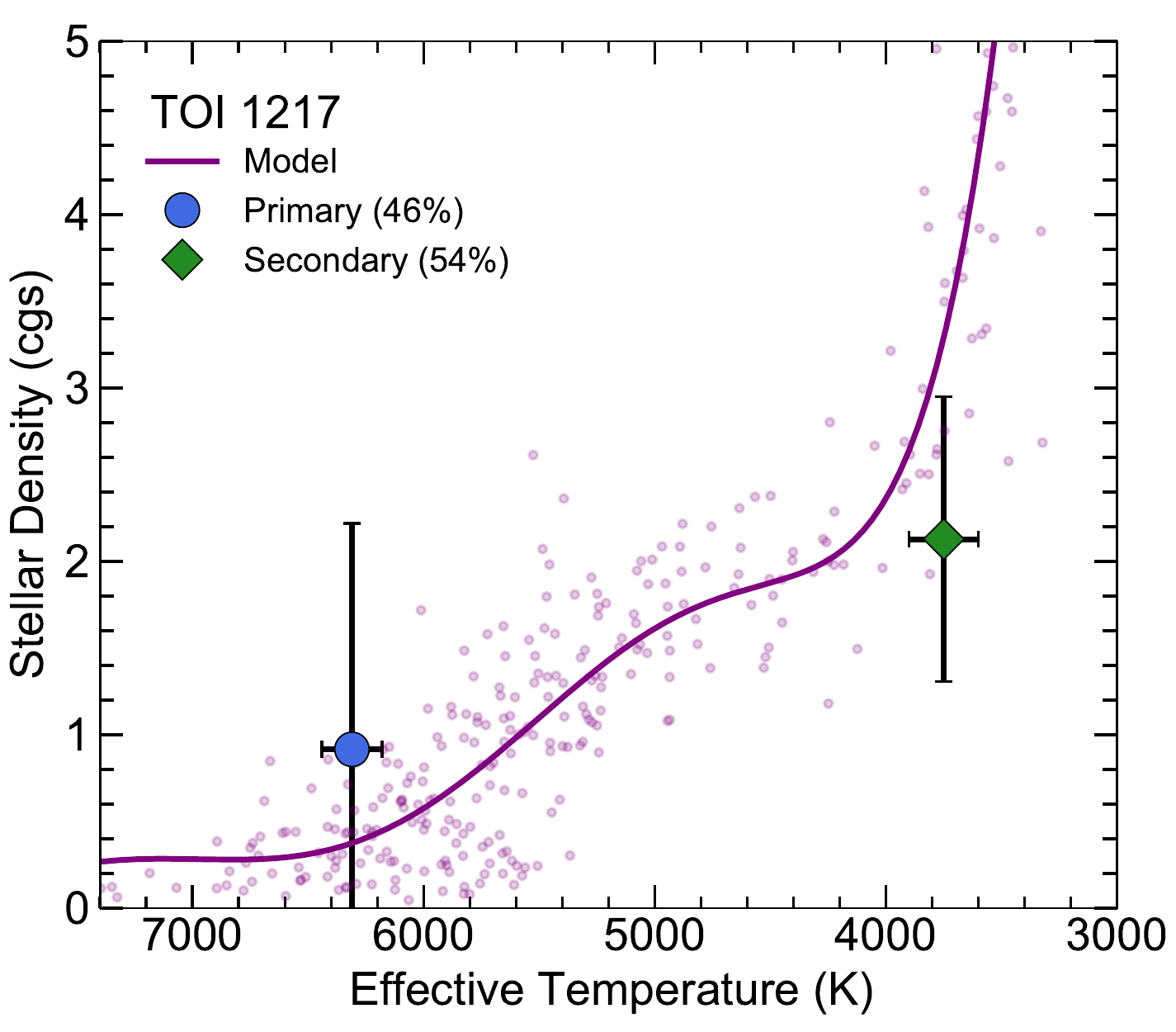}
\includegraphics[width=0.329\textwidth]{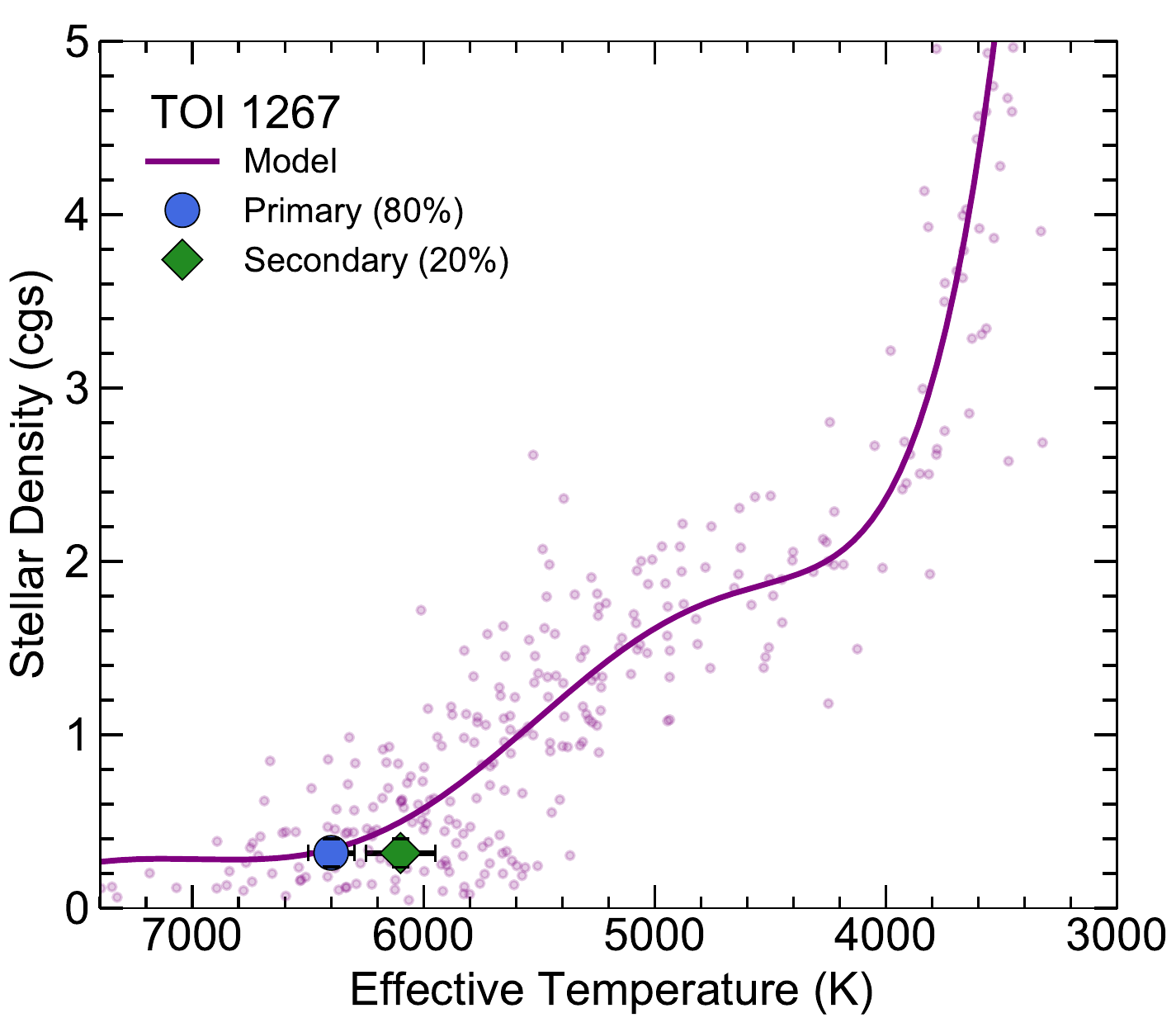}
\includegraphics[width=0.329\textwidth]{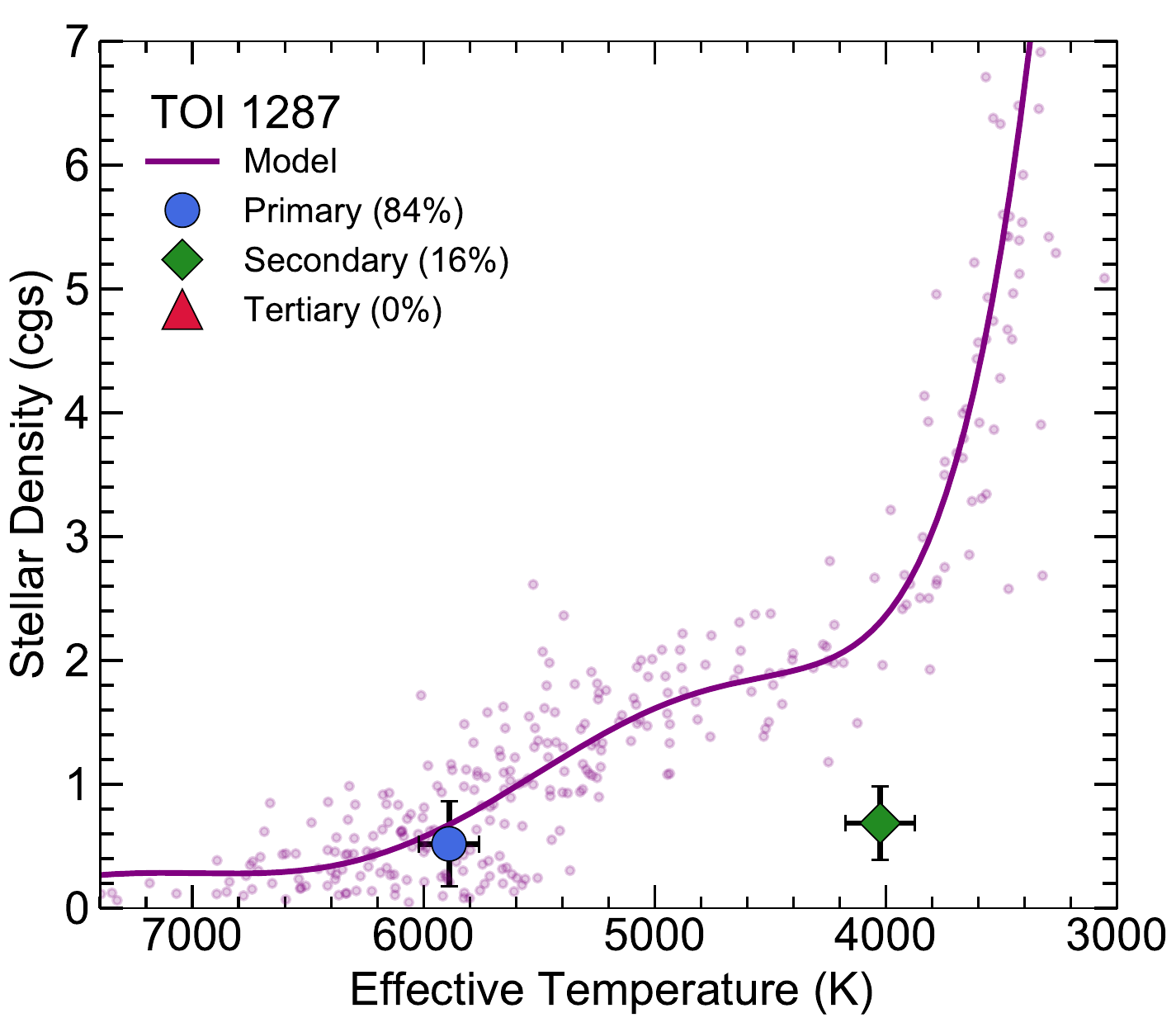}
\includegraphics[width=0.329\textwidth]{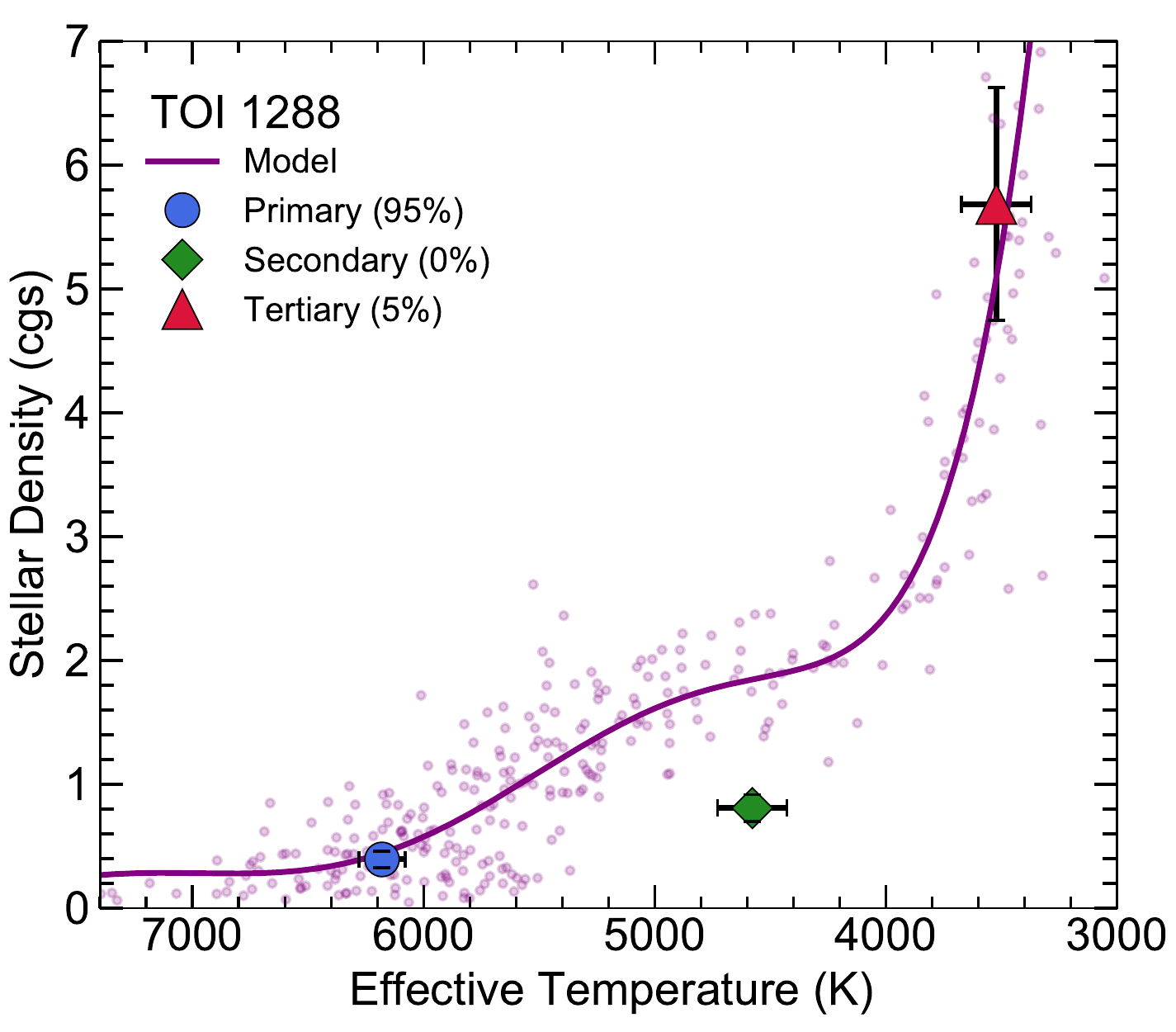}
\includegraphics[width=0.329\textwidth]{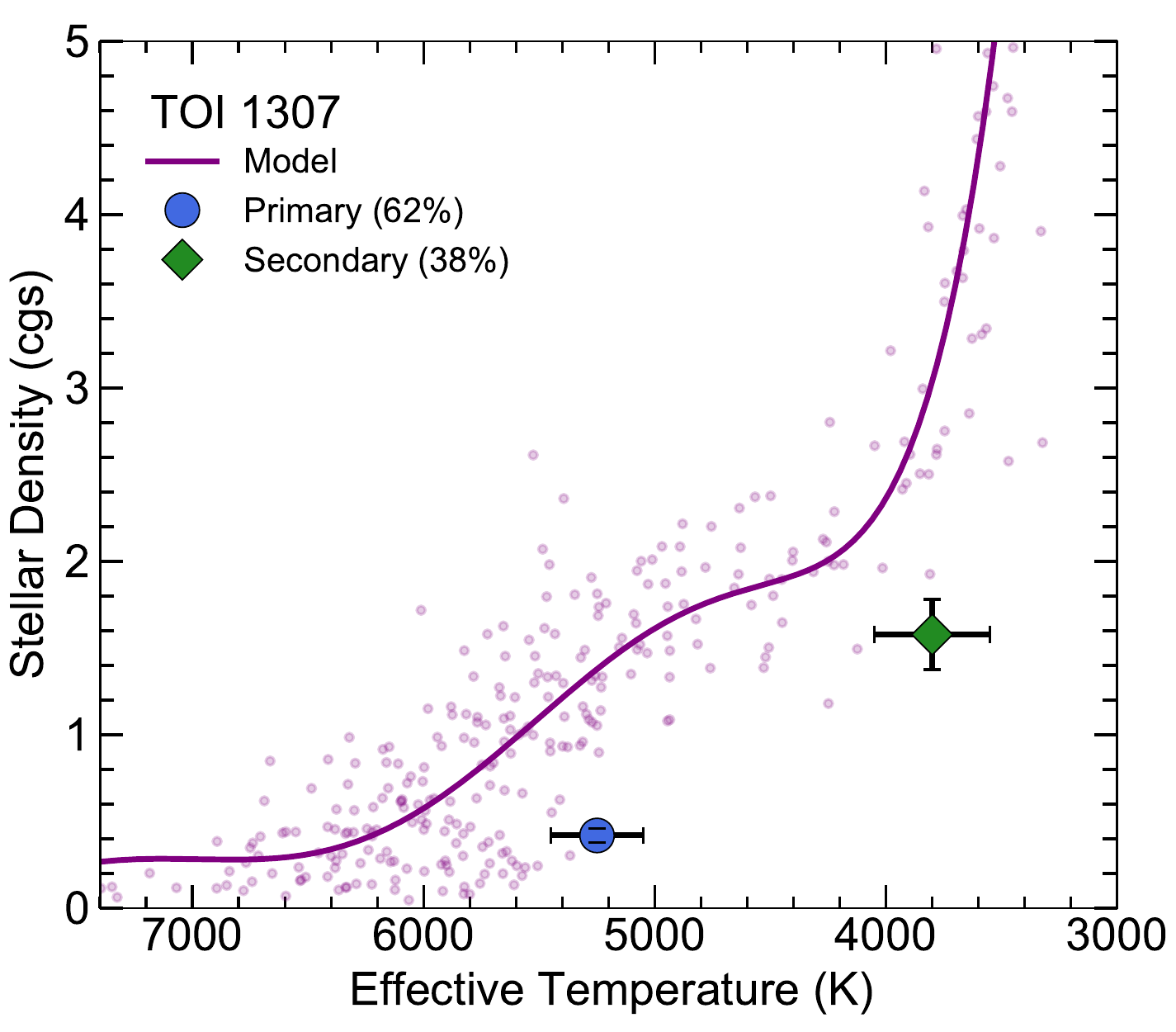}
\includegraphics[width=0.329\textwidth]{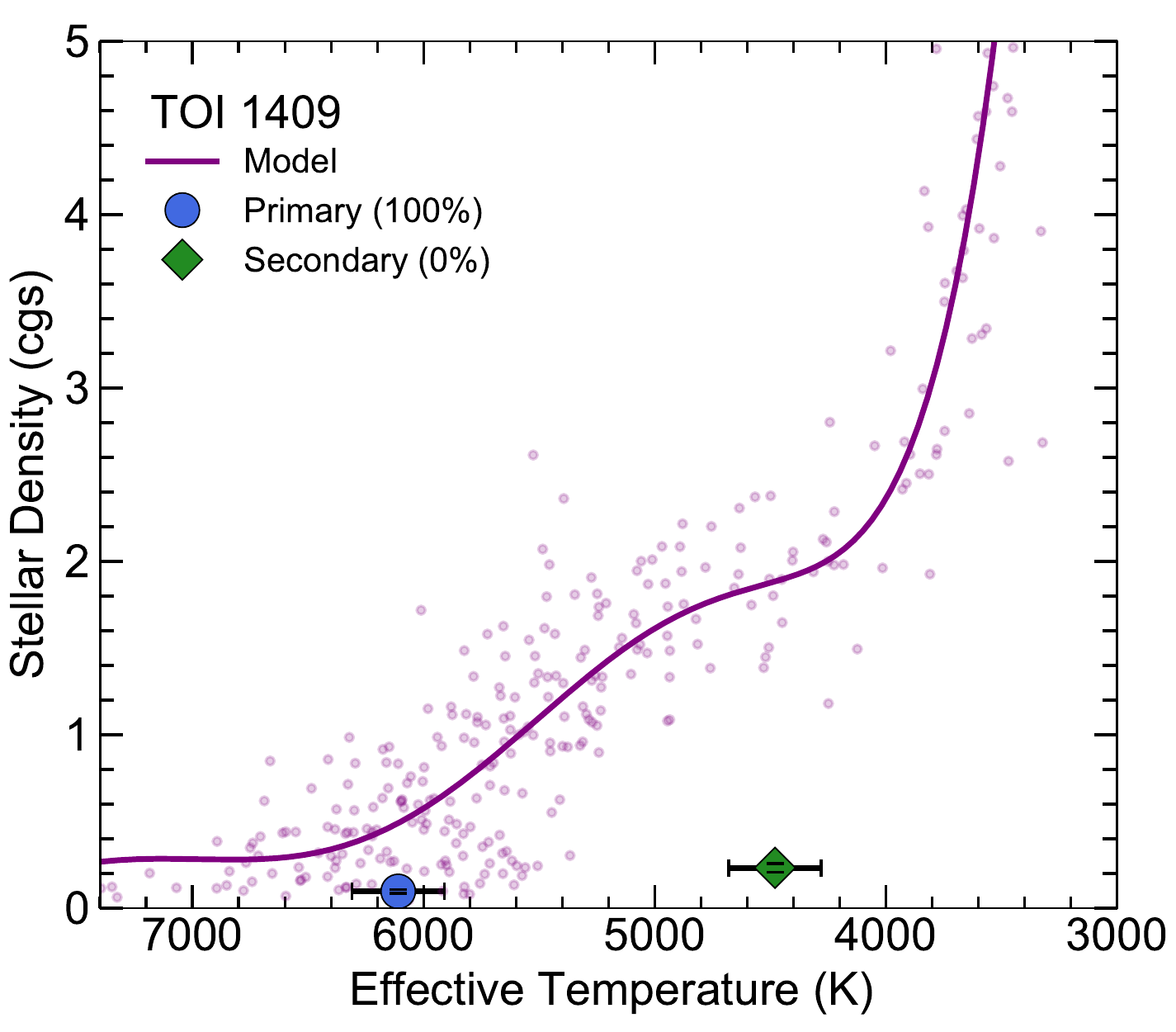}  
\includegraphics[width=0.329\textwidth]{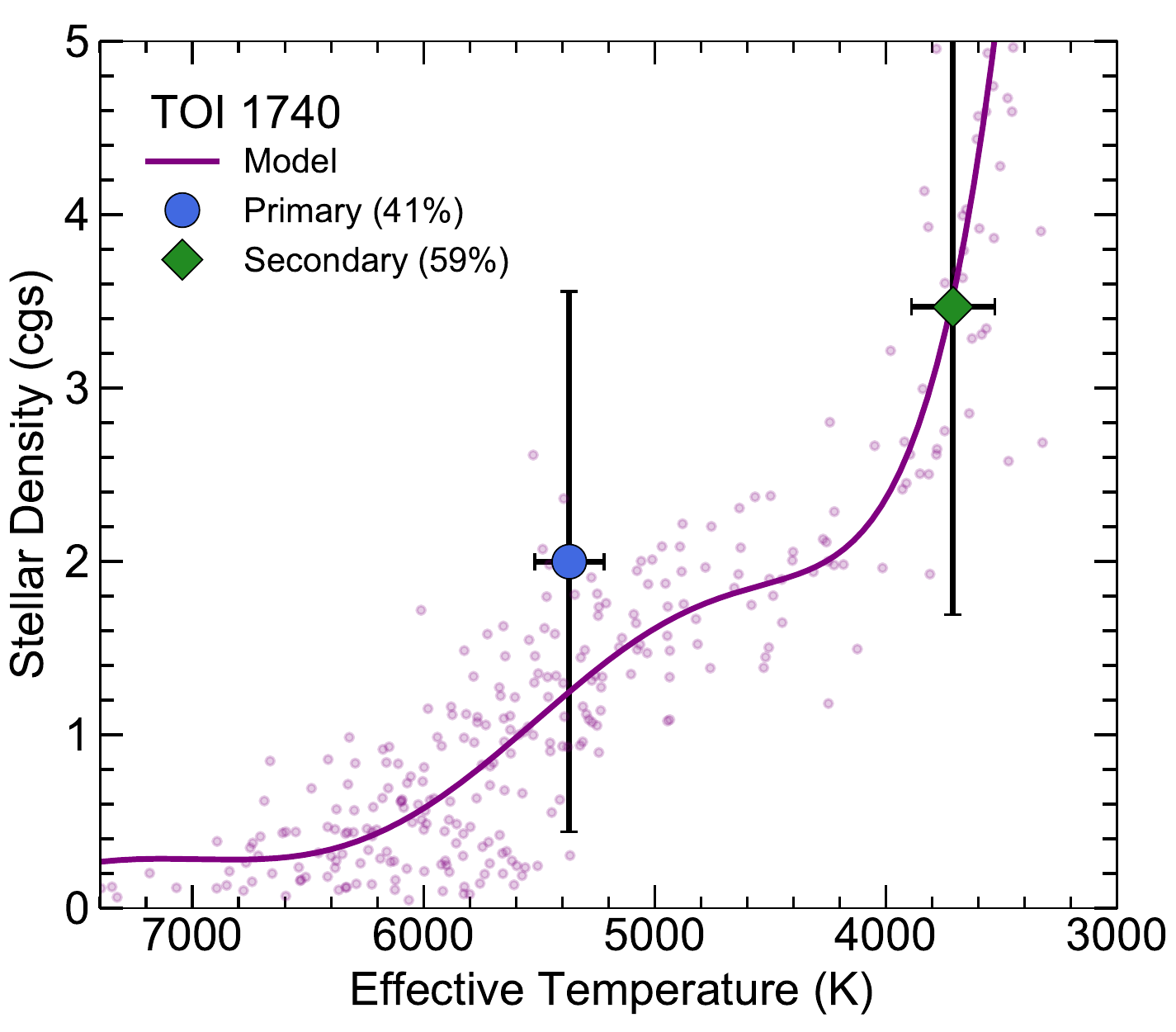}
\caption{Stellar density vs. effective temperature plots for TOI 601 -- 1740. The dervied value for the primary component is shown as the blue circle, the secondary component is shown as the green diamond, and any tertiary components are shown as red triangles.  The solid purple line shows the  model density-$T_{\rm eff}$ relation built from the single TOI's (purple points).  The host probabilities of each component are also listed.
\label{group4}}
\end{figure*}

We compared the observed stellar densities to model values to determine which component provides the best match. To build our model density-$T_{\rm eff}$ curve, we started with TOI’s found to be single stars with Gemini speckle interferometry \citep{lester21}. Speckle with Gemini is more sensitive to companions than with WIYN or SOAR due to the larger aperture and better angular resolution, so this sample would be the least contaminated by unresolved binaries. We pulled the TIC parameters for each TOI, kept only stars with uncertainties in radius, temperature, and mass less than 20\%, and calculated the mean density for each system. Following the method of \citet{payne18}, we then created a model density-$T_{\rm eff}$ curve by fitting a 6th-order polynomial to these temperatures and densities. Figures~\ref{group3} and \ref{group4} show the stellar density vs. effective temperature plots for all of the TOI's in our sample.

Next, we calculated the distance of the observed stellar parameters from this density-$T_{\rm eff}$ model (in terms of $\sigma$) for each of the 10,000 iterations used in the previous section.  Finally, we tallied how many times each component was closest to the model and adopted this percentage as the host star probability ($P_{1,2,3}$). We classified each system system as ``likely primary host" for $P_1 \ge 0.7$, ``likely secondary host" for $P_2 \ge 0.7$, ``likely tertiary host" for $P_3 \ge 0.7$, and ``uncertain host" for $0.3 < P_{1} < 0.7$. The host probability of each binary component and classifications are listed in Table~\ref{densities}, and we discuss each system in detail in the next section. 

We also calculated the planet radius correction factor ($X_R$) using the equations in \citet{ciardi15} for the systems where the host component was identified (listed in Table~\ref{densities}). The radius correction for planets orbiting the primary star depends only on the binary magnitude difference, while the radius corrections for planets orbiting the secondary star depends on the magnitude difference and the radius ratio of the stars. We do not have individual colors or age information for our binaries, so we estimated the radii of each component by interpolating the values in the Modern Mean Dwarf Stellar Color and Effective Temperature Sequence \citep{pecaut13} at the effective temperatures of each component. As a self check, we also estimated the radius ratio using a 1~Gyr MIST isochrone \citep{mist1, mist2} and found similar results to within 5\%. The planet radius corrections are typically small ($\overline{X_R} = 1.07$) when the planet orbits the primary star but large ($\overline{X_R} = 4.97$) when the planet orbits the secondary star.

\input{table_densities.txt}

\subsection{Notes on Individual Systems}

\textbf{TOI 141} -- 
The observed parameters of the primary component match the model relation well;  TOI 141 has a 95\% probability that the primary star hosts the transiting planet. This is consistent with past studies of TOI 141, as \citet{espinoza20} confirmed the planetary nature of TOI 141 b and found that it orbits the primary star through spectroscopic follow-up. Because TOI 141 b likely orbits the primary star, the radius correction factor is only 1.01.

\textbf{TOI 159} -- 
The observed parameters of both components match the model relation equally well, so TOI 159 has a 42\% and 58\% probability that the primary and secondary star host the transiting planet, respectively. The transit geometry and stellar densities are well constrained, so perhaps the binary magnitude difference is causing a poor fit to the model relation. TOI 159 b creates a deep transit, so it's also possible that it could be a red dwarf companion instead. This would be a good candidate to follow-up to confirm the binary parameters and nature of the transiting object. This would also help constrain the planet candidate's radius, as the correction factor for TOI 159 b would be 1.02 if it orbits the primary and 10.66 if it orbits the secondary. However, such a large correction would cause the planet density to be unrealistically small \citep{furlan20}.

\textbf{TOI 172} -- 
The observed parameters of the primary component match the model relation well; TOI 172 has a 96\% probability that the primary star hosts the transiting planet. This is consistent with the past study confirming TOI 172 b as a planet by \citet{rodriguez19} through spectroscopic follow-up. Because TOI 141 b likely orbits the primary star, the radius correction factor is only 1.01.

\textbf{TOI 194} -- 
TOI 194 has a 72\% and 28\% probability that the primary and secondary star host the transiting planet, respectively. A primary component host star is consistent with the study confirming TOI 194 b as a planet (WASP-20b) by \citet{anderson15} through radial velocity analysis, but this study did not list a stellar companion so the derived planet radius is underestimated by a factor of 1.18.

\textbf{TOI 264} -- 
The observed parameters of the primary component match the model relation well; TOI 264 has a 100\% probability that the primary star hosts the transiting planet. This is consistent with the past study confirming TOI 264 b as a planet (WASP-72b) by \citet{gillon2013}. This study did not list a stellar companion, but the radius correction factor is only 1.01, so the planet radius would not change within the uncertainties.

\textbf{TOI 295} -- 
The observed parameters of the primary component match the model relation well; TOI 295 has a 100\% probability that the primary star hosts the transiting planet. This is consistent with the past spectroscopic analysis to confirm TOI 295 b as a planet (WASP-146b) by \citet{anderson18}. This study did not list a stellar companion, but the radius correction factor is only 1.01, so the planet radius would not change within the uncertainties.

\textbf{TOI 309} -- 
Due to the large impact parameter found by the transit fit, the stellar densities are poorly constrained. We found that TOI 309 has a 31\% and 69\% probability that the primary and secondary star host the transiting planet, respectively, so the host star is uncertain. If the planet orbits the primary, the radius correction factor would be 1.18, but if the planet orbits the secondary, the correction factor would be 2.20 and largely change the derived planet density and classification. Additional follow-up is needed to rule out any background eclipsing binaries or false positive scenarios and to better characterize the system before a host star can be identified.

\textbf{TOI 322} -- 
The observed parameters of the primary component match the model relation slightly better than that of the secondary component. We found that TOI 322 has an 80\% and 20\% probability that the primary and secondary star host the transiting planet, respectively, so the radius correction would be a factor of 1.23.

\textbf{TOI 462} -- 
The observed parameters of the primary component match the model relation slightly better than that of the secondary component. We found that TOI 462 has an 73\% and 27\% probability that the primary and secondary star host the transiting planet, respectively, so the planet radius correction would be a factor of 1.27.

\textbf{TOI 487} -- 
The observed parameters of the primary component match the model relation well; TOI 487 has a 93\% and 7\% probability that the primary and secondary star host the transiting planet, respectively. The radius correction factor for TOI 487 b would be 1.15.

\textbf{TOI 489} -- 
The observed parameters of the primary component match the model relation well; TOI 489 has an 88\% and 12\% probability that the primary and secondary star host the transiting planet, respectively. This is consistent with the past study confirming TOI 489 b as a planet (HAT-P-35b) by \citet{bakos12} through spectroscopic analysis. This study did not list a stellar companion, but the radius correction factor is only 1.01, so the planet radius would not change within the uncertainties.

\textbf{TOI 492} -- 
The observed parameters of the primary component match the model relation well; TOI 492 has a 100\% probability that the primary star hosts the transiting planet. This is consistent with the past study confirming TOI 492 b as a planet (HAT-P-50b) by \citet{hartman15}.  This study did not list a stellar companion, but did test light curve fits that included an unseen companion and could not rule out scenarios with a faint secondary star (as found later by speckle). However, the planet radius correction factor is only 1.01, so the derived planet radius would not change within the uncertainties.

\textbf{TOI 601} -- 
The observed parameters of the secondary component match the model relation better than those of the primary component, making TOI 601 one of the few ``likely secondary host" systems in our sample. TOI 601 has a 19\% and 81\% probability that the primary and secondary stars host the transiting planet, respectively. However, the stellar density uncertainties are rather large due to the low SNR transit, so we recommend additional follow-up to better characterize this system, constrain the stellar densities, and confirm the identity of the host star. This would also help constrain the planet radius, as the correction factor for TOI 601 b would be minor (1.03) if it orbits the primary but quite large (8.08) if it orbits the secondary. However, such a large correction would cause the planet density to be unphysically small \citep{furlan20}.

\textbf{TOI 640} -- 
The observed parameters of the primary component match the model relation well; TOI 640 has a 90\% probability that the primary star hosts the transiting planet. TOI 640 b was confirmed to be a planet by \citet{rodriguez21}, but they reported that no stellar companions were found by SOAR, even though the source paper did report a 0.25" companion \citep{ziegler20}. Nonetheless, the secondary star is 4.8~mag fainter than the primary and unlikely to be seen in spectroscopic follow-up, so the radial velocity signal detected by \citet{rodriguez21} would also confirm the primary star hosts the planet. The radius correction factor for TOI 640 b is only 1.01, so the planet radius would not be largely affected.

\textbf{TOI 697} -- 
The observed parameters of the primary component match the model relation well. TOI 697 has a 87\%, 10\%, and 2\% probability that the primary, secondary, and tertiary stars host the transiting planet, respectively. Because the planet likely orbits the primary, the radius correction is a factor of only 1.01.

\textbf{TOI 931} -- 
Neither the primary or secondary components fit the model relation very well, but the secondary  matches slightly better. We found that TOI 931 has a 19\% and 81\% probability that the primary and secondary stars host the transiting planet, respectively. The radius correction factor for TOI 931 b would be 1.86.

\textbf{TOI 1217} -- 
The observed parameters of both components match the model relation due to the large uncertainties. We found that TOI 1217 has a 46\% and 54\% probability that the primary and secondary stars host the transiting planet, respectively, so the host star is uncertain until additional photometric observations can better constrain the transit geometry.  This would also help constrain the planet radius, as the correction factor for TOI 1217 b would be minor (1.01) if it orbits the primary but very large (26.99). However, such a large correction would cause the planet density to be unphysically small \citep{furlan20}.

\textbf{TOI 1267} -- 
The observed parameters of the primary component match the model relation slightly better than those of the secondary. We found that TOI 1267 has a 80\% and 20\% probability that the primary and secondary stars host the transiting planet, respectively. This is consistent with the study confirming TOI 492 b as a planet (Kepler-14b) through spectroscopic follow-up by \citet{buchave11}, who concluded the planet orbits the primary star using a centroid analysis of the Kepler data and corrected the derived planet parameters accordingly.

\textbf{TOI 1287} -- 
The observed parameters of the primary component match the model relation well; TOI 1287 has an 84\% and 16\% probability that the primary and secondary stars host the transiting planet, respectively. The radius correction for TOI 1287 b would therefore be a factor of 1.02.

\textbf{TOI 1288} -- 
The observed parameters of the primary component match the model relation well. TOI 1288 has a 95\%, 0\%, and 5\% probability that the primary, secondary, and tertiary stars host the transiting planet, respectively. The radius correction for TOI 1288 b would therefore be a factor of 1.05.

\textbf{TOI 1307} -- 
Neither the primary or secondary components fit the model relation very well, so the host star is uncertain. We found that TOI 1307 has a 62\% and 38\% probability that the primary and secondary stars host the transiting planet, respectively. If the planet orbits the primary star, the radius correction factor would be only 1.03, but if it orbits the secondary star, the correction would be a factor of 6.54. The transit geometry is well constrained, so the poor fit to the model relation is likely due to the primary star's evolutionary state. The ExoFOP website notes that TOI 1307 is slightly evolved, which would cause the observed density to be lower than the model (made from mostly main sequence TOI's).

\textbf{TOI 1409} -- 
The observed parameters of the primary component match the model relation well; we found that TOI 1409 has a 100\% probability that the primary stars hosts the transiting planet. The planet radius correction would therefore be a factor of 1.04.

\textbf{TOI 1740} -- 
The observed parameters of both components match the model relation due to the large uncertainties in the density, so the host star is uncertain. We found that TOI 1740 has a 41\% and 59\% probability that the primary and secondary stars host the transiting planet, respectively. If the planet orbits the primary star, the radius correction factor would be only 1.02, but if it orbits the secondary star, the correction would be a factor of 9.37. Additional photometric observations are needed to better constrain the transit geometry and identify the planet host.

\subsection{Overall Results}

\begin{figure*}
\centering
\includegraphics[width=0.8\textwidth]{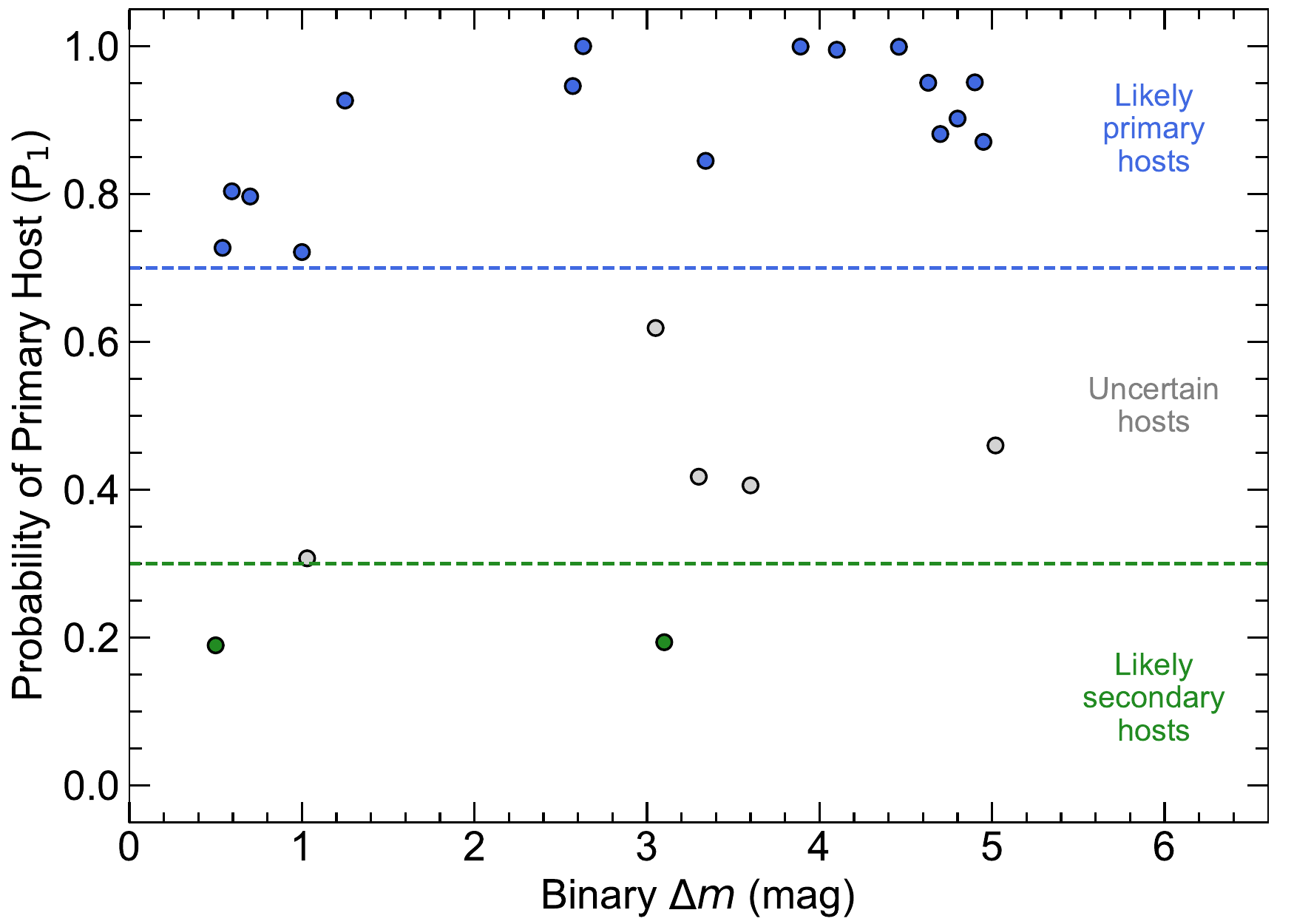}
\caption{The probability of a primary star host ($P_1$) vs. the binary magnitude difference ($\Delta m$) for each binary in our sample. The ``likely primary host" systems are plotted in blue, the ``likely secondary host" systems are plotted in green, and the ``uncertain host" systems are plotted in grey. The preference for primary star hosts in systems with faint companions is easily visible.
\label{dmagplot}}
\end{figure*}

Here are the results of our study:
\begin{itemize}
\setlength\itemsep{0em}
\item Likely primary host -- 16 out of \nstars systems (70\%)
\item Likely secondary host --  2 out of \nstars systems (9\%)
\item Likely tertiary host -- 0 out of 3 systems (0\%)
\item Uncertain host -- 5 out of \nstars systems (21\%)
\end{itemize}
We found that most transiting planets in our sample (70\%) orbit the primary star in their binary system. Figure~\ref{dmagplot} shows the primary host star probability as a function of the the binary magnitude difference. While we do not see the strong trend found by \citet{payne18}, where the host component in equal brightness pairs were uncertain due to the similarity in temperature and density, the equal brightness systems in our sample do have slightly lower primary host probabilities that unequal brightness pairs. Furthermore, the systems with faint secondaries ($\Delta m \ge$ 4~mag) are even more likely to have a primary star host (89\%), as seen in Figure~\ref{dmagplot}. This is because the light from the primary star washes out the transit signal from a planet orbiting the secondary (or tertiary) star until it becomes too difficult to detect \citep[e.g.,][]{lester21}. This result is consistent with that of \citet{furlan17b}, who found that faint secondary host stars would result in planet densities that are too low to be physically realistic. 
Finally, we did not find any trends with planet radius or binary separation due to the small number of secondary host systems.

Note that our results apply to transiting planets in binary systems, not to all planets. The main selection effects of our study are that speckle imaging cannot detect very close binary companions ($<$20 mas) and that our transit signal-to-noise limit may exclude small planets. However, past studies \citep[e.g.,][]{kraus16, moe21, howell21} have found that binary host separation peaks around 100~AU, which is well within the speckle angular resolution limits for most TESS stars, and that the number of companions rapidly decreases towards small separations. Next, the transits of small planets ($\le 2 R_E$) in binary systems are not seen by TESS due to the observational bias from third light contamination \citep{lester21}, but the transits of giant planets commonly detected with TESS can still be seen in multistar systems. While a full completeness analysis is beyond the scope of this paper, our observational sample broadly represents the binary planet hosts found by TESS.

\section{Conclusions \label{disc}} 
Using the transit light curve from TESS and binary parameters from speckle imaging, we determined the stellar densities of each component in \nstars binary exoplanet host systems. By comparing these observed densities with models, we determined which binary component was more likely to host the planet. We found that 70\% of the TESS transiting planets in our sample are likely to orbit the primary star in their binary host system. If this holds for all transit observations, it would be reasonable in future studies to assume that the primary star hosts the transiting planet when no additional observations are available. The planets in these systems would still need a small radius correction, as the fainter companions do slightly dilute the planet transit \citep{ciardi15}.

However, we found several systems have a non-negligible chance that the planet orbits the secondary star, which would greatly affect the resulting planet parameters. Further follow-up is recommended for systems with an uncertain host star or poor density fits – likely due to low SNR transits, small number of transits, or small binary $\Delta m$. This could include additional transit observations with TESS or groundbased photometry, spectroscopic observations to search for the planet's radial velocity signal, or multi-wavelength photometric follow-up to further refine the stellar effective temperatures.

Identifying which binary component hosts the transiting planet is essential for determining the correct transit depth, the true planet radius and mean density, and the host star's properties and habitable zone \citep{ciardi15,  hirsch17, furlan17b, furlan20}. Studies of exoplanet demographics must take into account whether each planet host resides in a multi-star system to ensure the derived planet properties and occurrence rates are corrected accordingly \citep{bouma18, saval20}. Knowing which planet hosts are binary systems and which component hosts the planet is also important for target selection of upcoming coronagraph observations (e.g., by Roman Space Telescope or LUVOIR). In order to directly image planets in well-resolved binary systems, one must know ahead of time which star to block with the coronagraph or if it is possible to block the light from both stars \citep{bendek21}.

\acknowledgments
We would like to thank the anonymous referee for their helpful comments.
KVL is supported by an appointment to the NASA Postdoctoral Program at the NASA Ames Research Center, administered by Oak Ridge Associated Universities under contract with NASA.
This paper includes data collected with the TESS mission obtained from the MAST data archive at the Space Telescope Science Institute (STScI). Funding for the TESS mission is provided by the NASA Explorer Program. STScI is operated by the Association of Universities for Research in Astronomy, Inc., under NASA contract NAS 5–26555.
This work also made use of the Exoplanet Follow-up Observation Program website \citep{exofop}  and \texttt{EXOFAST} \citep{exofast} tool provided by the NASA Exoplanet Archive, which is operated by the California Institute of Technology under contract with the National Aeronautics and Space Administration under the Exoplanet Exploration Program.

\facilities{TESS, SOAR, Gemini North, Gemini South, WIYN}


\end{document}

%% file: table_atmospar.txt
\begin{deluxetable*}{lccccccccccc}
\tablewidth{0pt}
\tabletypesize{\footnotesize}
\tablecaption{Stellar Parameters \label{atmospar}    }
\tablehead{
\colhead{TOI} 
& \colhead{TIC} 
& \colhead{$T_{\rm eff 1}$ (K)} 
& \colhead{$T_{\rm eff 2}$ (K)} 
& \colhead{$T_{\rm eff 3}$ (K)} 
& \colhead{[Fe/H]} 
& \colhead{$r_{12}$ (")} 
& \colhead{$r_{13}$ (")}
& \colhead{$\Delta m_{12}$ (mag)} 
& \colhead{$\Delta m_{13}$ (mag)} 
& \colhead{Ref.}
}
\startdata 
141	 & 403224672 & $5800\pm110$ & $3640\pm150$ & $3470\pm150$ & \nodata   	    & 0.49  & 1.33     & $4.63\pm0.10$  &  $5.60\pm0.49$  & 1,2 \\
159	 & 394657039 & $6980\pm150$ & $4760\pm150$ & \nodata  	  & $0.17\pm0.12$   & 0.64  & \nodata  & $3.30\pm0.20$  &  \nodata        & 1,3 \\
172	 & 29857954  & $5760\pm130$ & $3510\pm150$ & \nodata  	  & $0.15\pm0.08$   & 1.13  & \nodata  & $5.36\pm0.20$  &  \nodata        & 2,4 \\
194	 & 211438925 & $5940\pm150$ & $5320\pm200$ & \nodata  	  &  \nodata        & 0.27  & \nodata  & $1.00\pm0.20$  &  \nodata        & 1,3 \\
264	 & 122612091 & $6250\pm100$ & $3859\pm150$ & \nodata  	  & $-0.06\pm0.09$  & 0.65  & \nodata  & $4.46\pm0.20$  &  \nodata        & 2,5 \\
295	 & 327952677 & $5860\pm120$ & $3800\pm150$ & \nodata  	  & \nodata   	    & 0.98  & \nodata  & $4.10\pm0.20$  &  \nodata        & 1,3 \\
309	 & 228507250 & $5330\pm250$ & $5140\pm250$ & \nodata  	  & \nodata   	    & 0.34  & \nodata  & $1.03\pm0.20$  &  \nodata        & 1,2 \\
322	 & 198077394 & $6500\pm200$ & $6080\pm200$ & \nodata  	  & \nodata   	    & 0.18  & \nodata  & $0.70\pm0.20$  &  \nodata        & 1,3 \\
462	 & 420049884 & $5500\pm150$ & $5240\pm150$ & \nodata  	  & \nodata   	    & 0.17  & \nodata  & $0.54\pm0.20$  &  \nodata        & 3,7 \\
487	 & 31852980  & $5450\pm130$ & $4685\pm150$ & \nodata  	  & \nodata   	    & 0.52  & \nodata  & $1.25\pm0.20$  &  \nodata        & 1,2 \\
489	 & 455096220 & $6100\pm100$ & $3730\pm150$ & \nodata  	  & $0.11\pm0.08$   & 0.94  & \nodata  & $4.70\pm0.20$  &  \nodata        & 8,9,10 \\
492	 & 17746821  & $6280\pm100$ & $4043\pm150$ & \nodata  	  & $0.18\pm0.08$   & 0.67  & \nodata  & $3.89\pm0.20$  &  \nodata        & 2,9 \\
601	 & 141363913 & $6220\pm140$ & $4370\pm150$ & \nodata  	  & \nodata   	    & 0.15  & \nodata  & $3.10\pm0.20$  &  \nodata        & 1,3 \\
640	 & 147977348 & $6440\pm130$ & $3880\pm150$ & \nodata  	  & \nodata   	    & 0.25  & \nodata  & $4.80\pm0.20$  &  \nodata        & 1,3 \\
697	 & 77253676  & $5450\pm130$ & $3468\pm150$ & $3427\pm150$ & \nodata   	    & 1.17  & 1.27     & $4.95\pm0.20$  &  $5.09\pm0.20$  & 1,2 \\
931	 & 206474443 & $5930\pm150$ & $5640\pm200$ & \nodata      & \nodata   	    & 0.12  & \nodata  & $0.50\pm0.10$  &  \nodata        & 1,2 \\
1217 & 248092710 & $6310\pm130$ & $3750\pm150$ & \nodata	  &	\nodata         & 1.17  & \nodata  & $5.02\pm0.20$  & \nodata 	  	  & 1,2 \\
1267 & 158561566 & $6400\pm100$ & $6100\pm150$ & \nodata	  &	$0.12\pm0.06$   & 0.29  & \nodata  & $0.59\pm0.20$  & \nodata 	  	  & 11,12 \\
1287 & 352764091 & $5890\pm130$ & $4025\pm150$ & \nodata	  &	\nodata         & 0.10  & \nodata  & $3.34\pm0.20$  & \nodata 	  	  & 1,12  \\
1288 & 365733349 & $6180\pm100$ & $4578\pm150$ & $3523\pm150$ & \nodata   	    & 0.07  & 1.12     & $2.57\pm0.20$  &  $5.90\pm0.20$  & 1,12 \\
1307 & 236815160 & $5250\pm200$ & $3800\pm250$ & \nodata      & \nodata   	    & 0.33  & \nodata  & $3.05\pm0.20$  &  \nodata        & 7,12 \\
1409 & 115771549 & $6110\pm200$ & $4480\pm200$ & \nodata      & \nodata   	    & 0.46  & \nodata  & $2.63\pm0.20$  &  \nodata        & 1,2 \\
1740 & 174041208 & $5370\pm150$ & $3710\pm180$ & \nodata      & \nodata   	    & 0.24  & \nodata  & $3.60\pm0.20$  &  \nodata        & 1,2 \\
\enddata  
\tablerefs{
1. \citet{tic81}, 
2. \citet{lester21}, 
3. \citet{ziegler20},
4. \citet{rodriguez19}, 
5. \citet{gillon2013}, 	
6. \citet{dumusque19},  
7. Spectroscopic follow-up analysis on ExoFOP, 
8. \citet{stassun17},    
9. \citet{bonomo17},
10. \citet{ziegler21},
11. \citet{buchave11}, 
12. \citet{howell21}
}
\end{deluxetable*}

%% file: table_transitpar.txt
\begin{deluxetable*}{lccccccc}
\tablewidth{0pt}
\tabletypesize{\small}
\tablecaption{Transit Parameters \label{transitpar}    }
\tablehead{
\colhead{Planet} 
& \colhead{$P$ (d)} 
& \colhead{$\delta$ (ppt)\tablenotemark{*}}
& \colhead{$t_T$ (d)}
& \colhead{$b$} 
& \colhead{$R_{pl}$ ($R_J$)\tablenotemark{*}} 
& \colhead{SNR}
}
\startdata 
TOI 141 b	 &  $ 1.007933\pm0.000173$  &  	$ 0.20 \pm 0.01 $    &  $0.066\pm0.003$   &  $0.65\pm0.04$  &   $0.16\pm0.01$    & 6.9     \\
TOI 159 b	 &  $ 3.762837\pm0.000002$  &  	$10.00 \pm 0.09 $    &  $0.169\pm0.002$   &  $0.28\pm0.08$  &   $1.80\pm0.77$    & 87.4    \\   
TOI 172 b	 &  $ 9.476893\pm0.000022$  &  	$ 3.11 \pm 0.07 $    &  $0.196\pm0.002$   &  $0.23\pm0.10$  &   $0.97\pm0.03$    & 7.8     \\
TOI 194 b	 &  $ 4.899643\pm0.000003$  &  	$ 9.76 \pm 0.11 $    &  $0.140\pm0.001$   &  $0.70\pm0.02$  &   $1.29\pm0.04$    & 63.9     \\
TOI 264 b	 &  $ 2.216735\pm0.000001$  &  	$ 4.35 \pm 0.06 $    &  $0.139\pm0.003$   &  $0.56\pm0.01$  &   $1.07\pm0.05$    & 57.6    \\
TOI 295 b	 &  $ 3.397068\pm0.000293$  &  	$10.20 \pm 0.43 $    &  $0.099\pm0.003$   &  $0.80\pm0.03$  &   $1.36\pm0.52$    & 18.8    \\
TOI 309 b	 &  $ 6.156681\pm0.000807$  &  	$19.60 \pm 5.71 $    &  $0.118\pm0.006$   &  $0.92\pm0.32$  &   $2.16\pm3.44$    & 24.9    \\   
TOI 322 b	 &  $ 3.990584\pm0.000055$  &  	$ 3.33 \pm 0.12 $    &  $0.154\pm0.002$   &  $0.75\pm0.06$  &   $1.03\pm0.11$    & 27.3    \\
TOI 462	b	 &  $ 4.107456\pm0.000027$  &  	$ 0.79 \pm 0.10 $    &  $0.132\pm0.008$   &  $0.42\pm0.33$  &   $0.46\pm0.24$    & 4.2     \\
TOI 487	b	 &  $ 7.414321\pm0.000480$  &  	$ 0.35 \pm 0.04 $    &  $0.158\pm0.007$   &  $0.39\pm0.32$  &   $0.33\pm0.18$    & 2.9     \\
TOI 489	b	 &  $ 3.646659\pm0.000004$  &  	$ 8.05 \pm 0.27 $    &  $0.163\pm0.003$   &  $0.39\pm0.22$  &   $1.29\pm0.15$    & 24.2    \\
TOI 492	b	 &  $ 3.122008\pm0.000003$  &  	$ 6.19 \pm 0.18 $    &  $0.153\pm0.004$   &  $0.73\pm0.06$  &   $1.31\pm0.05$    & 27.1    \\
TOI 601	b	 &  $ 3.477689\pm0.000708$  &  	$ 1.16 \pm 0.19 $    &  $0.092\pm0.008$   &  $0.60\pm0.41$  &   $0.51\pm0.25$    & 7.0     \\
TOI 640	b	 &  $ 5.003812\pm0.000112$  &  	$ 7.70 \pm 0.15 $    &  $0.152\pm0.001$   &  $0.89\pm0.01$  &   $1.78\pm0.29$    & 55.4    \\
TOI 697	b	 &  $ 8.607870\pm0.000055$  &  	$ 0.50 \pm 0.08 $    &  $0.147\pm0.007$   &  $0.50\pm0.34$  &   $0.42\pm0.23$    & 5.8     \\
TOI 931	b	 &  $ 3.245232\pm0.000179$  &  	$ 6.21 \pm 0.22 $    &  $0.116\pm0.002$   &  $0.28\pm0.20$  &   $0.78\pm0.09$    & 18.8    \\
TOI 1217 b	 &  $41.485403\pm0.000492$  &  	$ 0.95 \pm 0.17 $    &  $0.256\pm0.021$   &  $0.68\pm0.45$  &   $0.89\pm0.54$    & 7.1     \\ 
TOI 1267 b	 &  $ 6.790125\pm0.000029$  &  	$ 2.10 \pm 0.08 $    &  $0.247\pm0.004$   &  $0.31\pm0.20$  &   $0.89\pm0.06$    & 7.1     \\
TOI 1287 b	 &  $ 9.597430\pm0.000242$  &  	$ 0.42 \pm 0.07 $    &  $0.214\pm0.010$   &  $0.61\pm0.38$  &   $0.34\pm0.28$    & 4.7     \\
TOI 1288 b	 &  $ 2.699840\pm0.000009$  &  	$ 2.73 \pm 0.08 $    &  $0.109\pm0.002$   &  $0.82\pm0.03$  &   $0.86\pm0.16$    & 41.9    \\
TOI 1307 b	 &  $ 2.534599\pm0.000053$  &  	$ 6.19 \pm 0.18 $    &  $0.100\pm0.002$   &  $0.87\pm0.01$  &   $1.16\pm0.06$    & 37.1    \\  
TOI 1409 b	 &  $ 4.717965\pm0.000321$  &  	$ 3.00 \pm 0.07 $    &  $0.192\pm0.003$   &  $0.87\pm0.01$  &   $1.51\pm0.09$    & 33.3    \\
TOI 1740 b	 &  $19.428487\pm0.000918$  &  	$ 1.01 \pm 0.22 $    &  $0.154\pm0.012$   &  $0.58\pm0.38$  &   $0.26\pm0.23$    & 3.1     \\ 
\enddata   
\tablenotetext{*}{Not corrected for flux dilution from the stellar companion(s). Radius corrections are discussed in Section~\ref{results}.}
\end{deluxetable*} 

%% file: table_densities.txt
\begin{deluxetable*}{lcccccclc}
\tablewidth{0pt}
\tabletypesize{\small}
\tablecaption{Stellar Densities \& Host Probabilities \label{densities}    }
\tablehead{
\colhead{TOI} 
& \colhead{$\rho_1$ (cgs)} 
& \colhead{$\rho_2$ (cgs)} 
& \colhead{$\rho_3$ (cgs)}
& \colhead{ \ \ $P_1$ \ \ } 
& \colhead{ \ \ $P_2$ \ \ } 
& \colhead{ \ \ $P_3$ \ \ } 
& \colhead{Result} 
& \colhead{$X_R$} 
}
\startdata 
141		& $1.07\pm0.19$	  & $1.70\pm0.27$   & $2.18\pm0.46$  &  0.95	&	0.03	&	0.02       &  Likely primary host     & 1.01     \\
159		& $0.58\pm0.04$	  & $1.43\pm0.15$   & \nodata        &  0.42	&	0.58	&	\nodata    &  Uncertain host          & \nodata  \\
172		& $0.84\pm0.06$	  & $3.38\pm0.40$   & \nodata        &  0.96	&	0.04	&	\nodata    &  Likely primary host     & 1.01     \\
194		& $0.73\pm0.04$	  & $0.97\pm0.06$   & \nodata        &  0.72	&	0.28	&	\nodata    &  Likely primary host     & 1.18     \\
264		& $0.40\pm0.03$	  & $1.46\pm0.19$   & \nodata        &  1.00	&	0.00	&	\nodata    &  Likely primary host     & 1.01     \\
295		& $0.96\pm0.16$	  & $6.83\pm1.21$   & \nodata        &  1.00	&	0.00	&	\nodata    &  Likely primary host     & 1.01     \\
309		& $2.38\pm3.47$	  & $3.62\pm7.09$   & \nodata        &  0.31	&	0.69	&	\nodata    &  Uncertain host          & \nodata  \\
322		& $0.31\pm0.07$	  & $0.36\pm0.08$   & \nodata        &  0.80	&	0.20	&	\nodata    &  Likely primary host     & 1.23     \\
462		& $1.06\pm0.32$	  & $1.09\pm0.32$   & \nodata        &  0.73	&	0.27	&	\nodata    &  Likely primary host     & 1.27     \\
487		& $1.08\pm0.29$	  & $1.16\pm0.29$   & \nodata        &  0.93	&	0.07	&	\nodata    &  Likely primary host     & 1.15     \\
489		& $0.62\pm0.21$	  & $2.60\pm0.57$   & \nodata        &  0.88	&	0.12	&	\nodata    &  Likely primary host     & 1.01     \\
492		& $0.28\pm0.06$	  & $1.16\pm0.18$   & \nodata        &  1.00	&	0.00	&	\nodata    &  Likely primary host     & 1.01     \\
601		& $1.98\pm1.53$	  & $2.56\pm0.91$   & \nodata        &  0.19	&	0.81	&	\nodata    &  Likely secondary host   & 8.08     \\
640		& $0.22\pm0.02$	  & $3.37\pm0.56$   & \nodata        &  0.90	&	0.10	&	\nodata    &  Likely primary host     & 1.01     \\
697		& $1.45\pm0.62$	  & $2.41\pm0.67$   & $2.52\pm0.65$  &  0.87	&	0.10	&	0.02       &  Likely primary host     & 1.01     \\
931		& $1.60\pm0.33$	  & $1.73\pm0.35$   & \nodata        &  0.19	&	0.81	&	\nodata    &  Likely secondary host   & 1.86     \\
1217	& $0.92\pm1.30$	  & $2.13\pm0.82$   & \nodata        &  0.46	&	0.54	&	\nodata    &  Uncertain host          & \nodata  \\
1267	& $0.32\pm0.08$	  & $0.32\pm0.08$   & \nodata        &  0.80	&	0.20	&	\nodata    &  Likely primary host     & 1.31     \\
1287	& $0.52\pm0.34$	  & $0.69\pm0.30$   & \nodata        &  0.84	&	0.16	&	\nodata    &  Likely primary host     & 1.02     \\
1288	& $0.39\pm0.07$	  & $0.81\pm0.11$   & $5.69\pm0.94$  &  0.95	&	0.00	&	0.05       &  Likely primary host     & 1.05     \\
1307	& $0.42\pm0.04$	  & $1.58\pm0.20$   & \nodata        &  0.62	&	0.38	&	\nodata    &  Uncertain host          & \nodata  \\
1409	& $0.10\pm0.01$	  & $0.23\pm0.03$   & \nodata        &  1.00	&	0.00	&	\nodata    &  Likely primary host     & 1.04     \\
1740	& $2.00\pm1.56$	  & $3.47\pm1.78$   & \nodata        &  0.41	&	0.59	&	\nodata    &  Uncertain host          & \nodata  \\
\enddata                                                                                                      
\end{deluxetable*}

%% file: ms.bbl
\begin{thebibliography}{}

\bibitem[Anderson et al.(2015)]{anderson15} Anderson, D.~R., Collier Cameron, A., Hellier, C., et al.\ 2015, \aap, 575, A61. doi:10.1051/0004-6361/201423591

\bibitem[Anderson et al.(2018)]{anderson18} Anderson, D.~R., Bouchy, F., Brown, D.~J.~A., et al.\ 2018, \href{https://ui.adsabs.harvard.edu/abs/2018arXiv181209264A}{arXiv:1812.09264}

\bibitem[Astropy Collaboration et al.(2013)]{astropy1} Astropy Collaboration, Robitaille, T.~P., Tollerud, E.~J., et al.\ 2013, \aap, 558, A33. doi:10.1051/0004-6361/201322068

\bibitem[Astropy Collaboration et al.(2018)]{astropy2} Astropy Collaboration, Price-Whelan, A.~M., Sip{\H{o}}cz, B.~M., et al.\ 2018, \aj, 156, 123. doi:10.3847/1538-3881/aabc4f

\bibitem[Bakos et al.(2012)]{bakos12} Bakos, G. {\'A}., Hartman, J.~D., Torres, G., et al.\ 2012, \aj, 144, 19. doi:10.1088/0004-6256/144/1/19

\bibitem[Bendek et al.(2021)]{bendek21} Bendek, E.~A., Belikov, R., Sirbu, D., et al.\ 2021, \procspie, 11823, 1182311. doi:10.1117/12.2594992

\bibitem[Bonomo et al.(2017)]{bonomo17} Bonomo, A.~S., Desidera, S., Benatti, S., et al.\ 2017, \aap, 602, A107. doi:10.1051/0004-6361/201629882


\bibitem[Bouma et al.(2018)]{bouma18} Bouma, L.~G., Masuda, K., \& Winn, J.~N.\ 2018, \aj, 155, 244. doi:10.3847/1538-3881/aabfb8

\bibitem[Buchhave et al.(2011)]{buchave11} Buchhave, L.~A., Latham, D.~W., Carter, J.~A., et al.\ 2011, \apjs, 197, 3. doi:10.1088/0067-0049/197/1/3

\bibitem[Choi et al.(2016)]{mist2} Choi, J., Dotter, A., Conroy, C., et al.\ 2016, \apj, 823, 102. doi:10.3847/0004-637X/823/2/102

\bibitem[Ciardi et al.(2015)]{ciardi15} Ciardi, D.~R., Beichman, C.~A., Horch, E.~P., et al.\ 2015, \apj, 805, 16




\bibitem[Dotter(2016)]{mist1} Dotter, A.\ 2016, \apjs, 222, 8. doi:10.3847/0067-0049/222/1/8

\bibitem[Dumusque et al.(2019)]{dumusque19} Dumusque, X., Turner, O., Dorn, C., et al.\ 2019, \aap, 627, A43. doi:10.1051/0004-6361/201935457


\bibitem[Eastman et al.(2013)]{exofast} Eastman, J., Gaudi, B.~S., \& Agol, E.\ 2013, \pasp, 125, 83. doi:10.1086/669497

\bibitem[Espinoza et al.(2020)]{espinoza20} Espinoza, N., Brahm, R., Henning, T., et al.\ 2020, \mnras, 491, 2982. doi:10.1093/mnras/stz3150

\bibitem[ExoFOP(2019)]{exofop} ExoFOP. 2019. Exoplanet Follow-up Observing Program - TESS. IPAC. doi: 10.26134/EXOFOP3


\bibitem[Foreman-Mackey et al.(2017)]{celerite1} Foreman-Mackey, D., Agol, E., Ambikasaran, S., et al.\ 2017, \aj, 154, 220. doi:10.3847/1538-3881/aa9332

\bibitem[Foreman-Mackey(2018)]{celerite2} Foreman-Mackey, D.\ 2018, Research Notes of the American Astronomical Society, 2, 31. doi:10.3847/2515-5172/aaaf6c

\bibitem[Furlan et al.(2017)]{furlan17} Furlan, E., Ciardi, D.~R., Everett, M.~E., et al.\ 2017, \aj, 153, 71. doi:10.3847/1538-3881/153/2/71

\bibitem[Furlan \& Howell(2017)]{furlan17b} Furlan, E. \& Howell, S.~B.\ 2017, \aj, 154, 66. doi:10.3847/1538-3881/aa7b70

\bibitem[Furlan \& Howell(2020)]{furlan20} Furlan, E. \& Howell, S.~B.\ 2020, \apj, 898, 47. doi:10.3847/1538-4357/ab9c9c



\bibitem[Gillon et al.(2013)]{gillon2013} Gillon, M., Anderson, D.~R., Collier-Cameron, A., et al.\ 2013, \aap, 552, A82. doi:10.1051/0004-6361/201220561

\bibitem[Guerrero et al.(2021)]{TOIcatalog} Guerrero, N.~M., Seager, S., Huang, C.~X., et al.\ 2021, \apjs, 254, 39. doi:10.3847/1538-4365/abefe1

\bibitem[Harris et al.(2020)]{numpy} Harris, C.~R., Millman, K.~J., van der Walt, S.~J., et al.\ 2020, \nat, 585, 357. doi:10.1038/s41586-020-2649-2

\bibitem[Hartman et al.(2015)]{hartman15} Hartman, J.~D., Bhatti, W., Bakos, G. {\'A}., et al.\ 2015, \aj, 150, 168. doi:10.1088/0004-6256/150/6/168

\bibitem[Hirsch et al.(2017)]{hirsch17} Hirsch, L.~A., Ciardi, D.~R., Howard, A.~W., et al.\ 2017, \aj, 153, 117. doi:10.3847/1538-3881/153/3/117








\bibitem[Howell et al.(2021)]{howell21} Howell, S.~B., Matson, R.~A., Ciardi, D.~R., et al.\ 2021, \aj, 161, 164. doi:10.3847/1538-3881/abdec6

\bibitem[Hunter(2007)]{matplotlib} Hunter, J.~D.\ 2007, Computing in Science and Engineering, 9, 90. doi:10.1109/MCSE.2007.55


\bibitem[Jenkins et al.(2016)]{jenkins16} Jenkins, J.~M., Twicken, J.~D., McCauliff, S., et al.\ 2016, \procspie, 9913, 99133E. doi:10.1117/12.2233418



\bibitem[Kraus et al.(2016)]{kraus16} Kraus, A.~L., Ireland, M.~J., Huber, D., et al.\ 2016, AJ, 152, 8

\bibitem[Lester et al.(2021)]{lester21} Lester, K.~V., Matson, R.~A., Howell, S.~B., et al.\ 2021, \aj, 162, 75.  doi:10.3847/1538-3881/ac0d06



\bibitem[Matson et al.(2018)]{matson18} Matson, R.~A., Howell, S.~B., Horch, E.~P., et al.\ 2018, AJ, 156, 31

\bibitem[Matson et al.(2019)]{matson19} Matson, R.~A., Howell, S.~B., \& Ciardi, D.~R.\ 2019, AJ, 157, 211 

\bibitem[Moe \& Kratter(2021)]{moe21} Moe, M. \& Kratter, K.~M.\ 2021, \mnras, 507, 3593. doi:10.1093/mnras/stab2328



\bibitem[Payne et al.(2018)]{payne18} Payne, A.~N., Ciardi, D.~R., Kane, S.~R., et al.\ 2018, \aj, 156, 209. doi:10.3847/1538-3881/aae310


\bibitem[Pecaut \& Mamajek(2013)]{pecaut13} Pecaut, M.~J. \& Mamajek, E.~E.\ 2013, \apjs, 208, 9



\bibitem[Ricker et al.(2014)]{tess} Ricker, G.~R., Winn, J.~N., Vanderspek, R., et al.\ 2014, \procspie, 9143, 914320. doi:10.1117/12.2063489


\bibitem[Rodriguez et al.(2019)]{rodriguez19} Rodriguez, J.~E., Quinn, S.~N., Huang, C.~X., et al.\ 2019, \aj, 157, 191. doi:10.3847/1538-3881/ab11d9

\bibitem[Rodriguez et al.(2021)]{rodriguez21} Rodriguez, J.~E., Quinn, S.~N., Zhou, G., et al.\ 2021, \aj, 161, 194. doi:10.3847/1538-3881/abe38a

\bibitem[Savel et al.(2020)]{saval20} Savel, A.~B., Dressing, C.~D., Hirsch, L.~A., et al.\ 2020, \aj, 160, 287. doi:10.3847/1538-3881/abc47d

\bibitem[Seager \& Mall{\'e}n-Ornelas(2003)]{seager03} Seager, S. \& Mall{\'e}n-Ornelas, G.\ 2003, \apj, 585, 1038. doi:10.1086/346105

\bibitem[Stassun et al.(2017)]{stassun17} Stassun, K.~G., Collins, K.~A., \& Gaudi, B.~S.\ 2017, \aj, 153, 136. doi:10.3847/1538-3881/aa5df3


\bibitem[Stassun et al.(2019)]{tic81} Stassun, K.~G., Oelkers, R.~J., Paegert, M., et al.\ 2019, \aj, 158, 138. doi:10.3847/1538-3881/ab3467

\bibitem[Virtanen et al.(2020)]{scipy} Virtanen, P., Gommers, R., Oliphant, T.~E., et al.\ 2020, Nature Methods, 17, 261. doi:10.1038/s41592-019-0686-2

\bibitem[Ziegler et al.(2020)]{ziegler20} Ziegler, C., Tokovinin, A., Brice{\~n}o, C., et al.\ 2020, \aj, 159, 19

\bibitem[Ziegler et al.(2021)]{ziegler21} Ziegler, C., Tokovinin, A., Latiolais, M., et al.\ 2021, arXiv:2103.12076


\end{thebibliography}
